\documentclass[onecolumn,superscriptaddress,aps,pre,nofootinbib]{revtex4}
\usepackage{color}
\usepackage{amsmath}
\usepackage{amssymb}
\usepackage{graphicx}
\usepackage{subfigure}
\usepackage{ulem}

\usepackage{lscape}

\usepackage{lineno}
\usepackage{appendix}

\newcommand{\MM}{Methods}
\newcommand{\SI}{Supplementary Information}
\usepackage[font={small,it,singlespacing}]{caption}
\newcommand{\ud}{\mathrm{d}}

\newcommand{\mat}[1]{\boldsymbol{#1}}
\newcommand{\vect}[1]{\boldsymbol{#1}}

\DeclareMathOperator*{\vol}{\mathrm{vol}}
\DeclareMathOperator\erfc{erfc}

\usepackage{lscape}

\begin{document}

%\linenumbers

\title{The geometry of coexistence in large ecosystems}

\author{Jacopo Grilli}
\email{jgrilli@uchicago.edu}
\affiliation{Dept.~of Ecology \& Evolution, University of Chicago. Chicago, IL 60637, USA}

\author{Matteo Adorisio}
%\affiliation{SISSA (Scuola Internazionale Superiore di Studi Avanzati), I-34136 Trieste, Italy.}
\affiliation{Dept.~of Physics and Astronomy ``Galileo Galilei'', Universit\`a degli Studi di Padova, Padova, Italy}

\author{Samir Suweis}
\affiliation{Dept.~of Physics and Astronomy ``Galileo Galilei'', Universit\`a degli Studi di Padova, Padova, Italy}

\author{Gyuri Barab\'as}
\affiliation{Dept.~of Ecology \& Evolution, University of Chicago. Chicago, IL 60637, USA}

\author{Jayanth R. Banavar}
\affiliation{Department of Physics, University of Maryland, College Park, MD 20742, USA}

\author{Stefano Allesina}
\affiliation{Dept.~of Ecology \& Evolution, University of Chicago. Chicago, IL 60637, USA}
\affiliation{Computation Institute, University of Chicago}

\author{Amos Maritan}
\affiliation{Dept.~of Physics and Astronomy ``Galileo Galilei'', Universit\`a degli Studi di Padova, Padova, Italy}

\begin{abstract}
  The role of species interactions in controlling the interplay
  between the stability of an ecosystem and its biodiversity is still
  not well understood. The ability of ecological communities to
  recover after a small perturbation of the species abundances (local
  asymptotic stability) has been well studied, whereas the likelihood
  of a community to persist when the interactions are altered
  (structural stability) has received much less attention.  Our goal
  is to understand the effects of diversity, interaction strenghts and
  ecological network structure on the volume of parameter space
  leading to feasible equilibria, i.e., ones in which all populations
  have positive abundances. We develop a geometrical
  framework to study the range of conditions necessary for feasible
  coexistence in both mutualistic and consumer--resource systems. Using
  analytical and numerical methods, we show that feasibility is
  determined by just a handful of quantities describing the
  interactions, yielding a nontrivial \textit{complexity--feasibility}
  relationship. Analyzing more than 100 empirical networks, we show
  that the range of coexistence conditions in mutualistic systems can
  be analytically predicted by means of a null model of random interactions,
  whereas food webs are characterized by smaller coexistence domains
  than those expected by chance. Finally, we characterize the
  geometric shape of the feasibility domain, thereby identifying the
  direction of perturbations that are more likely to cause
  extinctions. Interestingly, the structure of mutualistic
  interactions leads to very heterogeneous responses to perturbations,
  making those systems more fragile than expected by chance.
\end{abstract}

\maketitle

Natural populations are faced with constantly varying environmental
conditions. Environmental conditions affect physiological parameters
(e.g., metabolic rates~~\cite{Gillooly2001}) as well as ecological ones
(e.g., the presence and strength of interactions between
populations~\cite{Walther2002, Tylianakis2008, Harmon2009,
  Tylianakis2010}). Therefore in order to persist, ecological
communities necessarily need, at the very least, to be able to cope
with small environmental changes.  Mathematically, this translates
into an argument on the robustness of the qualitative behavior of an
ecological dynamical system: to guarantee robust coexistence, a model
describing an ecological community needs at least to be
(qualitatively) insensitive to small perturbations of the parameters
~\cite{Meszenaetal2006, Barabasetal2014ELE}. This notion has been
formalized in the measure of 
robustness~\cite{Barabas2012} or ``structural
stability''~\cite{Rohr2014}, expressed as the volume of the parameter
space resulting in the coexistence of all populations in a community.

While the local asymptotic stability (the ability to recover after a
small change in the population abundances) of ecological communities
has been studied in small~\cite{May1975} and large~\cite{May1972,
  Allesina2012, suweis2014disentangling, Allesina2015a} systems, the
study of \textit{structural stability} (i.e., the ability of a community to retain the
same dynamical behavior if conditions are slightly altered)---despite
being proposed early on as a key feature in the context of the
diversity-stability debate~\cite{MacArthur1955, GILPIN1975,
  ROBERTS1974, Goh1977a}---has historically been restricted to the
case of small communities, with the first studies of larger
communities appearing only recently~\cite{Rohr2014,Saavedra2016}, and---because of
mathematical limitations---dealing exclusively with the case of large
mutualistic communities. Studies of structural stability have so far
focused on the effect of ecological network structure (who interacts
with whom) on the volume of parameter space leading to \textit{feasible}
equilibria, in which all populations have positive abundances.

Here we develop a geometrical framework for studying the feasibility
of large ecological communities. We overcome the limitations that have
hitherto prevented the study of consumer-resource networks, thereby
providing a unified view of feasibility in ecological systems. Using a
random matrix approach (which helped identify main drivers of local
asymptotic stability), we pinpoint the key quantities controlling the
volume of parameter space leading to feasible communities, as well as
its sensitivity to changes in these parameters. We then contrast these
expectations for randomly connected systems with simulations on
structured empirical networks, quantifying the effects of network
structure on feasibility.

\section*{Theoretical framework}

For simplicity, we consider a community composed of $S$ species whose
dynamics is determined by a system of autonomous ordinary differential
equations:
\begin{equation}
  \frac{d n_i}{dt} = n_i \left( r_i + \sum_{j=1}^S A_{ij} n_j \right)
\ ,
  \label{eq:lotkavolt}
\end{equation}
where $n_i$ is the density of population $i$, $r_i$ is its intrinsic
growth rate, and $A_{ij}$ (which in principle could depend on
$\vect{n}$; see \SI) measures the interaction strength between population $i$ and $j$. A fixed point
$\vect{n}^\ast$ (i.e., a vector of densities making the right side of
each equation zero) is \textit{feasible} if $n_i^\ast > 0$ for every
population. A fixed point is \textit{locally asymptotically stable} if,
following any sufficiently small perturbation of the densities, the
system returns to a small vicinity of the fixed point. The fixed point is \textit{globally}
asymptotically stable if the system eventually return to it, starting
from any positive initial condition within a finite domain. A system
with a fixed point is \textit{structurally stable} if, following a sufficiently small change in
the growth rates $r_i$, the new fixed point is still feasible and
stable.

To study the range of conditions leading to stable coexistence, we
need to disentangle feasibility and local stability. This problem is
well discussed in Rohr \textit{et al.}~\cite{Rohr2014}, where it was
solved for the case of one possible parameterization of mutualistic
interactions. If $\mat{A}$ is \textit{diagonally stable} or \textit{Volterra
dissipative} (i.e., there exists a positive diagonal matrix $\mat{D}$
such that $\mat{D}\mat{A}+\mat{A}^T\mat{D}$ is stable), then any
feasible fixed point is globally stable~\cite{Volterra1931,Goh1977}. Unfortunately, a general
characterization of this class of matrices is
unknown~~\cite{Logofet2005}.  We proceeded therefore by considering
only the matrices such that all the eigenvalues of $\mat{A}+\mat{A}^T$
are negative (i.e., the matrix $\mat{A}$ is negative definite in a generalized sense~~\cite{Johnson1970},
corresponding to $\mat{D}$ being equal to the identity matrix; see
\MM).  This choice reduces the number of parameterizations one can
analyze, as not all the diagonally stable matrices are
negative definite. However, as shown in the \SI, only very few
parameter combinations are excluded from this set. Moreover, the
effects of negative definitness are well--studied for random
matrices~~\cite{Tang2014}, and by using it we can extend the study of
feasibility to any ecological network, including food webs.

Our goal is to measure the fraction of growth rate combinations, out
of all possible combinations, that lead to the coexistence of all $S$
populations. Since we can separate stability and feasibility, we only
need to find those $r_i$ leading to feasible fixed points, and the
condition above ensures that these will be globally stable. As pointed
out before~\cite{Rohr2014}, the problem is not to find a particular set
of $r_i$ leading to coexistence, but rather to measure how flexibly
one may choose these rates. As shown in Fig.~\ref{fig:volumes}, this
quantity--indicated by $\Xi$ henceforth--can be thought of as a
volume, or more precisely a solid angle, in the space of growth
rates~\cite{Svirezhev1978} (see \MM).

To calculate $\Xi$, one might naively wish to perform direct numerical
computation of the fraction of growth rates leading to a feasible
equilibrium. While a direct calculation is viable when $S$ is
sufficiently small, this procedure becomes extremely inefficient for
large $S$~\cite{Rohr2014}.
% Previous approaches have therefore relied on indirect
% quantifications of this space~\cite{Rohr2014, Saavedra2014}, so that,
% rather than calculating $\Xi$ directly, one might correlate it with
% properties of interest.
In the \SI\ we introduce a method
that can be used to efficiently calculate $\Xi$ with arbitrary
precision, even for large $S$.  Using this method, we can accurately
measure the size of the feasibility domain, with larger values of
$\Xi$ corresponding to larger proportions of conditions (intrinsic
growth rates) compatible with stable coexistence. For reference, we
normalize $\Xi$ so that $\Xi = 1$ when populations are self--regulated
and not interacting (\MM), i.e., when the interaction matrix $\mat{A}$
is a negative diagonal matrix, and thus Eq.~\ref{eq:lotkavolt}
simplifies to $S$ independent logistic equations.

\section*{Results}

\subsection*{Feasibility is universal for random matrices.}

May's seminal work~\cite{May1972} pioneered the use of random matrices
as a reference, or null model, of ecological interactions. A
particularly interesting feature of random matrices is that the
distribution of their eigenvalues (determining local stability) is
\textit{universal}~\cite{Allesina2015}. This means that local stability
depends on just a few, coarse--grained properties of the matrix (i.e.,
the number of species and the first few moments of the distribution of
interaction strengths) and not on the finer details (e.g., the
particular distribution of interaction strengths; see \SI).  In fact,
these moments can be combined into just three parameters: $E_1$,
$E_2$, and $E_c$ (\MM). Together with $S$, they completely determine
local asymptotic stability.

We tested whether universality also applies to feasibility.  We
considered different random matrix ensembles obtained for different
connectance values and distributions from which the matrix entries
were drawn, but with constant values of $S$ and of $E_1$, $E_2$, and
$E_c$. We then checked whether the size $\Xi$ of the feasibility
domain depended only on these four quantities or also on finer
details.  Surprisingly, we found that the feasibility of random
matrices is also universal (\MM \ and \SI).  Two very different
(random) ecosystems, with completely different interaction types and
distributions of interaction strengths, but having the same number of
species $S$ and the same $E_1$, $E_2$, and $E_c$, have the same $\Xi$
in the large $S$ limit. This result has important theoretical
implications, as it indicates those moments as the drivers of
feasibility, but also very practical consequences, namely that the
parameter space one needs to explore is dramatically reduced.

\subsection*{An analytical complexity--feasibility relationship}

The universality of $\Xi$ suggests that it is amenable to analytical
treatment. As explained in the \SI \ and shown in
Fig.~\ref{fig:sstabtot}, when the mean and variance of interaction
strengths are not too large (\SI), we are able to derive the following
approximation for $\Xi$ for random interaction matrices $\mat{A}$:
\begin{equation}
  \Xi \sim \Bigl( 1 + \frac{1}{\pi} \frac{E_1 (2 d - S E_1) }{ d - S E_1^2  } \Bigr)^{S} \ ,
  \label{eq:prediction}
\end{equation}
where $S$ is is the number of species, $d$ is the mean of $\mat{A}$'s
diagonal entries, and $E_1 = C \mu$, the product of the connectance
$C$ and the average interaction strength $\mu$ (see \MM).  A more
accurate formula is presented in the \SI.

In analogy with the celebrated result of May~\cite{May1972} connecting
stability and complexity, Eq.~\ref{eq:prediction} can be considered as
a \textit{complexity--feasibility} relationship.  While in May's scenario
and in its generalizations~\cite{Allesina2012} the effect of complexity
and diversity on stability is always detrimental, it does depend on
the interaction type in the case of feasibility. Given that $d$ is
negative by construction (\SI), having more species or connections can
either increase ($E_1 > 0$) or shrink ($E_1 < 0$) the size of the
feasibility domain, as a function of the sign of interaction strenghts (see
Fig.~\ref{fig:sstabtot}).  It is important to stress that we computed
$\Xi$ under the assumption of $\mat{A}$ being negative definite. When
we consider how $\Xi$ depends on $S$ and other parameters, we need to
take into account the conditions making the matrix negative definite
(see \MM \ and \SI). In the case of positive interaction strengths,
this condition is $d + S C \mu < 0$, implying an upper bound for $\mu$
that depends on $S$.

\subsection*{Our analytical formula predicts feasibility of empirical mutualistic networks and overestimates that of food webs}

Having explored the feasibility of random networks, we proceed to
investigate the effects of incorporating empirical network
structure. Ecological networks are in fact
non--random~\cite{Cohenetal1990, Williams2000, Bascompte2003}, and many
studies have hypothesized that the structure of interactions could
increase the likelihood of coexistence~\cite{Bascompte2006,
  Bastolla2009, Thebault2010}.  Having an analytical prediction for
random matrices, we can study whether it predicts the size of the
feasibility domain for empirical networks as
well. Fig.~\ref{fig:sstabtot} shows the simulated values of $\Xi$ for
89 mutualistic networks and 15 food webs (\SI), parameterized multiple
times and compared with our analytical approximation (see \MM). We
find that $\Xi$ of empirical mutualistic networks is well predicted by
our formula, while it overestimates the feasibility domain of food
webs, indicating that their non--random structure has a strong negative
effect on feasibility.

In the \SI, we compare the effect of the empirical structure of
mutualistic networks with randomizations, by controlling for the
interaction strengths. We show that, in the absence of variability in
interaction strengths, the structure of empirical mutualistic networks
has a positive effect on feasibility, which is strongly reduced when
interaction strengths are allowed to vary. While this effect of
empirical mutualistic networks is statistically significant, its
effect on $\Xi$ is negligible compared to the effect of mean
interaction strengths, and can only be detected by controlling very
precisely for interaction strengths (see \SI). On a broader scale, as
shown in Fig.~\ref{fig:sstabtot}, the size of the feasibility domain
of empirical networks is well predicted by our analytical formula.

On the other hand, the negative effect of food web structure on $\Xi$
is substantial. In the \SI\ we compare each network with
randomizations and also with predictions of the cascade
model~\cite{Cohenetal1990}, which has recently been shown to predict
well the stability of empirical food webs~\cite{Allesina2015a}. By
analyzing different parameterizations we found that the feasibility
domain of empirical structures is consistently and significantly
smaller than that of both the randomizations and the cascade
model. For most of the webs, the prediction obtained from the cascade
model is better than that of randomizations, suggesting that the
directionality of empirical webs plays a role in reducing feasibility,
with other properties of the structure of empirical networks also
contributing significantly to feasibility.

\subsection*{The shape of the feasibility domain carries information on the response to perturbations, and can be analytically predicted for random interactions}

\begin{figure*}[hp]
\centering
  \includegraphics[width = 0.95\columnwidth]{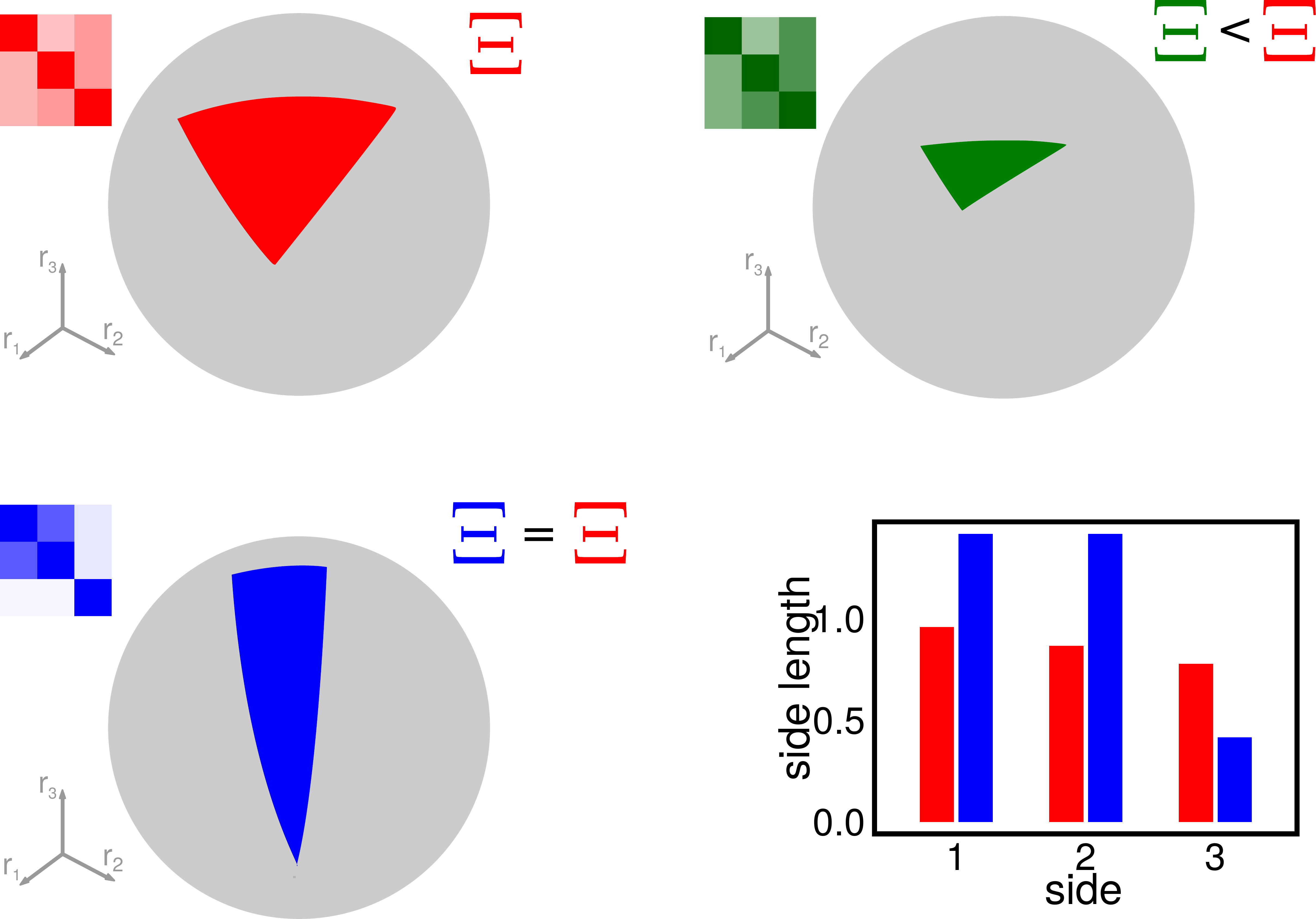}
  \caption{\textbf{Geometrical properties of feasibility.}  The panels
    show the size and shape of the feasibility domain for three
    interaction matrices, each defining the interactions between three
    populations. If $\mat{r}$ corresponds to a feasible equilibrium,
    so does $c\mat{r}$ for any positive $c$; one can therefore study
    the feasibility domain on the surface of a
    sphere~\cite{Svirezhev1978} (\SI). The gray sphere represents the
    $S=3$-dimensional space of growth rates, while the colored part
    corresponds to the combination of growth rates leading to stable
    coexistence.  The area (or volume for higher--dimensional systems)
    of the colored part is measured by $\Xi$. Larger values of $\Xi$
    correspond to a higher fraction of growth rate combinations
    leading to coexistence: the red interaction matrix is therefore
    more robust against perturbations of $\mat{r}$ than the green one.
    The size of this region (i.e., the value of $\Xi$) does not
    capture all the properties relevant for coexistence. The red and
    blue systems have the same $\Xi$, but the two regions--despite
    having the same area--have very different shapes, summarized in
    the bottom--right panel, where we show the length of each side for
    the red and blue systems. In the red system, the three sides have
    about the same length, and thus moving from the center in any
    direction will have about the same effect. In the blue system
    however, one side is much shorter than the other two, implying
    that even small perturbations falling along this direction may
    drive the system outside the feasibility domain. One of our main
    results is that, roughly speaking, if the red system corresponds
    to the random case, then the green one to food webs (having the
    same heterogeneity in side lengths as the random case but with a
    smaller $\Xi$ overall), and the blue one to empirical mutualistic
    networks ($\Xi$ rougly the same as in the random case but with the
    heterogeneity in side lengths much greater).}
\label{fig:volumes}
\end{figure*}

So far, we have focused on the volume of the parameter space resulting
in feasiblity. However, two systems having the same $\Xi$ can still
have very different responses to parameter perturbations, just as two
triangles having the same area need not to have sides of the same
length (Fig.~\ref{fig:volumes}). The two extreme cases correspond to
a) an isotropic system in which, if we start at the barycenter of the
feasibility domain, moving in any direction yields roughly the same
effect (equivalent to an equilateral triangle); b) anisotropic
systems, in which the feasibility domain is much narrower in certain
directions than in others (as in a scalene triangle). For our problem,
the domain of growth rates leading to coexistence is---once the growth
rates are normalized---the $(S-1)$--dimensional generalization of a
triangle on a hypersphere. For $S=3$, this domain is indeed a triangle
lying on a sphere as shown in Fig.~\ref{fig:volumes}. If all the
$S(S-1)/2$ sides of this (hyper--)triangle are about the same length,
then different perturbations will have similar effects on the
system. On the other hand, if some sides are much shorter than others,
then there will be changes of conditions which will more likely impact
coexistence than others. We therefore consider a measure of the
heterogeneity in the distribution of the side lengths
(Fig.~\ref{fig:volumes} and \SI). The larger the variance of this
distribution, the more likely it is that certain perturbations can
destroy coexistence, even when $\Xi$ is large and the perturbation
small. This way of measuring heterogeneity is particularly convenient
because it is independent of the initial conditions. Moreover, the
length of each side can be directly related to the similarity between
the corresponding pair of species (\SI), drawing a strong connection
between the parameter space allowing for coexistence and the
phenotypic space.
%%% add more here
As in the case of $\Xi$, this measure is a function of the interaction
matrix and corresponds to a geometrical property of the coexistence
domain.

While $\Xi$ is a universal quantity for random networks, the
distribution of side lengths is not: it depends on the full
distribution of interaction strengths (\SI). On the other hand, as
shown in the \SI, it is possible to compute it analytically in full
generality, i.e., for any distribution of interaction strengths and
any interaction types. In particular we are able to obtain an
expression for its mean and variance, which depend only on $S$, $E_1$,
$E_2$, and $E_c$ (\SI). Fig.~\ref{fig:sside} shows that the analytical
formula, in the case of random $\mat{A}$, matches the observed mean
and variance of side lengths of random networks perfectly.

\subsection*{Empirical network structures correspond to more heterogeneous shapes}

As we have done for $\Xi$, we can now test how non--random empirical
network topologies influence the distribution of side
lengths. Fig.~\ref{fig:sside} shows that empirical food webs and, in
particular, empirical mutualistic networks display a much larger
variation in side lengths than expected by chance. This result is
particularly relevant, indicating that even if the feasibility domains
of empirical mutualistic networks are larger than those of random
networks, their shapes are less regular than expected by chance, and
thus we expect perturbations in certain directions to quickly lead out
of the feasible domain of growth rates.

\section*{Discussion}

A classic problem in mathematical ecology is determining the response
of systems to perturbations of model parameters. In the community
context, one important application is getting at the range of
parameters allowing for species coexistence~\cite{ArmstrongMcGehee1976,
  ArmstrongMcGehee1980, Abrams1983}. Several methods exist for
evaluating this range~\cite{Yodzis1988, Dambacheretal2002,
  Aufderheideetal2013, Barabasetal2014ELE}, but they either rely on
raw numerical techniques, or else can only evaluate system response to
small parameter perturbations. Here, in the context of the general
Lotka--Volterra model, we have given a method for the global
assessment of all combinations of species' intrinsic growth rates
compatible with coexistence---what we have called the domain of
feasibility. Our geometrical approach can determine not only the total
size of the feasibility domain, but also its shape: it is always a
simply connected domain forming a convex polyhedral cone whose side
lengths can be evaluated from the interaction matrix. Applying our
method to empirical interaction networks, we were able to characterize
the region of parameter space compatible with coexistence; the
importance of this kind of information is underlined by a rapidly
changing environment that is expected to cause substantial shifts in
the parameters influencing these systems.

The geometrical framework we employed, pioneered by Svirezhev and
Logofet~\cite{Svirezhev1978}, allows for the formulation of a
\textit{complexity--feasibility} relationship. In analogy with the
celebrated complexity--stability relationship, it relates the size of
the feasibility domain with diversity, connectance and interaction
strengths of a random interacting community.  While communities are
not random, this relationship sets a null expectation for the scaling
of the proportion of feasible conditions. We obtain that the mean of
interaction strengths sets the behavior of feasibility with the number
of species. If the mean is negative (e.g., in case of competition or
predation with limited efficiency), the larger the system is, the
smaller is the set of conditions leading to coexistence, while for
positive mean (e.g., in the case of mutualism) the converse is true.

Several recent works have studied the effect of network structure on
coexistence in species--rich communities, with contrasting results
~\cite{Bascompte2006, Bastolla2009, Thebault2010, Suweis2013,
  Rohr2014}.  Here we have shown that the fraction of conditions
compatible with coexistence is mainly determined by the number and the
mean strength of interactions. In terms of network properties, the
relevant quantity is the connectance, with other properties (e.g.,
nestedness or degree distribution) having minimal effects. In
particular, once the connectance and mean interaction strength are
fixed, the matrices built using empirical mutualistic networks have
feasibility domains very similar to that expected for the random case,
as was also observed previously in a similar context~\cite{James2012}.

The empirical network structure of mutualistic networks has a
statistically significant effect on the size of the feasibility domain
(see \SI). Whether this effect is ecologically relevant depends on the
specific application at hand. For instance, the effect of structure
could be neglected to quantify how the feasibility domain would change
if a fraction of pollinators went extinct, and it could be evaluated
using our analytical result.  In contexts where the interaction
strengths are strongly constrained, structure would play an important
role.  Our method provides, in this respect, a direct way of
quantifying the importance of different factors, disentangling the way
different interaction properties affect feasibility.

For mutualistic interaction networks, our results clearly show which
properties determine the global health of the community, and therefore
indicate which properties should be measured in the field. While not
observing a link or measuring a wrong interaction coefficient could
have strong effects on ecosystem dynamics, they have very little
effect on how the community copes with environmental perturbations and
how likely extinctions are~\cite{BarabasAllesina2015}. The major role
is played by corse--grained statistical properties of the interactions,
such as connectance or the mean and variance of the interaction
strengths.

For food webs, on the other hand, empirical systems tend to have
feasibility domains smaller than either their random counterparts or
models conserving the directionality of interactions (cascade
model). It is an open question which properties of real food webs are
responsible for restricting the feasibility domain in this way. A
possible candidate is the group structure observed in food
webs~\cite{allesina2009food}, corresponding to larger similarity of how
certain species interact with the rest of the system than expected by
chance, which in turn reduces the size of the feasibility domain (see
\SI).

These results parallel those for the distribution of the side lengths
of the convex polyhedral cone delimiting the feasibility domain. The
variance of side lengths for empirical structures is much higher than
in random networks. This implies that even if the total size of the
feasibility domain is large, it will have a distorted shape that is
very stretched along some directions and shortened along others
(Fig.~\ref{fig:volumes}). Consequently, it will be possible to find
parameter perturbations of small magnitude that will drive the system
outside its feasibility domain~~\cite{Suweis2015}.

We have shown that each side of the feasibility domain corresponds to
a pair of species, with the length determined by how similarly the two
species interact with the rest of the system. As two species interact
more and more similarly (i.e., have a larger niche overlap), the
corresponding side becomes shorter and shorter, which in turns means
greater sensitivity to parameter perturbations. Consistently with
earlier results~\cite{Barabas2012, Barabasetal2014ELE}, this fact
establishes a relationship between niche overlap and the range of
conditions that lead to coexistence: greater niche overlap means a
more restricted parameter range allowing for coexistence, irrespective
of the details of the interactions.

These differences between the size and shape of the feasibility domain
shed light on the contrasting results obtained in the past on the
effect of network structure on persistence~\cite{Bascompte2006,
  Bastolla2009, Thebault2010, James2012, Suweis2013,Suweis2015}. Most of these
studies rely on numerical integration, and therefore strongly depend
on initial conditions. Given the difference in the shape of the
feasibility domains of random and empirical networks, different
initial conditions and their perturbations could result in markedly
different outcomes: the feasibility domain could appear to be large or
small depending on the direction in which perturbations are made.

% Having established the general geometrical properties of the
% feasibility domain, we are in a much better position to critically
% evaluate the feasibility domains of real ecological communities.

% In conclusion, we have provided, for the first time, a general and
% \textit{global} characterization of the feasibility domain in simple
% geometrical terms. We consider this a first step along the way of
% describing feasibility in more complex models and ecological
% scenarios.

\section*{Methods}

\subsection*{Disentangling stability and feasibility}
From Eq.~\ref{eq:lotkavolt}, a feasible fixed point, if it exists, is
given by the solution of
\begin{equation}
  \sum_{j=1}^S A_{ij} n^\ast_j = - r_i \ ,
\end{equation}
where the asterisk denotes equilibrium values. A fixed point is
locally asymptotically stable if all eigenvalues of the community
matrix
\begin{equation}
  M_{ij} = n_i^\ast A_{ij}
\end{equation}
have negative real parts.  As discussed in the \SI, if $\mat{A}$ is
diagonally stable or Volterra dissipative (i.e., there exists a
positive diagonal matrix $\mat{D}$ such that
$\mat{D}\mat{A}+\mat{A}^T\mat{D}$ is stable), then a feasible fixed
point is globally stable in $\mathbb{R}_+$.

A general characterization of diagonally stable matrices is unknown
for $S>3$~~\cite{Logofet2005}.  There exist
algorithms~~\cite{Redheffer1985} that reduce the problem of determining
if a $S \times S$ matrices is diagonally stable into two simultaneous
problems of $(S-1) \times (S-1)$ matrices. While this method can be
efficiently used to determine the diagonal stability of $4 \times 4$
matrices, it becomes computationally intractable for large $S$.

A matrix $\mat{A}$ is negative definite if
\begin{equation}
  \sum_j x_i A_{ij} x_j < 0 \ ,
\end{equation}
for any non--zero vector $\vect{x}$. A necessary and sufficient
condition for a real matrix $\mat{A}$ to be negative definite is that
all the eigenvalues of $\mat{A}+\mat{A}^T$ are
negative~\cite{Johnson1970}.  A negative definite matrix is also
diagonally stable, as the condition for diagonal stability holds with
$\mat{D}$ being the identity matrix.  Since it is extremely simple to
verify this condition and it has been characterized for random
matrices, we will study feasibility of negative definite matrices. In
the \SI \ we show that with this choice we are excluding only a small
region of the parameter space.

\subsection*{Size of the feasibility domain}
The quantity $\Xi$ is the proportion of intrinsic growth rates leading
to feasible equilibria. While a more rigourous definition is presented
in the \SI, with a slight abuse of notation, $\Xi$ can be thought of
as
\begin{equation}
  \Xi = 2^{S} \frac{\text{\# growth rate vectors leading to feasible
      equilibrium} }{\text{total \# growth rate vectors} } .
  \label{eq:xidef}
\end{equation}
The factor $2^S$ is an arbitrary choice that does not affect the
results. It has been introduced to have $\Xi = 1$ in absence of
interspecific interactions ($A_{ij}=0$ if $i\neq j$ in
Eq.~\ref{eq:lotkavolt}) and when all the species are self--regulated
($A_{ii}<0$ if $i\neq j$ in Eq.~\ref{eq:lotkavolt}).  Given the
geometrical properties of the feasibility domain, the proportion of
feasible growth rates can be calculated considering only growth rate
vectors of length one (Fig.~\ref{fig:volumes} and \SI), as this choice
does not affect the value given by Eq.~\ref{eq:xidef}. In the \SI \ we
provide an integral formula for $\Xi$~~\cite{Ribando2006,Gourion2010} which makes both numerical and
analytical calculations possible.

Our method is still valid if some of the species are not
self--regulated (i.e., $A_{ii} = 0$ for some $i$).  In the \SI \ we
explicitly discuss the properties of the feasibility domain of a
community with consumer--resource interactions.  In that case, $\Xi =
0$ either when the diversity of consumers exceeds the diversity of
resources or in the absence of interspecific interactions. Since
consumers are regulated by their resources, they cannot survive in
their absence and should therefore be characterized by negative
intrinsic growth rates. We observe indeed that a necessary condition
for an intrinsic growth rate vector to be contained in the feasibility
domain, is to have negative values for the components corresponding to
consumers.

\subsection*{Random matrices and moments}

$E_1$, $E_2$, and $E_c$ are moments of the random distribution for the
off--diagonal elements of the interaction matrix, and are simply and
directly related to the interaction strengths. They can be calculated
as
\begin{equation}
\begin{split}
  E_1 & = \frac{1}{S(S-1)} \sum_{i \neq j} A_{ij} \ , \\
  E_2^2 & = \frac{1}{S(S-1)} \sum_{i \neq j} A_{ij}^2 - E_1^2 \ , \\
  E_c & = \frac{1}{S(S-1)E_2^2} \sum_{i \neq j} A_{ij}A_{ji} -
  \frac{E_1^2}{E_2^2} \ .
\end{split}
\end{equation}
For random networks with connectance $C$, these expressions reduce
to~\cite{Allesina2015}
\begin{equation}
\begin{split}
  E_1 & = C \mu \ , \\
  E_2^2 & = C(1-C)\mu^2 + C \sigma^2 \ , \\
  E_c & = \frac{\rho\sigma^2 + (1-C)\mu^2}{\sigma^2 + (1-C)\mu^2 } \,
\end{split}
\end{equation}
where $\mu$ is the mean of the interaction strengths, $\sigma$ is
their variance, and $\rho$ is the average pairwise correlation between
the interaction coefficients of species pairs~\cite{Allesina2015}.

\subsection*{Universality of the size of the feasibility domain}

The size of the feasibility domain should, at least in principle,
depend on all the entries of the interaction matrix.  When these
elements are drawn from a distribution, the size $\Xi$ of the
feasibility domain is then expected to depend on all the moments of
that distribution. As $S$ increases, the dependence of $\Xi$ on some
of those moments and parameters might become less and less important.
$\Xi$ is universal if, in the limit of large $S$, it depends only on a
few properties of the interaction matrix (i.e., on just the first few
moments of the distribution).

Specifically, for each unique pair of species $(i,j)$, we set
$A_{ij}=0$ with probability $1-C$ and assign a random pair of
interaction strengths $(M_{ij}, M_{ji}) = (x,y)$ with probability
$C$. The pair $(x,y)$ is drawn from a bivariate distribution with
given mean $\mu$, variance $\sigma$, and correlation $\rho$ between
$x$ and $y$~~\cite{Allesina2015}. By considering different bivariate
distributions, we can analyze the effect of different sign patterns
(e.g., only $(+,-)$ or $(+,+)$ interactions) and different marginal
distributions (e.g., drawing elements from a uniform or a lognormal
distribution).

Non--universality of $\Xi$ would mean that it depends on all the fine
details of the parameterization:
\begin{equation}
  \Xi = f\left(S, \mu,\sigma, \rho,C, \text{sign pattern}, \dots \right)  \ ,
\end{equation}
where $f(\cdot)$ is an arbitrary function. The dependence on $\mu$,
$\sigma$, and $\rho$ can, without loss of generality, be expressed in
terms of $E_1$, $E_2$, and $E_c$:
\begin{equation}
  \Xi = g\left(S, E_1,E_2,E_c,C, \text{sign pattern}, \dots \right)  \ .
\end{equation}

However, if $\Xi$ \textit{is} universal, then for large $S$, it is
possible to express it as a function of $E_1$, $E_2$, and $E_c$ only:
\begin{equation}
  \Xi = h\left(S, E_1,E_2,E_c \right)  \ .
\end{equation}
To verify this conjecture, we calculated $\Xi$ for matrices with the
same values of $E_1$, $E_2$, and $E_c$ that differed for the values of
the other parameters. As extensively shown in the \SI, $\Xi$ is
uniquely determined by $S$, $E_1$, $E_2$, and $E_c$
(Eq.~\ref{eq:prediction}).

\subsection*{Parameterization of mutualistic networks}
 
The 89 mutualistic networks (59 pollination networks and 30
seed--dispersal networks) were obtained from the Web of Life dataset
(\texttt{www.web-of-life.es}), where references to the original works
can be found. Empirical networks are encoded in terms of adjacency
matrices $\mat{L}$: $L_{ij} = 1$ if species $j$ interact with species
$i$ and $0$ otherwise.  When the original network was not fully
connected, we considered the largest connected component.

In the case of mutualistic networks, the adjacency matrix $\mat{L}$ is
bipartite, i.e., it has the structure
\begin{equation}\label{eq:adjblock}
  \mat{L} = \left(
    \begin{array}{cc}
      0 & \mat{L}_b \\
      \mat{L}_b^T & 0
    \end{array}
  \right) \ ,
\end{equation}
where $\mat{L}_b$ is a $S_A\times S_P$ matrix ($S_A$ and $S_P$ being
the number of animals and plants, respectively).  The adjacency matrix
contains information only about the interactions between animals and
plants, but not about competition within plants or animals.

We parameterized the interaction matrix in the following way:
\begin{equation}\label{eq:interblock}
  \mat{A} = \left(
    \begin{array}{cc}
      \mat{W}^A & \mat{L}_b \circ \mat{W}^{AP} \\
      \mat{L}_b^T \circ \mat{W}^{PA} & \mat{W}^P
    \end{array}
  \right) \ ,
\end{equation}
where the symbol $\circ$ indicates the Hadamard or entrywise product
(i.e., $(\mat{A}\circ \mat{B})_{ij} = A_{ij} B_{ij}$), while
$\mat{W}^A$, $\mat{W}^{AP}$, $\mat{W}^{PA}$, and $\mat{W}^{P}$ are all
random matrices. $\mat{W}^A$ and $\mat{W}^P$ are square matrices of
dimension $S_A\times S_A$ and $S_P\times S_P$, while $\mat{W}^{AP}$
and $\mat{W}^{PA}$ are rectangular matrices of size $S_A\times S_P$
and $S_P\times S_A$. The diagonal elements $W^A_{ii}$ and $W^P_{ii}$
are set to $-1$, while the pairs $(W^{A}_{ij},W^{A}_{ji})$ and
$(W^{P}_{ij},W^{P}_{ji})$ are drawn from a bivariate normal
distribution with mean $\mu_{-}$, variance $\sigma^2_+ = c \mu_{-}^2$,
and correlation $\rho \sigma_{+}^2$. Since these two matrices
represent competitive interactions, $\mu_{-}<0$.  The pairs
$(W^{AP}_{ij},W^{PA}_{ji})$ were extracted from a bivariate normal
distribution with mean $\mu_{+}$, variance $\sigma_{-}^2 = c
\mu_{+}^2$, and correlation $\rho \sigma_{-}^2$, where
$\mu_{+}>0$. For each network and parametrization we computed the size
of the feasibility domain $\Xi$.

We considered different values of $\mu_{-}$, $\mu_{+}$, $c$, and
$\rho$. Their values cannot be chosen arbitrarily, since $\mat{A}$
must be negative definite.  For a choice of $c$, $\rho$, and a ratio
$\mu_{-}/\mu_{+}$, the largest eigenvalue of $(\mat{A}+\mat{A}^T)/2$
is linear in $\mu_{+}$ (as an arbitrary $\mu_{+}$ can be obtained by
multiplying $A$ by $\mu_{+}$ and then shifting the diagonal).  Given
the values of $\mu_{-}/\mu_{+}$, $c$, and $\rho$, one can therefore
determine $\mu_{\text{max}}$, the maximum value of $\mu_{+}$ still
leading to a negative definite $\mat{A}$ (i.e., the value of $\mu_{+}$
such that the largest eigenvalue of $(\mat{A}+\mat{A}^T)/2$ is equal
to $0$).  Fig.~\ref{fig:sstabtot} was obtained by considering more
than 1000 parameterizations. Both the ratio $\mu_{-}/\mu_{+}$ and the
coefficient of variation $c$ could assume the values $0.5$ or $2$,
while the correlation $\rho$ assumed values from the set
$\{-0.9,0.5,0,0.5,0.9\}$. The value of $\mu_+$ was set equal to $0.25
\mu_{\text{max}}$ and $ 0.75 \mu_{\text{max}}$.

\subsection*{Parameterization of food webs}

In the case of food webs the adjacency matrix $L$ is not symmetric,
$L_{ij}=1$ indicating that species $j$ consumes species $i$. We
removed all cannibalistic loops.  Since $L_{ij}$ and $L_{ji}$ are
never simultaneously equal to one (there are no loops of length two),
we parameterized the offdiagonal entries of $\mat{A}$ as
\begin{equation}\label{eq:interfw}
  A_{ij} = W^{+}_{ij}  L_{ij} + W^{-}_{ji} L_{ji} \ ,
\end{equation}
while the diagonal was fixed at $-1$. Both $ \mat{W}^{+}$ and
$\mat{W}^{-}$ are random matrices, where the pairs
$(W^{+}_{ij},W^{-}_{ij})$ are drawn from a bivariate normal
distribution with marginal means $(\mu_{+},\mu_{-})$ and correlation
matrix
\begin{equation}\label{eq:corrmatfw}
  \left(
    \begin{array}{cc}
      c \mu_{+}^2 & \rho c \mu_{+}^2 \\
      \rho c \mu_{-}^2 & c \mu_{-}^2
    \end{array}
  \right) \ .
\end{equation}

\begin{figure*}[hp]
\centering
  \includegraphics[width = 0.95\columnwidth]{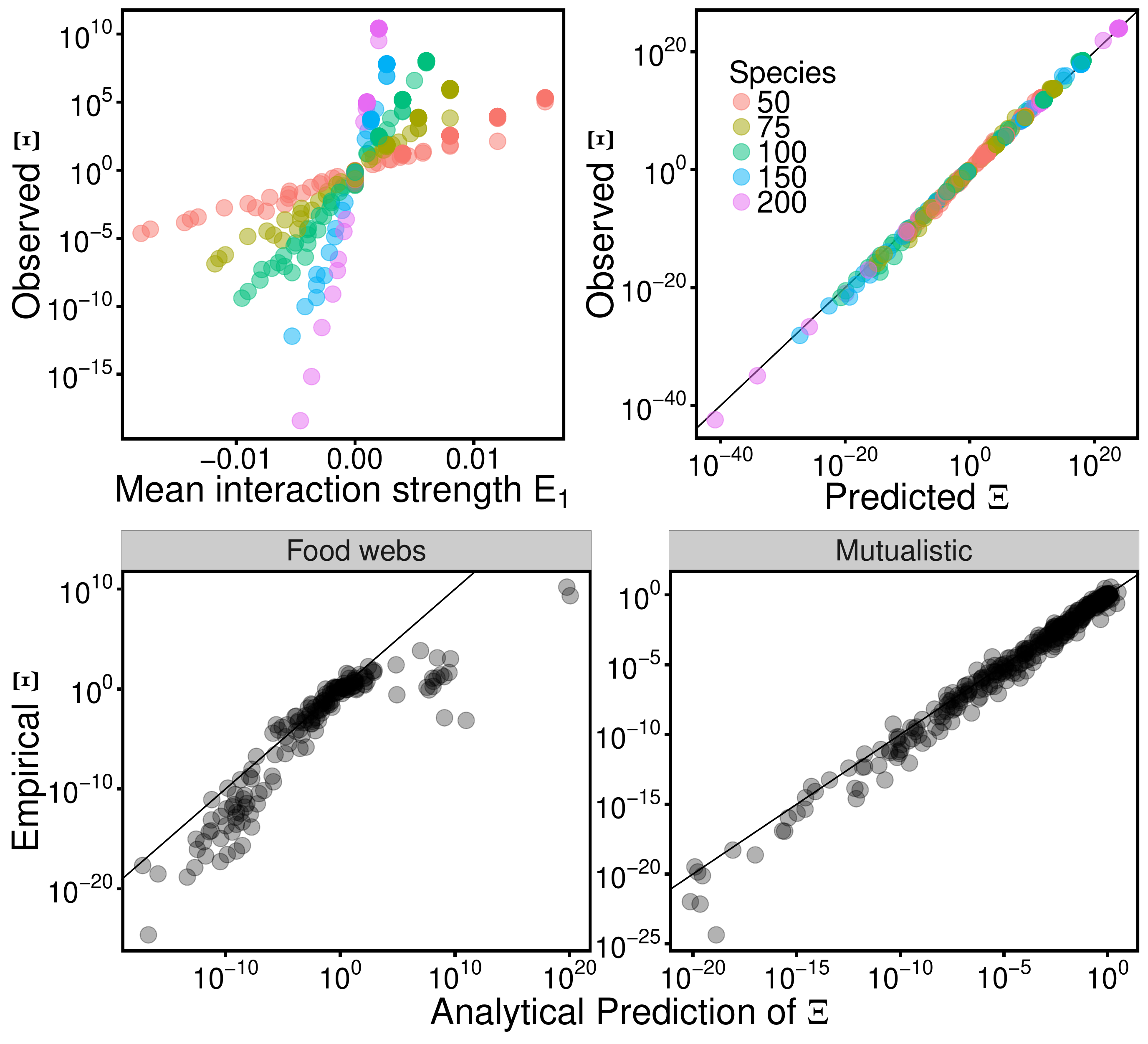}
  \caption{\textbf{Feasibility domain in random and empirical webs.}
    The top two panels show $\Xi$, the size of the domain of growth
    rates leading to coexistence, in the case of random networks. The
    left panel shows the dependence of $\Xi$ on $E_1 = C\mu$ (where
    $C$ is the connectance and $\mu$ is the mean interaction
    strength), and the number of species $S$. The right panel shows
    the match between our analytical prediction
    (Eq.~\ref{eq:prediction} and \SI) and the numerical value of
    $\Xi$. The bottom panels show a comparison between $\Xi$ computed
    for empirical webs (89 mutualistic networks on the right, and 15
    network was parameterized with different distributions of
    interaction strengths (\MM). Mutualistic networks have values of
    $\Xi$ comparable with random networks with similar interactions
    ($R^2=0.98$), indicating that their structure has little effect on
    the size of the feasibility domain. Food webs have lower values of
    $\Xi$ than their random counterparts ($R^2=0.80$). Empirical
    networks were parameterized extracting interaction strengths from
    a bivariate normal distribution with different means, variances,
    and correlations (\SI).}
\label{fig:sstabtot}
\end{figure*}

\begin{figure*}[hp]
\centering
  \includegraphics[width = 0.9\textwidth]{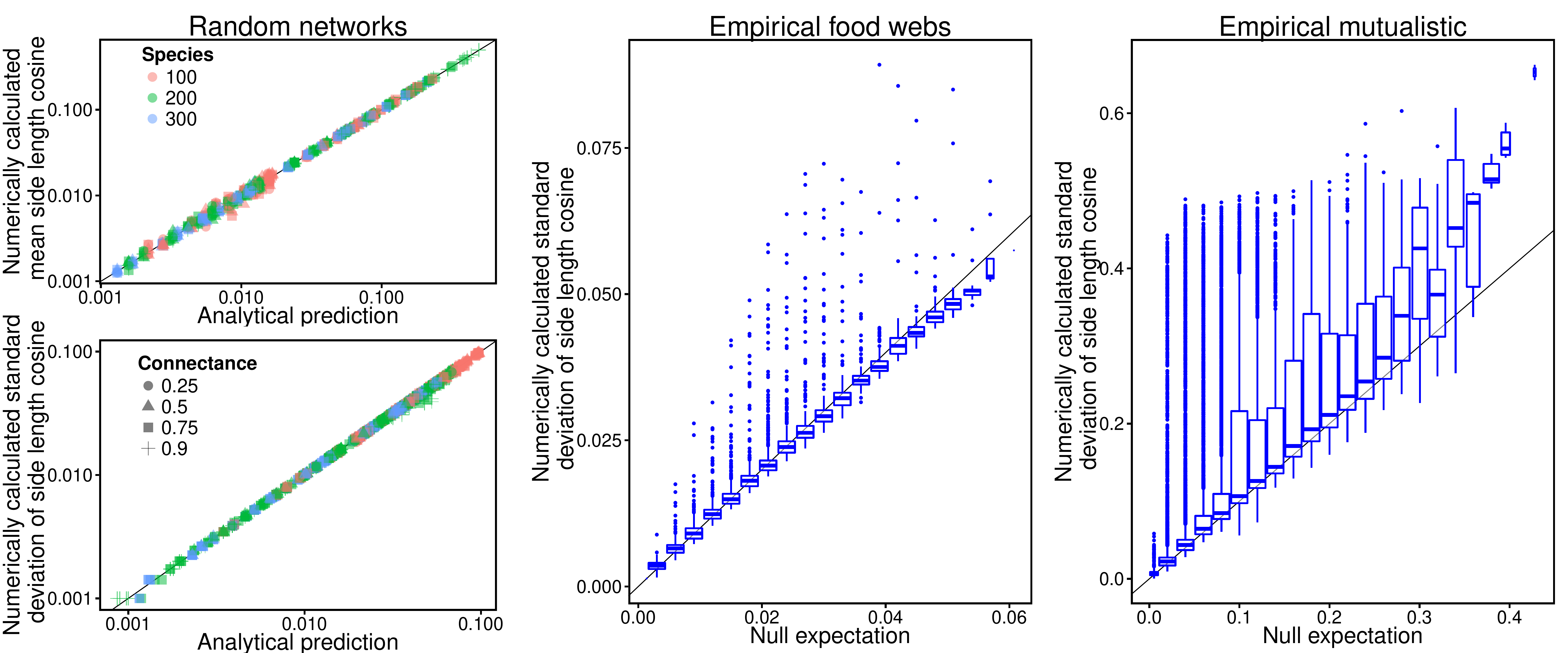}
  \caption{\textbf{Distribution of side lengths in random, structured,
      and empirical networks.}  Left panels show the mean and the
    standard deviation of $\cos(\eta)$, where $\eta$ is the side
    length.  Analytical predictions for the first two moments of
    $\cos(\eta)$ (\SI) perfectly match the numerical simulations. The
    two panels on the right show the standard deviation of
    $\cos(\eta)$ for mutualistic and food webs compared to the
    expectations for the randomized cases. Both trophic and
    mutualistic interactions show larger fluctuations of side lengths,
    suggesting the existence of perturbation directions to which the
    system is more sensitive than to others. This effect is
    particularly pronounced and relevant for mutualistic networks.
    While mutualistic and random networks have a similar feasibility
    domain size $\Xi$, this result implies that the response of
    mutualistic networks to perturbations is in fact more
    heterogeneous than those of their random counterparts.}
\label{fig:sside}
\end{figure*}

We considered considering
different values of $\mu_{-}$, $\mu_{+}$, $c$, and $\rho$.  As
explained above, given the values of $\mu_{-}/\mu_{+}$, $c$, and
$\rho$, one can determine $\mu_{\text{max}}$, the maximum value of
$\mu_{+}$ still corresponding to a negative definite $\mat{A}$.
Fig.~\ref{fig:sstabtot} was obtained by considering more that 350
parameterizations. Both the ratio $\mu_{-}/\mu_{+}$ and the
coefficient of variation $c$ could assume the values $0.5$ or $2$,
while the correlation $\rho$ assumed either the value $-0.5$ or
$0.5$. The value of $\mu_+$ was set either to $0.25 \mu_{\text{max}}$
or $ 0.75 \mu_{\text{max}}$.

%% Put the bibliography here, most people will use BiBTeX in
%% which case the environment below should be replaced with
%% the \bibliography{} command.

%\begin{thebibliography}{1}
%\bibitem{dummy} Articles are restricted to 50 references, Letters
%to 30.
%\bibitem{dummyb} No compound references -- only one source per
%reference.
%\end{thebibliography}
%\bibliographystyle{naturemag}
%\bibliography{StructStab}

%% Here is the endmatter stuff: Supplementary Info, etc.
%% Use \item's to separate, default label is "Acknowledgements"

%\begin{scilastnote}
%\item J.G.\ funded by the Human Frontier Science Program, S.A.\ and
%  G.B.\ supported by NSF--1148867.  We thank J.\ Hidalgo, D.\ Logofet,
%  M.J.\ Michalska--Smith, R.\ Rohr, S.\ Saavedra and M.\ Wilmes for
%  comments.
%\end{scilastnote}

%%
%% TABLES
%%
%% If there are any tables, put them here.
%%

\newpage

%\bibliographystyle{unsrt}
%\bibliography{StructStab}

%\begin{scilastnote}
%\item JG funded by the Human Frontier Science Program, SA by NSF DEB-1148867, XXX. We thank XXX and YYY
%  for comments.
%\end{scilastnote}

\newpage

%%%%%%%%%%%%% SUPPLEMENTARY FIGURES %%%%%%%%%%%%%%%%%%

%%% PUT THIS IN \include FILE

\clearpage

\renewcommand{\thesection}{S\arabic{section}}
\setcounter{section}{0}
\setcounter{figure}{0}
\renewcommand{\figurename}{Supplementary Figure}
\renewcommand{\thefigure}{S\arabic{figure}}

\setcounter{table}{0}
\renewcommand{\tablename}{Supplementary Table}
\renewcommand{\thetable}{S\arabic{table}}
\setcounter{equation}{0}
\renewcommand{\theequation}{S\arabic{equation}}

\begin{center}
  {\Large \bf The geometry of coexistence in large ecosystems \\ Supplementary Information}
\end{center}

  \tableofcontents

\newpage

\section{Community dynamics, feasibility, and stability}

We consider an ecological community composed of $S$ populations, whose
dynamics is described by the following equations:
\begin{equation}\label{eq:dynmodel_LV}
  \displaystyle
  \frac{\ud n_i}{\ud t} = n_i \left( r_i + \sum_{j=1}^S A_{ij} n_j \right) 
  \ ,
\end{equation}
where $n_i$ is the population abundance of species $i$ and $r_i$ is its
intrinsic growth rate, and $A_{ij}$ is the effect of a unit change in
species $j$'s density on species $i$'s per capita growth rate. For
notational convenience, we collect the coefficients $A_{ij}$ into the
interaction matrix $\mat{A}$, and $n_i$ and $r_i$ into the vectors
$\vect{n}$ and $\vect{r}$, respectively.

In principle, the interaction matrix $\mat{A}$ may depend on
$\vect{n}$. We discuss this more general case in
section~\ref{sec:nonlin}. In the following, we consider the simpler
case of $\mat{A}$ being independent of $\vect{n}$; then, equation
\eqref{eq:dynmodel_LV} is a general system of Lotka--Volterra population
equations.

A vector $\vect{n}^\ast$ is a fixed point (equilibrium) if
\begin{equation}\label{eq:fixed_point}
  \displaystyle
  0 = n^\ast_i \Bigl( r_i + \sum_{j=1}^S A_{ij} n^\ast_j \Bigr) \qquad (i = 1, 2, \ldots, S)
  \ .
\end{equation}
A fixed point is feasible if $n^\ast_i > 0$ for all $i$. A feasible
fixed point (if it exists) is then a solution to the equation
\begin{equation}\label{eq:fixed_point_eq}
  \displaystyle
  r_i = - \sum_{j=1}^S A_{ij} n^\ast_j 
  \ ,
\end{equation}
and therefore, assuming $\mat{A}$ is invertible,
\begin{equation}\label{eq:fixed_point_eq2}
  \displaystyle
  n^\ast_i = - \sum_{j=1}^S (A^{-1})_{ij} r_j  
  \ .
\end{equation}
A fixed point $n^\ast_i$ is locally stable if the system returns to it
following any sufficiently small perturbation of the population
abundances. Introducing $n_i = n^\ast_i + \delta n_i$ in
equation~\ref{eq:dynmodel_LV} and assuming that $\delta n_i$ is small,
we obtain, by expanding around $\delta n_i = 0$,
\begin{equation}\label{eq:linearization}
  \displaystyle
  \frac{\ud \delta n_i}{\ud t} = \sum_{j=1}^S M_{ij} \delta n_j  
  \ ,
\end{equation}
where $M_{ij}$ is the $(i,j)$th entry of the Jacobian evaluated at the fixed point (also
called the community matrix), which, in the case of
equation~\ref{eq:dynmodel_LV}, reduces to
\begin{equation}\label{eq:community_matrix}
  \displaystyle
  M_{ij} = n_i^\ast A_{ij} = - \left( \sum_{k=1}^S (A^{-1})_{ik} r_k  \right) A_{ij} 
  \ .
\end{equation}
Substituting into equation~\ref{eq:linearization}, we get
\begin{equation}\label{eq:linearization2}
  \displaystyle
  \frac{\ud \delta n_i}{\ud t} = - \sum_{j=1}^S \left( \sum_{k=1}^S (A^{-1})_{ik} r_k  \right) A_{ij} \delta n_j  
  \ .
\end{equation}
There are two possible scenarios for the dynamics of
equation~\ref{eq:linearization}. If all eigenvalues of $\mat{M}$ have
negative real parts, then the perturbation $\delta \vect{n}$
decays exponentially to zero and $n_i^\ast$ is locally stable. If at least one eigenvalue of $\mat{M}$
has a positive real part, then there exists an infinitesimal
perturbation such that the system does not return to equilibrium.  If
we order the eigenvalues $\lambda_i$ of $\mat{M}$ according to their real parts,
i.e., $\Re(\lambda_1) > \Re(\lambda_2) > \dots > \Re(\lambda_S)$, then
stability depends exclusively on $\Re(\lambda_1)$: if it is negative,
$n_i^\ast$ is dynamically locally stable; otherwise, it is unstable~\cite{Svirezhev1978}.

A fixed point is globally stable if it is the final outcome of the
dynamics from any initial condition involving strictly positive
population abundances.

\section{Disentangling stability and feasibility}
\label{sec:stabfeas}

As we can see from equations~\ref{eq:fixed_point_eq2}
and~\ref{eq:linearization2}, both feasibility and stability depend on
both $\vect{r}$ and $\mat{A}$ and, at least in principle, a
fixed point can be stable or unstable, independently of the fact that it is feasible or not.
 
We want to study the proportion of conditions (i.e., the number of
combinations of the growth rates $\vect{r}$ out of all possible
combinations) leading to coexistence, i.e., leading to stable and
feasible equilibria. Therefore in principle we should, for a fixed
matrix $\mat{A}$, look for growth rates $\vect{r}$ that satisfy
both stability and feasibility. In probabilistic terms, we want to
measure the likelihood that a random combination of the intrinsic
growth rates corresponds to a stable and feasible solution.

In the case of equation~\ref{eq:dynmodel_LV}, it is possible to
disentangle feasibility and stability by applying a mild condition on
the interaction matrix $\mat{A}$. To this end, we introduce some
terminology \cite[section~2.1.2]{Kaszkurewicz2000}:
\begin{itemize}
\item \textbf{Stability}. A real matrix $\mat{B}$ is stable if all its
  eigenvalues have negative real parts.
\item \textbf{\textit{D}-stability}. A real matrix $\mat{B}$ is D-stable if
  $\mat{D} \, \mat{B}$ is stable for any diagonal matrix  $\mat{D}$
with strictly
  positive diagonal entries.
\item \textbf{Diagonal stability}. A real matrix $\mat{B}$ is
  diagonally stable if there exists a positive diagonal matrix
  $\mat{D}$ such that $\mat{D} \,
  \mat{B} + \mat{B}^T \, \mat{D}$ is stable (where $\mat{B}^T$ is the transpose of $\mat{B}$).
\end{itemize}
We also consider
\begin{itemize}
\item \textbf{Negative definiteness} (in a generalized sense). A real matrix $\mat{B}$ is negative definite if $\sum_{ij} x_i B_{ij} x_j < 0$
for any non-zero vector $\vect{x}$~\cite{Johnson1970}. 
\end{itemize}

These properties are closely related to each
other~\cite{berman1983matrix,Kaszkurewicz2000}:
\begin{equation}
  \text{Negative definiteness} \implies \text{Diagonal stability} \implies \text{\textit{D}-stability} \implies \text{Stability}
\end{equation}
\begin{itemize}
\item \textbf{Negative definiteness $\implies$ Diagonal stability}.
    A matrix $\mat{B}$ is negative definite if and only if
    all the eigenvalues of
  $\mat{B} +\mat{B}^T$ are negative~\cite{Johnson1970}.
 If this condition hold, then the positive diagonal matrix satisfying
  the definition of diagonal stability is simply the identity matrix.
\item \textbf{Diagonal stability $\implies$ \textit{D}-stability}. See the
  book by Kaszkurewicz \& Bhaya for the proof \cite[lemma
  2.1.4]{Kaszkurewicz2000}.
\item \textbf{\textit{D}-stability $\implies$ Stability}. This follows from the
  definition of \textit{D}-stability when $\mat{D}$ is the identity matrix.
\end{itemize}

In the case of equation~\ref{eq:dynmodel_LV}, those conditions applied
to the matrix $\mat{A}$ are related to the stability of the system.
One can use the definition of the community matrix
(equation~\ref{eq:community_matrix}) to show that \textbf{\textit{D}-stability
  of $\mat{A}$ implies the local asymptotic stability of \emph{any}
  feasible fixed point}. This is because the community matrix with entries $M_{ij}
= n_i^\ast A_{ij}$ can be written as $\mat{N} \, \mat{A}$, where
$\mat{N}$ is the diagonal matrix with $N_{ii}=n^\ast_i$. If the fixed
point is feasible and $\mat{A}$ is D-stable, then local asymptotic
stability is guaranteed. Moreover it is possible to show
\cite{Goh1977, Rohr2014} that \textbf{diagonal stability of $\mat{A}$
  $\implies$ global stability}.

Thus, we have a condition on $\mat{A}$ that makes it possible to
disentangle the problems of stability and feasibility:
\textbf{$\mat{A}$ is negative definite $\implies$ global stability of the
  feasible fixed point}~\cite{Volterra1931}. Therefore, if we assume $\mat{A}$ is
negative definite, then feasibility of the equilibrium is sufficient to
guarantee its global stability as well, i.e., feasibility guarantees
globally stable coexistence. Consistently with this, it is known that
the largest eigenvalue of $(\mat{A} + \mat{A}^T)/2$ is always larger
than or equal to the real part of $\mat{A}$'s leading eigenvalue
\cite{Tang2014}, i.e. negative definiteness implies stability. While this was indeed
observed before, it is important to underline that, in the case of
ref.~\cite{Tang2014}, this property was considered on the community
matrix $\mat{M}$ (which also depends on the fixed point's position in
phase space) and not on the interaction matrix $\mat{A}$.

Since we are interested in studying how interactions (i.e., the matrix
$\mat{A}$) determine coexistence, and which properties of the former
determine the latter, we will restrict our analysis to negative definite
matrices $\mat{A}$ and focus only on the problem of feasibility. This
condition has the advantage of being analytically computable for large
random matrices (see section~\ref{sec:spectrumrnd}).

\section{Geometrical properties of the feasibility domain}
\label{sec:geomprop}

In section~\ref{sec:stabfeas} we showed how to separate feasibility
and stability, i.e., we have a sufficient condition on the interaction matrix
that guarantees (global) stability of the feasible fixed point. The
problem of determining the size of the coexistence domain is therefore
reduced to that of determining the size of the feasibility domain.
% In ref.~\cite{Rohr2014} feasibility has been introduced as the
% volume of the domain of intrinsic growth rates leading to a feasible
% and stable equilibrium point.
The ecological interpretation of this volume is the proportion of
different conditions leading to feasible equilibria out of all
possible conditions.  The larger this volume is, the higher the
probability that the system is able to sustain biodiversity.  In terms
of equation~\ref{eq:dynmodel_LV}, we want to quantify the proportion of
growth rate vectors $\vect{r}$ corresponding to a feasible fixed
point. 
%\rem{Thoughts about biologically realistic values?}

This geometrical approach was pioneered in~\cite{Svirezhev1978} where
the space of feasible solution was studied for dissipative systems, and
the size of that domain was computed in the case $S=3$ (see section~\ref{sec:feas3D}).

At this point, it is important to observe that if a vector
$\vect{r}$ corresponds to a feasible solution, then $c
\vect{r}$, $c$ being an arbitrary positive constant, also
corresponds to a feasible solution.  This is because the equilibrium
solution $n_i^\ast$ is given by equation~\ref{eq:fixed_point_eq2},
which is linear in $r_i$. Therefore, the equilibrium corresponding to
$c r_i$ is simply $c n_i^\ast$, and since $c$ is positive, $c
n_i^\ast$ is also feasible.

This fact implies that, given a large number of growth rate vectors
$\vect{r}$, the expected proportion of vectors corresponding to a
feasible fixed point is independent of $\vect{r}$'s norm. In
other words, $\vect{r}$ is feasible if and only if
$\vect{r}/\|\vect{r}\|$ is feasible, where $\| \vect{r}
\| = \sqrt{\sum_i r_i^2}$ is the Euclidean norm of $\vect{r}$.
The proportion of feasible growth rates among all possible ones is
therefore equal to the proportion of feasible growth rates calculated
using only growth rate vectors with $\|\vect{r}\| = 1$; i.e.,
those lying on the unit sphere.

Before proceeding with the mathematical definition of the size of the
feasibility domain, we discuss the geometrical interpretation of
equation \ref{eq:fixed_point_eq2}. From this equation, the feasibility
condition reads
\begin{equation}
  \sum_{j=1}^S (A^{-1})_{ij} r_j < 0 \ .
\label{eq:feasrcond}
\end{equation}
This equation defines a convex polyhedral cone in the $S$-dimensional
space of growth rates.  A convex polyhedral
cone~\cite{Rockafellar1997} is a subset of $\mathbb{R}^S$ whose
elements $\vect{x}$ can be written as positive linear
combinations of $N_G$ different $S$-dimensional vectors
$\vect{g}^k$ called the generators of the cone:
\begin{equation}
  \vect{x} = \sum_{k = 1}^{N_G} \vect{g}^k \lambda_k \ ,
\label{eq:conedef}
\end{equation}
where the $\lambda_k$ are arbitrary positive
constants. Due to this arbitrariness, if $\vect{g}^k$ is a
generator of a given convex polyhedral cone, then also $c
\vect{g}^k$ (where we rescale just the $k$th generator with the
positive constant $c$, leaving the others unchanged) will be a
generator of the \emph{same} cone~\cite{Svirezhev1978}.  In the case of
equation~\ref{eq:fixed_point_eq}, each and every growth rate vector
belonging to the feasibility domain can be written as
\begin{equation}
  r_i = - \sum_{k = 1}^S A_{ik} n^\ast_k \ ,
\label{eq:coneour}
\end{equation}
where, by definition, $n^\ast_k$ is feasible and therefore a positive
constant. One can easily see that this equation corresponds to
equation~\ref{eq:conedef} where the number of generators $N_G$ is
equal to $S$ and the $i$th component of the vector $\vect{g}^k$
is proportional to $-A_{ik}$. As the lengths of the generators can be
set to any positive value, we will normalize them to one, i.e.,
\begin{equation}
  \displaystyle
  g^k_i(\mat{A}) = \frac{-A_{ik}}{ \sqrt{ \sum_{j=1}^S (A_{jk})^2 } } \ .
\label{eq:generatormat}
\end{equation}
The generators completely define the feasibility domain in the space
of growth rates. A growth rate vector corresponds to a feasible
equilibrium if and only if it can be written as a linear combination
of the generators with positive coefficients.  Biologically the
generators correspond to the growth rate vectors that bound the
coexistence domain. They correspond to nonfeasible equilibria with just one
species with positive abundance (and all the others with zero
abundance), such that there exist arbitrarily small perturbations of
the growth rate vector that make the equilibrium feasible.

The set of all the growth rate vectors leading to a feasible
equilibrium is therefore a convex polyhedral cone, defined by
\begin{equation}
  \displaystyle
  K(\mat{A}) = \{ \vect{r} \in \mathbb{R}^S | \sum_{j=1}^S (A^{-1})_{ij} r_j < 0 \} \ .
\label{eq:convexcone}
\end{equation}
Equivalently, it can be defined in terms of the generators:
\begin{equation}
  \displaystyle
  K(\mat{A}) = \{ \vect{r} \in \mathbb{R}^S | \exists \lambda_1,\lambda_2,\dots,\lambda_k > 0 , \vect{r} = \sum_{k=1}^S \vect{g}^k(\mat{A}) \lambda_k \} \ ,
\end{equation}
where the generators $\vect{g}^k(\mat{A})$ are defined in
equation~\ref{eq:generatormat}.
 In section~\ref{sec:feas3D} we
show explicitly how these concepts pan out in the case of $S=3$.

This geometrical definition and characterization of the feasibility
domain allows us to identify classes of matrices having the exact same
feasibility domain: they are simply matrices having the same set of
generators. In particular, there are two basic transformations of the
matrix $\mat{A}$ (and their combinations) that leave the set of
generators unchanged: permutations and positive rescaling. A square
matrix $\mat{P}$ is a permutation matrix if each row and column has
one and only one nonzero entry and the value of that entry is equal to one. A
positive rescaling is performed by a positive diagonal matrix
$\mat{D}$. The set of generators of $\mat{A}$ is the same as those of
$\mat{P} \, \mat{A}$ and $\mat{D} \, \mat{A}$. This can be seen by
observing that a permutation of the rows just changes the order of the
generators but not the generators themselves. In the same way, a
generator with the same direction but different length generates the
same cone, and so any positive constant that rescales a row of the
matrix leaves the feasibility domain unchanged. It is important to
note however that these two transformations do not leave the
properties of the matrix $\mat{A}$ unchanged: both exchanging rows of
a matrix and rescaling rows by different constants will in general
change the structure of the matrix.

Using this geometrical framework, one can easily identify the center
of the feasibility domain (also known as structural vector~\cite{Rohr2014}).
There are several possible ways to define
the center of a hypervolume and, without additional assumptions, all
the definitions are different.  One natural choice is the barycenter
(``center of mass'') of the domain of feasible intrinsic growth
rates. Any plane passing through the barycenter divides the volume
into two subvolumes of equal size. The barycenter is equivalent to the center of
mass of the volume (in the case of constant density).
Then, the vector $\vect{x}^b$ pointing from
the origin to the barycenter is given by
\begin{equation}
  \displaystyle
  \vect{x}^b = \int_{K(A) \cap \mathbb{S}_S } \ud^S \vect{y} \ \  \vect{y} \ ,
\label{eq:barycenter}
\end{equation}
where $\cap$ is the intersection of two sets, and $\mathbb{S}_S = \{
\vect{r} \in \mathbb{R}^S | \|\vect{r}\| = 1 \}$ represents
the surface of the $S$-dimensional unit sphere. The variable
$\vect{y}$ is therefore integrated over the feasibility domain
restricted to the unit sphere's surface. All points in the feasibility
domain are positive linear combinations of the generators, i.e.,
\begin{equation}\label{eq:arbitrarypoint}
  \vect{y} = \sum_k \lambda^k \vect{g}^k   \
  \ ,
\end{equation}
where the $\lambda^k$ are positive constants. The fact that we
consider only the points lying on the unit sphere, i.e.,
$\|\vect{y}\|=1$, can be expressed as a constraint on
$\vect{\lambda}$ (the vector of $\lambda$s). Thus, we can write
equation~\ref{eq:barycenter} as
\begin{equation}
  \displaystyle
  \vect{x}^b = \int \ud^S \vect{\lambda} \, q(\vect{\lambda}) \sum_k \lambda^k \vect{g}^k \ ,
\label{eq:barycenter2}
\end{equation}
where $q$ is an appropriate distribution, introduced to take into
account three different constraints: all the components of
$\vect{\lambda}$ must be positive; the vector $\sum_k \lambda^k
\vect{g}^k$ must lie on the unit sphere; and those vectors must
be sampled uniformly on the feasibility domain.
One can show that the distribution $q(\vect{\lambda})$ 
has the following form
\begin{equation}
  \displaystyle
q(\vect{\lambda}) \propto \exp\left( - \sum_{i,j} \lambda^i (\vect{g}^i\cdot\vect{g}^j) \lambda^j  \right) 
\prod_k \Theta(\lambda^k) \ ,
\end{equation}
where the proportionality constant is given by the normalization.
Therefore, by defining,
\begin{equation}
  \displaystyle
\int \ud^S \vect{\lambda} \,  q(\vect{\lambda}) \ \lambda^k =: \langle \lambda^k \rangle \ ,
\end{equation}
we obtain
\begin{equation}
  \displaystyle
  \vect{x}^b = \sum_k \vect{g}^k  \int \ud^S \vect{\lambda} \,  q(\vect{\lambda}) \ \lambda^k =  \sum_k \vect{g}^k \langle \lambda^k \rangle  \ .
\label{eq:barycenter3}
\end{equation}

\section{Definition and calculation of $\Xi$}
\label{sec:numint}

As explained in section~\ref{sec:geomprop}, the proportion of feasible
growth rates can be calculated considering only growth rate vectors of
length one, i.e., $\|\vect{r}\|=1$. This proportion can be
interpreted as the volume of the intersection of a convex cone and the
surface of a sphere. Equivalently, it is the solid angle of the convex polyhedral
cone~\cite{Ribando2006,Gourion2010}. 
%\rem{Define ``solid angle''.}

We define the quantity $\Xi$ as
\begin{equation}
  \Xi = 2^{S} \, \frac{\text{\# growth rate vectors corresponding to a feasible fixed point} }{\text{total \# growth rate vectors} } 
  \ .
\label{eq:xidef}
\end{equation}
The factor $2^S$ that appears in this equation is an arbitrary choice,
and it has been introduced to have $\Xi = 1$ when species are not
interacting ($A_{ij}=0$ if $i\neq j$).  In this case
equation~\ref{eq:dynmodel_LV} reduces to $S$ independent logistic
equations with equilibrium densities $n_i^\ast = - r_i /
A_{ii}$. Taking each $A_{ii}$ to be negative (otherwise each species
would have an unstoppable positive feedback on itself), this equilibrium is
feasible if and only if each $r_i$ is positive.  For a single species
then, the probability of randomly drawing a feasible (i.e., positive)
growth rate out of all possible growth rates is one half. For two
species, both growth rates must have the correct sign to have the two
species with positive abundance, and therefore the proportion of growth
rate vectors satisfying this condition is $1/4$. For $S$ species the
combinations of the growth rates leading to a feasible fixed point is
$2^{-S}$. $\Xi$, defined as in equation~\ref{eq:xidef}, is therefore
equal to one when species do not interact.

In terms of geometrical properties and the convex polyhedral cone,
$\Xi$ can be defined as
\begin{equation}
  \Xi = 2^{S} \frac{ \vol_{S-1}( K(A) \cap \mathbb{S}_S ) }{ \vol_{S-1}( \mathbb{S}_S ) } 
  \ ,
\label{eq:xidefcone}
\end{equation}
where $K(A)$ is defined in equation~\ref{eq:convexcone},
$\mathbb{S}_S$ is the unit sphere in $\mathbb{R}^S$, while
$\vol_{S}(\cdot)$ means volume in $S$ dimensions. This definition is
equivalent to the one in equation~\ref{eq:xidef}\cite{Ribando2006,Gourion2010}.

These two equivalent definitions can be expressed in terms of an
integral in the space of the growth rate vectors:
\begin{equation}
  \Xi = \frac{2^{S}}{\vol_{S-1}(\mathbb{S}_S)} \int_{\mathbb{R}^S} \ud^S \vect{r} \ 2 \| \vect{r} \| \, \delta( \| \vect{r} \|^2 - 1) \
  \prod_{i=1}^S \Theta( n_i^\ast(\vect{r}) ) \ ,
\label{eq:sstab}
\end{equation}
where $\vol_{S-1}(\mathbb{S}_S)$ is the volume of the unit sphere's
surface in $S$ dimensions, $\Theta(\cdot)$ is the Heaviside function
(equal to $1$ is the argument is positive and to zero otherwise), and
$\delta(\cdot)$ is the Dirac delta function.
In this expression, we integrate over the surface of
the $S$-dimensional unit sphere. The integral of a function
$f(\vect{x})$ on the unit sphere is given by
\begin{equation}
  \int_{\mathbb{S}_S}  \ud^S \vect{x}  \ f(\vect{x}) = \int_{\mathbb{R}^S}  \ud^S \vect{x} \ 2 \| \vect{x} \| \delta( \| \vect{x} \|^2 - 1) \ f(\vect{x}) \ ,
\end{equation}
where the term $\delta( \| \vect{x} \|^2 - 1)$ that appears in
the integration constrains $\vect{x}$ on the surface of the unit
sphere, and the factor $2 \| \vect{x} \|$ is the derivative of
the delta function's argument, which is needed because the Dirac delta
is nonlinear in $\| \vect{r} \|$. The factor
$\vol_{S-1}(\mathbb{S}_S)$, the surface of sphere in $S$ dimensions, can be obtained by setting $f(x) = 1$:
\begin{equation}
  \mathrm{vol}_{S-1}( \mathbb{S}_S ) = \int \ud^S \vect{x} \ 2 \| \vect{x} \| \delta( \| \vect{x} \|^2 - 1) = \frac{2 \pi^{S/2}}{\Gamma(S/2)} \ , 
\label{eq:volumesphere}
\end{equation}
where $\Gamma(\cdot)$ is the Gamma function. Finally, the term
$\prod_{i=1}^S \Theta( \vect{n}_i^\ast(\vect{r}) )$ in
equation~\ref{eq:sstab} expresses the constraint of all $n_i^\ast$
having to be positive: this product is equal to $1$ if the equilibrium
$\vect{n}^\ast(\vect{r})$ is feasible and zero
otherwise. The equilibrium $\vect{n}^\ast(\vect{r})$ is a
function of $\vect{r}$ via equation~\ref{eq:fixed_point_eq2}.

Equation~\ref{eq:sstab} defines $\Xi$ as the volume of the domain of
growth rates leading to feasible solutions. Using the results of
section~\ref{sec:stabfeas}, we know that if the interaction matrix
$\mat{A}$ is negative definite then a feasible fixed point is globally
stable. In this case $\Xi$ is the volume of the domain of intrinsic
growth rates leading to feasible and (globally) stable solutions.

Unfortunately, direct numerical computation of $\Xi$ is inefficient
when the number of species $S$ is large. To evaluate the integral in
equation~\ref{eq:sstab}, e.g., via Monte Carlo integration, we should
draw intrinsic growth rates at random and count how many of them, out
of the total, lead to a feasible equilibrium. In order to have a
reliable estimate of this proportion, we should sample the space in such
a way that the number of feasible growth rates found is large. This
goal requires an exponentially increasing sampling effort as $S$
increases. In this section we provide an alternative, much faster and
reliable, way of estimating $\Xi$.

The equilibrium solution and the growth rates are linearly related via
$r_i = - \sum_{i=1}^S A_{ij} n_j^\ast$
(equation~\ref{eq:fixed_point_eq}). Our strategy is to use this to
perform a change of variables in equation~\ref{eq:sstab}, and
integrate over $\vect{n}^\ast$ instead of $\vect{r}$. Since
$\mat{A}$ is negative definite (and thus stable and not singular), it is
invertible, and so it is always possible to perform this change of
variables. Note that, more generally, the change of variables can be performed if $A$ is nonsingular (i.e., $\det(A)=0$).
We then obtain
\begin{equation}
  \Xi = \frac{2^{S} \ \Gamma(S/2) \ |\det(\mat{A})|}{2\pi^{S/2}} \int_{\mathbb{R}^S} \ud^S \vect{n}^\ast \ 2 \delta\left( \sum_{i,j,k} n_{i}^\ast A_{ki} A_{kj} n_{j}^\ast - 1 \right) \
  \prod_{i=1}^S \Theta( n_i^\ast ) \ ,
\label{eq:sstab_change}
\end{equation}
where $|\det(\mat{A})|$ is the determinant of $\mat{A}$, which is also
the Jacobian of the change of variables. After the change of
variables, the integration is now performed over the feasible
equilibrium points and so the condition of feasibility is
automatically implemented.

It is still difficult to evaluate the previous expression numerically,
because of the constraint that appears in the delta function. We can
further simplify it by introducing polar coordinates. In particular,
we write the vector $\vect{n}$ as $\vect{n} = n
\vect{u}$, where $n=\|\vect{n}\|$ and $\vect{u}$ is a
vector of unit length. We can perform a new change of variables,
passing from $\vect{n}$ to $n$ and
$\vect{u}$. Specifically, for any function $f(\vect{n})$, we
can write
\begin{equation}
  \int_{\mathbb{R}^S} \ud^S \vect{n} \ f(\vect{n}) = \int_0^{\infty} \ud n \ n^{S-1} \ \int_{\mathbb{R}^S} \ud^S \vect{u} \ 2 \delta( \| \vect{u} \|^2 - 1 ) f(n\vect{u}) = \int_0^{\infty} \ud n \ n^{S-1} \ \int_{\mathbb{S}_S} \ud^S \vect{u} \ f(n\vect{u})
  \ .
\label{eq:polarintegral}
\end{equation}
Using this expression in equation~\ref{eq:sstab_change}, we obtain
\begin{equation}
  \Xi = \frac{2^{S} \ \Gamma(S/2) \ \det(\mat{A}) }{2\pi^{S/2}}  \int_0^\infty \ud n \ n^{S-1} \ \int_{\mathbb{S}_S} \ud^S \vect{u} \ 2 \delta\left( n^2 \sum_{i,j} u_{i} G_{ij} u_{j}- 1\right) \ \prod_{i=1}^S \Theta( u_i ) \ ,
\label{eq:sstab_changesec}
\end{equation}
where we used the fact that $\Theta(n_i)=\Theta(u_i)$ (since $n_i = n
u_i$, and $n$ is positive by definition), and we have introduced the
matrix $G_{ij} = \sum_k A_{ki} A_{kj}$.  We can now perform the
integration over $n$, obtaining
\begin{equation}
\begin{split}
  \displaystyle \int_0^{\infty} \ud n & \ n^{S-1} \ 2 \ \delta\left(
    n^2 \sum_{i,j} u_{i} G_{ij} u_{j} - 1\right) \\ = & \int_0^{\infty} \ud
  n \ n^{S-1} \ 2 \ \delta\left( n - \frac{1}{\sqrt{ \sum_{i,j} u_{i}
        G_{ij} u_{j} }}\right)  \ \frac{1}{ 2 n \sum_{i,j}
    u_{i} G_{ij} u_{j} } = \displaystyle \left( \sum_{i,j} u_{i}
    G_{ij} u_{j} \right)^{-S/2} \ ,
\end{split}
\label{eq:sstab_rintegral}
\end{equation}
and therefore the integral of equation~\ref{eq:sstab} finally reads
\begin{equation}
  \Xi = \frac{2^{S} \ \Gamma(S/2) \ \sqrt{\det(\mat{G})} }{2\pi^{S/2}} \int_{\mathbb{S}_S} \ud^S \vect{u} \ \prod_{i=1}^S \Theta( u_i ) \ 
  \left( \sum_{i,j} u_{i} G_{ij} u_{j}  \right)^{-S/2}
  \ ,
\label{eq:sstab_final}
\end{equation}
where we have used the fact that $\det(\mat{G})=\det(\mat{A}^T
\mat{A})=\det(\mat{A})^2$. In terms of the interaction matrix, the
equation reads
\begin{equation}
  \Xi = \frac{2^{S} \ \Gamma(S/2) \ |\det(\mat{A})| }{2 \pi^{S/2}} \int_{\mathbb{S}_S} \ud^S \vect{u} \ \prod_{i=1}^S \Theta( u_i ) \ 
  \left( \sum_{i,j,k} u_{i} A_{ki} A_{kj} u_{j}  \right)^{-S/2}
  \ .
\label{eq:sstab_final2}
\end{equation}

Equation~\ref{eq:sstab_final} shows explicitly the role of the
generators. The matrix $\mat{G}$ can indeed be rewritten as
\begin{equation}
  G_{ik} = \sum_j g^i_j g^k_j c_i c_k = c_i c_k \vect{g}^i \cdot \vect{g}^k \ ,
  \label{eq:Gmat}
\end{equation}
where $g^k_j$ are the generators of the convex cone defined in
equation~\ref{eq:generatormat} and $c_i$ are arbitrary positive constants. Their presence, which can be
seen as a change of the
normalization of the vectors $\vect{g}^k$, does not affect the
form of equation~\ref{eq:sstab_final} and its dependence on $\mat{G}$ (see section~\ref{sec:geomprop}).
This property can be checked explicitly from equation~\ref{eq:sstab_final}, by introducing an explicit
dependence on $c_i$ and showing that $\Xi$ is independent of their values.
%% ADD HERE EXPLICIT DEMONSTRATION?

%In section~\ref{sec:geomprop} we showed that, since the domain of
%feasible growth rates is a convex polyhedral cone, any permutation or
%rescaling of the interaction matrix leaves the domain unchanged. This
%can be checked directly from equation~\ref{eq:sstab_final2}, observing
%that if $\mat{P}$ is a permutation and $\mat{D}$ a positive diagonal
%matrix, then $\mat{A}$ and $\mat{A} \, \mat{P} \, \mat{D}$ have the
%same value of $\Xi$. \rem{Is this obvious?}

Unfortunately, the integral in equation~\ref{eq:sstab_final} cannot be
computed analytically.  As mentioned before, when the integral is
written in the form of equation~\ref{eq:sstab} it is impractical to
evaluate it numerically, since it would require an exponentially
increasing sampling to get a reasonable precision.  Fortunately, this
is not the case when the integral is written as in
equations~\ref{eq:sstab_final} and~\ref{eq:sstab_final2}.  The main
difference is that, after changing variables, we are directly sampling
the space of feasible solutions, without losing computational time in
randomly exploring the space of intrinsic growth rates looking for
feasible solutions.

To evaluate the integral, we use the usual approach of Monte Carlo
algorithms. In particular, it is possible to write the integral as an
average over random points:
\begin{equation}
  \frac{ 1 }{T} \sum_{a=1}^T \Bigl( \sum_{i,j} u^a_{i} G_{ij} u^a_{j}  \Bigr)^{-S/2} \to
  \frac{ \Gamma(S/2) }{2\pi^{S/2}} \int d^S u \ \prod_{i=1}^S \Theta( u_i )  \ 2 \delta( \| u \|^2 - 1 ) \ 
  \Bigl( \sum_{i,j} u_{i} G_{ij} u_{j}  \Bigr)^{-S/2}
\label{eq:sstab_final_MC}
\end{equation}
when $T\to\infty$.  In this expression $\vect{u}^a$ are
independently drawn random vectors uniformly distributed on the unit
sphere and with only positive components. These two conditions are
introduced to satisfy the constraints $\prod_{i=1}^S \Theta( u_i )$
and $ 2 \delta( \| u \|^2 - 1 )$ that appear in the integral.  $T$ is
the sample size, and the average on the left hand side of
equation~\ref{eq:sstab_final_MC} converges to the right hand side in
the large $T$ limit.

One always has a finite sample size $T$, used to approximate the
integral.  It is therefore important to have an estimate of the error
made due to $T < \infty$.  Since the left hand side of
equation~\ref{eq:sstab_final_MC} is an average of a function over
random vectors, this error can be estimated by simply using the
variance of the function's values. In particular, the error
$\sigma_{\text{MC}}$ is defined as
\begin{equation}
\sigma_{\text{MC}} = \frac{1}{\sqrt{T}} \sqrt{\frac{ 1 }{T} \sum_{a=1}^T \left( \sum_{i,j} u^a_{i} G_{ij} u^a_{j}  \right)^{-S} - \left(\frac{ 1 }{T} \sum_{a=1}^T \left( \sum_{i,j} u^a_{i} G_{ij} u^a_{j}  \right)^{-S/2} \right)^2 }
 \ .
\label{eq:sstab_final_MCerror}
\end{equation}

The numerical simulation presented in the work where obtained were obtained with
different sampling effort $T$. Instead of fixing $T$ a priori, we determined a precision
goal, that we measured in terms of the relative error $\sigma_{MC}/\Xi$. We ran the simulations
until $\sigma_{MC}/\Xi < 0.05$. In order to avoid artificially small samples and to have enough statistical power
not to undershoot to much $\sigma_{\text{MC}}$, we ran $10\times S$ Monte Carlo
steps before checking the condition for the first time.

\section{Stability, negative definiteness, and feasibility in random matrices}

Random matrices are a useful tool in ecology, and have been studied
since May's seminal paper~\cite{May1972}. Mostly, they have been used
to model the community matrix~\cite{May1972, Allesina2012}.  In the
context of this work, we use random matrices to model interaction
matrices $\mat{A}$. We consider random matrices constructed in the
following way:
\begin{itemize}
\item $A_{ii} = -d$ where $d$ is a positive constant.
\item Each pair $(A_{ij},A_{ji})$ is set equal to a pair of random
  variables drawn from a joint distribution with probability density
  function $q(x,y)$.
\item The random variables are exchangeable---i.e., the probability
  distribution function is symmetric in its arguments:
  $q(x,y)=q(y,x)$---and all the moments are finite.
\end{itemize}

We show that the three most important quantities for our problem are
the moments
\begin{equation}
  E_1 = \int \ud x \ \ud y \ x q(x,y) = \int \ud x \ \ud y \ y q(x,y) 
  \ ,
\label{eq:E1def}
\end{equation}
\begin{equation}\displaystyle
  E_2 = \sqrt{\int \ud x \ \ud y \ (x-E_1)^2 q(x,y)}  = \sqrt{ \int \ud x \ \ud y \ (y-E_1)^2 q(x,y) } 
  \ ,
\label{eq:E2def}
\end{equation}
\begin{equation}
  E_c = \frac{1}{ E_2^2 } \int \ud x \ \ud y \ (x-E_1)(y-E_1) q(x,y) \ .
\label{eq:Ecdef}
\end{equation}
In the limit of large $S$, they can be computed as proper sample means
of $\mat{A}$'s entries:
\begin{equation}
E_1 = \frac{1}{S(S-1)} \sum_{i=1}^S\sum_{j \neq i} A_{ij} 
 \ ,
\label{eq:E1def2}
\end{equation}
\begin{equation}\displaystyle
E_2 = \sqrt{\frac{1}{S(S-1)} \sum_{i=1}^S\sum_{j \neq i} (A_{ij})^2 - E_1^2 } 
 \ ,
\label{eq:E2def2}
\end{equation}
\begin{equation}\displaystyle
  E_c = \frac{1}{ E_2^2 } \left( \frac{1}{S(S-1)} \sum_{i=1}^S\sum_{j \neq i} A_{ij}A_{ji} - E_1^2 \right) \ .
\label{eq:Ecdef2}
\end{equation}

The parameterization used by May~\cite{May1972} would correspond to
\begin{equation}\displaystyle
  q_{\text{May}}(x,y) = \Bigl( (1-C) \delta(x) + C p(x) \Bigr) \Bigl( (1-C) \delta(y) + C p(y) \Bigr)
  \ ,
\label{eq:mayparams}
\end{equation}
where $\delta(\cdot)$ is the Dirac delta function and $p(x)$ is an
arbitrary distribution with mean zero and variance $\sigma^2$. The
connectance $C$ sets the probability that each entry is equal to zero
(with probability $1-C$) or randomly drawn from the probability
distribution $p(x)$ with probability $C$. In this case $E_1=E_c=0$, while $E_2^2 = C
\sigma^2$.

In the following, we summarize known results on the spectra,
negative definiteness conditions, and properties of $\Xi$ for these matrices.

\subsection{Known results on the spectra of random matrices}
\label{sec:spectrumrnd}

Under the assumptions of the previous section, the eigenvalues of
$\mat{A}$ in the limit of large $S$ are uniformly distributed in an
ellipse in the complex plane.  If $E_1 \neq 0$ there is always an
eigenvalue $\lambda_m$ whose value is approximately
\begin{equation}\displaystyle
  \lambda_m \approx -d + S E_1 
  \ ,
\label{eq:lambdam}
\end{equation}
independently of the rest of the eigenvalue distribution. The ellipse
is centered at $-d - E_1$, its axes are aligned with the real
and imaginary axes, and their lengths are
\begin{equation}\displaystyle
  a = \sqrt{S} E_2 (1+E_c)
\label{eq:realaxis}
\end{equation}
and
\begin{equation}\displaystyle
  b = \sqrt{S} E_2 (1-E_c)
  \ .
\label{eq:imagaxis}
\end{equation}

If $\lambda_m = 0$, the eigenvalue with the largest real part(s) is approximated by the
rightmost point of the ellipse. The system is stable if its real part
is negative. In the most general case, this condition is equivalent to
\begin{equation}\displaystyle
-d +  \max \left\{ S E_1 , - E_1  +  \sqrt{S} E_2 (1+E_c) \right\}  < 0
  \ .
\label{eq:stabilityrnd}
\end{equation}

In section~\ref{sec:stabfeas} we introduced the concept of
negative definiteness. In particular, we showed that when the matrix is
negative definite then it is possible to disentangle stability and
feasibility. The matrix is negative definite if the eigenvalues of $\mat{A}
+ \mat{A}^T$ are all negative. This condition reads~\cite{Tang2014}
\begin{equation}\displaystyle
-d +  \max \left\{ S E_1 , - E_1  +  \sqrt{2S(1+E_c)} E_2  \right\} < 0
  \ .
\label{eq:reactivityrnd}
\end{equation}

\begin{figure}
  \begin{center}
    \includegraphics[width = 0.5\linewidth]{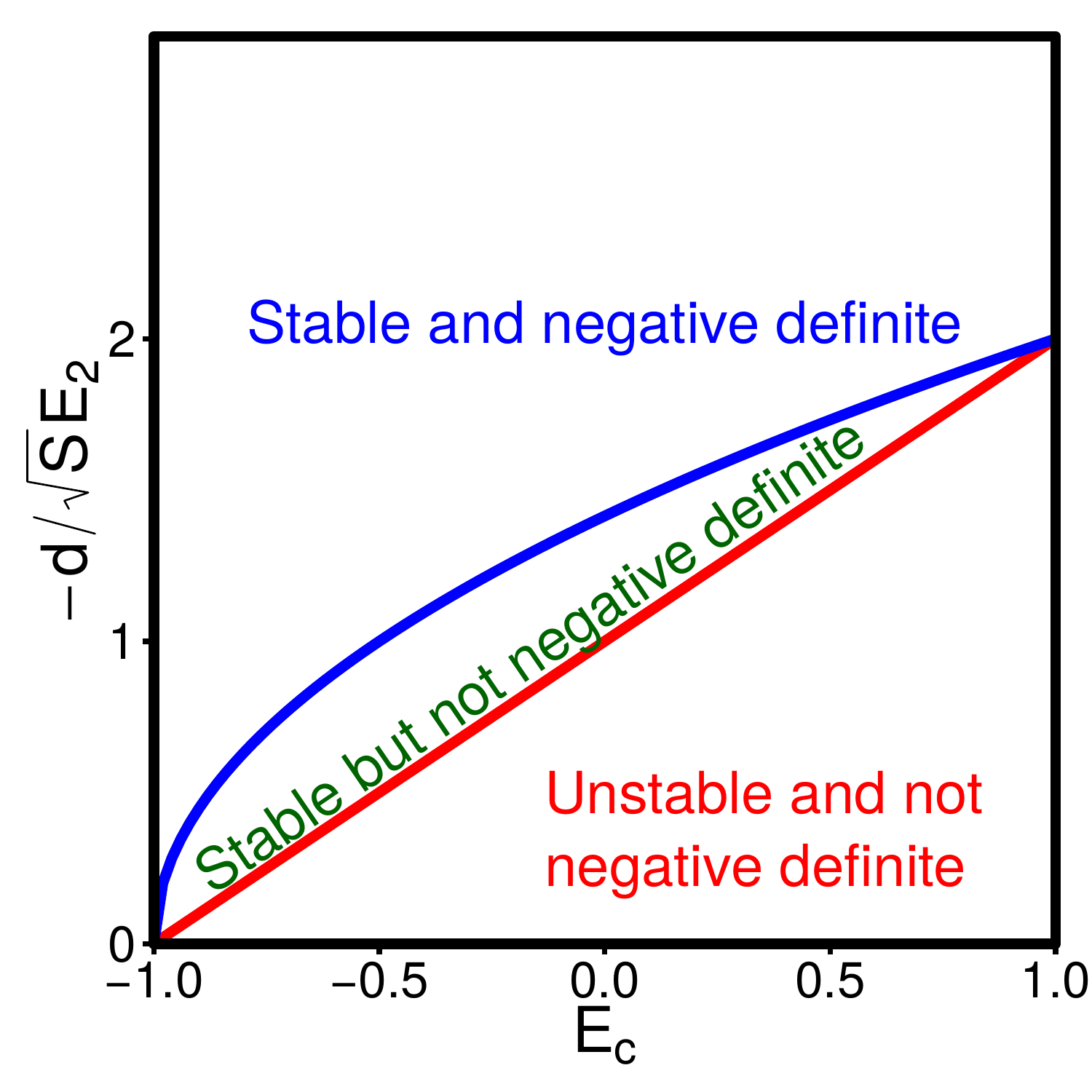}
  \end{center}
  \caption{Negative definiteness and stability for random matrices in the case
    $E_1=0$.  The red curve describes the condition for stability
    (equation~\ref{eq:stabilityrnd}), while the blue curve corresponds
    to the negative definiteness condition (equation~\ref{eq:reactivityrnd}). The
    region above the blue curve corresponds to matrices that
    are both stable and negative definite, while the region below the red
    curve corresponds to unstable and non-negative definite matrices.  The
    parameterizations that may still lead to stable and feasible
    points but we are not considering are in the region between the
    two curves. The shape of this region does not change substantially
    if $S$ and $E_2$ are changed or if $E_1<0$. For $E_1>0$ the
    not negative definite but stable region is always smaller and eventually disappears
    (i.e., the blue and the red curve become the same) when $S$ is
    large enough.  }
  \label{fig:reactivity}
\end{figure}

Figure~\ref{fig:reactivity} shows the values of parameters leading to
the possible combinations of stability and negative definiteness in random
matrices for the case $E_1 = 0$. Since we imposed that $\mat{A}$
is negative definite, the region of parameters we
explore is the one above the negative definiteness line.  One can see that in
this way we are missing some parameterizations, corresponding to those
that lead to a stable but not negative definite matrices. From
equations~\ref{eq:stabilityrnd} and~\ref{eq:reactivityrnd} one can see
that the case $E_1<0$ is very similar to the case $E_1 = 0$. More
interestingly, for $E_1 > 0$, the conditions for stability and
negative definiteness converge in the large $S$ limit, implying that we are
considering all the possible cases.

What is remarkable in these conditions and in the distribution of
eigenvalues is that they are \emph{universal}~\cite{Tao2010, Tao2010a,
  Tao2011, Allesina2015}.  Universality means that they depend only on
$S$, $E_1$, $E_2$, and $E_c$ (and $d$, but via a trivial
dependence). The spectrum of eigenvalues does not depend on the
detailed form of the distribution $q(x,y)$.

For instance, consider the case $q(x,y)=p(x)p(y)$, where the upper and
lower triangular entries $A_{ij}$ and $A_{ji}$ are independent random
variables. In this case $E_c=0$ and $E_1$ and $E_2$ are the mean and
standard deviation of the distribution $p(x)$. The distribution of
eigenvalues and the conditions for stability and negative definiteness are the
same for \emph{any} probability distribution $p(x)$ as long as their
mean $E_1$ and standard deviation $E_2$ are the same (provided some
mild conditions on higher moments hold). For instance, a Lognormal
distribution, a Gaussian distribution and an exponential distribution,
having same mean and standard deviation, produce the same eigenvalue
distribution, and therefore the same conditions for stability~\cite{ReviewRMT}.

From an ecological perspective, one can consider different interaction
matrices corresponding to different interaction types.  The
interaction type is given by the signs of the pairs $(A_{ij},A_{ji})$:
competitive interactions will have both entries with a negative sign,
while in trophic interactions the entries will have opposite sign.
The interaction pairs $(A_{ij},A_{ji})$ for competitive interactions
can for instance be obtained from the following distribution:
\begin{equation}\displaystyle
  q_{\text{comp}}(x,y) = (1-C) \delta(x)\delta(y) + C h_{-}(x) h_{-}(y)
  \ ,
\end{equation}
where $h_{-}$ is a probability distribution function with support on
the negative axis (i.e., the random variables are always negative),
and $C$ is the connectance (a pair is different from zero with
probability $C$). In the case of trophic interactions we could
consider
\begin{equation}\displaystyle
  q_{\text{troph}}(x,y) = (1-C) \delta(x)\delta(y) + \frac{C}{2} p_{-}(x) p_{+}(y) + \frac{C}{2} p_{+}(x) p_{-}(y)
  \ ,
\end{equation}
where $p_{+}$ and $p_{-}$ are two probability distribution functions
with positive and negative support, respectively.  Suppose that the
moments of $h_{-}$, $p_{+}$, and $p_{-}$ are chosen in such a way that
$q_{\text{comp}}(x,y)$ and $q_{\text{troph}}(x,y)$ have the same
values of $E_1$, $E_2$, and $E_c$. The interaction matrices will still
look very different in the two cases: one describes a foodweb and the
other a competitive system.  Despite this difference, the two will
have the same stability properties.  In other words, different
interaction types influence the stability properties of the system
only via $E_1$, $E_2$ and $E_c$.

\subsection{Universality of $\Xi$}

In this section we show that, apart from their spectral distribution,
$\Xi$ is also a universal quantity in large random matrices. That is, in the large $S$ limit,
its value does not depend on the entire distribution of the
coefficients, but only on the three moments $E_1$, $E_2$, and
$E_c$. It is important to remark that this result applies to the large $S$ limit: the sub-leading
corrections depend in principle on all the moments.

In order to show that $\Xi$ is universal, we parameterized random
networks with different distributions and checked whether $\Xi$
depends only on $E_1$, $E_2$, $E_c$, and $S$, but not on other
properties.  To do this, we constructed several $S \times S$
matrices. Each individual matrix had its entries drawn from some fixed
distribution, but the shape of the distribution was different across
matrices. However, regardless of the distribution's shape, their
moments were fixed at $E_1$, $E_2$, and $E_c$. We then checked whether
these matrices led to the same value of $\Xi$.

In our simulations we considered a distribution of the pairs $(A_{ij},A_{ji})$ of the form
\begin{equation}\displaystyle
  q(x,y) = (1-C) \delta(x)\delta(y) + C p(x,y)
  \ ,
  \label{eq:simdist}
\end{equation}
where the connectance $C$ is the probability that two species $i$ and
$j$ interact. The probability distribution $p(x,y)$ in
equation~\ref{eq:simdist} depends on three parameters $\mu$, $\sigma$,
and $\rho$, which define the mean, variance, and correlation of the
pairs drawn from $p(x,y)$. Given the values of $E_1$, $E_2$, and
$E_c$, we can arbitrary choose $C$ and tune $\mu$, $\sigma$, and
$\rho$ to obtain any desired $E_1$, $E_2$, and $E_c$. If $\Xi$ is
universal, then different matrices built with different values of $C$,
$\mu$, $\sigma$, and $\rho$ but the same values of $E_1$, $E_2$, and
$E_c$ will lead to the same $\Xi$.

We considered five parameterizations of the distribution $p(x,y)$:
\begin{itemize}
\item Random signs, normal distribution:
  \begin{equation}
    p(x,y) = BN(x,y|\mu,\sigma,\rho) \ .
  \end{equation}
  The distribution $BN(x,y|\mu,\sigma,\rho)$ is a bivariate normal
  distribution with marginal means equal to $\mu$, marginal variances
  equal to $\sigma^2$, and correlation equal to $\rho\sigma^2$. The
  pairs can in principle assume all possible combinations of signs.
\item Random signs, four corners:
  \begin{equation}
    \begin{split}
      p(x,y) = & \frac{q}{2} \delta( x - \mu - \sigma )\delta( y - \mu
      - \sigma ) +
      \frac{q}{2} \delta( x - \mu + \sigma )\delta( y - \mu + \sigma ) \\
      + & \frac{1-q}{2} \delta( x - \mu - \sigma )\delta( y - \mu +
      \sigma ) + \frac{1-q}{2} \delta( x - \mu + \sigma )\delta( y -
      \mu - \sigma ) \ .
    \end{split}
  \end{equation}
  The pairs $(x,y)$ can take on only four different, discrete values,
  potentially corresponding to all combinations on signs. The
  probability distribution depends on three parameters $\mu$ and
  $\sigma^2$ are means an variances of the distribution, while the
  correlation $\rho \sigma^2$ can be obtained from $\rho = 2q-1$.
\item $(+,+)$, Lognormal:
  \begin{equation}
    p(x,y) = LBN(x,y|\mu,\sigma,\rho) \ .
  \end{equation}
  The distribution $LBN(x,y|\mu,\sigma,\rho)$ is a bivariate lognormal
  distribution with marginal means equal to $\mu>0$, marginal
  variances equal to $\sigma^2$, and correlation equal to
  $\rho\sigma^2$. The pairs can in principle assume only positive
  signs. Note that not all values of $\rho$ between $-1$ and $1$ can
  be obtained when a Lognormal distribution is considered.
\item $(-,-)$, Lognormal:
  \begin{equation}
    p(x,y) = LBN(-x,-y|-\mu,\sigma,\rho) \ .    
  \end{equation}
  This distribution takes the values drawn from a bivariate lognormal
  distribution, times $-1$.  It has marginal means equal to $\mu < 0$,
  marginal variances equal to $\sigma^2$, and correlation equal to
  $\rho\sigma^2$. The pairs assume only negative signs. Note that not
  all values of $\rho$ between $-1$ and $1$ can be obtained when a
  Lognormal distribution is considered.
\item $(+,-)$, Lognormal:
  \begin{equation}
    \begin{split}
      p(x,y) = & \frac{1}{2} LN(x|\mu_1,(1+\rho)\sigma)LN(-y|-\mu_2,(1+\rho)\sigma) \\
      + & \frac{1}{2}
      LN(y|\mu_1,(1+\rho)\sigma)LN(-x|-\mu_2,(1+\rho)\sigma) \ .
    \end{split}
  \end{equation}
  The distribution $LN(x|\mu,\sigma)$ is Lognormal distribution with
  mean $\mu_1 + \mu_2$ (where $\mu_1>0$ and $\mu_2<0$), variance
  $\sigma^2$, and correlation $\rho\sigma^2$.  The pairs assume only
  values with opposite signs $(+,-)$ or $(-,+)$.
\end{itemize}
In ecological terms, the first two distributions correspond to a
random community (where the signs of the interaction strength are
random), the $(+,+)$ case corresponds to a mutualistic community,
$(-,-)$ to a competitive community, while $(+,-)$ corresponds to a
food web. The mutualistic/competitive matrices can lead only to
positive/negative means $E_1$, respectively, while the other settings
can produce arbitrarily values of $E_1$.

\begin{figure}
  \begin{center}
    \includegraphics[width =
    0.9\linewidth]{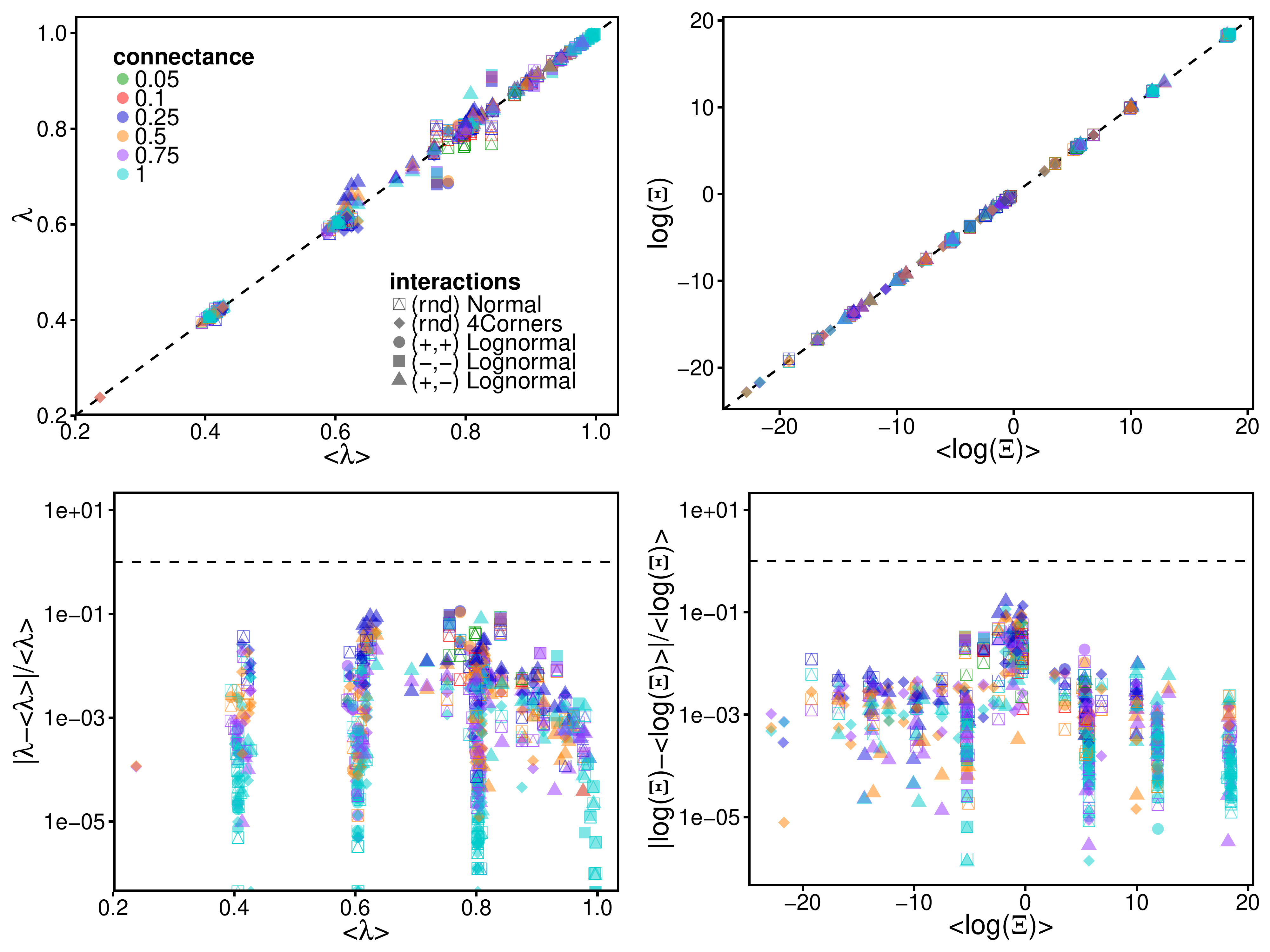}
  \end{center}
  \caption{Universality of $\lambda$ and $\Xi$ in random matrices. The
    two left panels refer to the eigenvalue with the largest real part
    $\lambda$ of the interaction matrix $\mat{A}$, while the right
    ones to the size $\Xi$ of the feasibility domain. We consider
    different values of the connectance (colors) and different
    distributions (shape), such that there were multiple combination
    of connectances and distributions having the same values of $E_1$,
    $E_2$, and $E_c$. We computed the averages $\big<\lambda\big>$ and
    $\big<\log(\Xi)\big>$ over all realizations of the matrices having
    the same values of $E_1$, $E_2$, and $E_c$. If the value of
    $\lambda$ and $\Xi$ are universal, then they depend only on $E_1$,
    $E_2$, and $E_c$, and therefore their values are equal to the
    mean: universality holds if $\lambda = \big<\lambda\big>$ and
    $\log(\Xi) = \big<\log(\Xi)\big>$. The top panels show that these
    two quantities are equal and the bottom panels quantify their
    deviations. We know that $\lambda$ is universal, and since $\Xi$
    has a similar behavior, we conclude that $\Xi$ is also universal.}
  \label{fig:checkuniversality}
\end{figure}

Figure~\ref{fig:checkuniversality} shows the value of $\Xi$ and of the
largest eigenvalue $\lambda$ for interaction matrices constructed with
different connectances $C$ and distributions, but with the same values
of $E_1$, $E_2$, and $E_c$. As seen from the figure, the values of
$\Xi$ and $\lambda$ in any particular case match up precisely with the
average values over several different realizations, demonstrating that
these two quantities are indeed universal.

\section{Mean-field approximation of $\Xi$}
\label{sec:approxxi}

The goal of this section is to compute an approximation for $\Xi$ in
the limit of large $S$.  The volume $\Xi$ is defined (see
section~\ref{sec:numint}) as
\begin{equation}
  \Xi = \frac{2^{S} \ \Gamma(S/2) \ \sqrt{\det(\mat{G})} }{2\pi^{S/2}} \int_{\mathbb{S}_S} \ud^S \vect{u} \ \prod_{i=1}^S \Theta( u_i ) \ 
  \left( \sum_{i,j} u_{i} G_{ij} u_{j}  \right)^{-S/2}
  \ ,
  \label{eq:sstab_final_conto}
\end{equation}
where the matrix $\mat{G}$ can be obtained from the generators of the
polytope (see equations~\ref{eq:generatormat} and~\ref{eq:Gmat}), and
therefore from the interaction matrix $\mat{A}$.

We can introduce a Gaussian function in
equation~\ref{eq:sstab_final_conto} using the fact that, for any
positive constant $c$,
\begin{equation}
  c^{-S/2} = \frac{2}{\Gamma(S/2)} \int_0^\infty \ud r \ r^{S-1} \exp( - c r^2 ) \ .
  \label{eq:sstab_Gauss}
\end{equation}
Introducing this Gaussian integral in
equation~\ref{eq:sstab_final_conto} by letting $c = \sum_{i,j} u_i
G_{ij} u_j$, we obtain
\begin{equation}
  \Xi = \sqrt{\det(\mat{G})} \left(\frac{2 }{\sqrt{\pi}}\right)^S  \int_0^\infty \ud r \ r^{S-1}
  \int_{\mathbb{S}_S} \ud^S \vect{u} \ \left( \prod_{i=1}^S \Theta( u_i ) \right) \ 
  \exp\left( - r^2 \sum_{i,j} u_{i} G_{ij} u_{j}  \right)
  \ ,
  \label{eq:sstab_final_conto2}
\end{equation}
which can be rewritten as
\begin{equation}
  \Xi = \sqrt{\det(\mat{G})}  \Bigl(\frac{2 }{\sqrt{\pi}}\Bigr)^S 
  \int_{\mathbb{R}^S} \ud^S \vect{z} \ \Bigl(  \prod_{i=1}^S \Theta( z_i )\Bigr) \ 
  \exp\Bigl( - \sum_{i,j} z_{i} G_{ij} z_{j}  \Bigr)
  \ ,
  \label{eq:sstab_final_conto3}
\end{equation}
where $z_i = r u_i$. We can rewrite this equation as
\begin{equation}
  \begin{split}
    \Xi = & \sqrt{\det(\mat{G})} \left(\frac{2 }{\sqrt{\pi}}\right)^S
    \int_{\mathbb{R}^S} \ud^S \vect{z} \ \prod_{i=1}^S \left(
    \Theta( z_i ) \ e^{-z_i^2 } \exp\left( - \sum_{j\neq i} z_{i}
    G_{ij} z_{j} \right) \right) \ ,
    \label{eq:sstab_final_conto4}
  \end{split}
\end{equation}
where we used the fact that the diagonal entries of $\mat{G}$, when
expressed in terms of the normalized generators, are equal to one.

The reader familiar with statistical mechanics will notice that
equation~\ref{eq:sstab_final_conto4}, which can be written as
\begin{equation}
  \begin{split}
    \Xi \propto \int_{\mathbb{R}^S} \ud^S \vect{z} \
    q(\vect{z}) \ \prod_{i=1}^S \left( \exp\left( - \sum_{j\neq
      i} z_{i} G_{ij} z_{j} \right) \right) \ ,
    \label{eq:partitionf}
  \end{split}
\end{equation}
has the form of a partition function. For instance one can recover the
Ising model~\cite{Parisi1998} with the choice
$q(\vect{z})=\prod_i\delta(z_i^2 = 1)$ or the spherical
model~\cite{Berlin1952} when $q(\vect{z})=\delta(S - \sum_i z_i^2
)$. The term $z_{i} G_{ij} z_{j}$ in particular plays the role of the
interactions of the system.

Integrals of the form~\ref{eq:partitionf} are the most studied objects
of statistical mechanics, and yet in most cases are not analytically
solvable. There are, on the other hand, many techniques that can be
used to obtain good approximations to~\ref{eq:partitionf}. The most
celebrated one is probably the mean-field
approximation~\cite{Parisi1998} and it is the one we are using in this
section. In particular, the idea of the mean-field approximation is to
replace the interactions of an entity (spins in the case of the Ising
model or species in our case) with an average ``effective''
interaction. This reduces a many-body problem, where all interactions
of spins or populations are coupled, into an effective one-body
problem. 

If the system is large enough (in our case if $S \to \infty$),
the mean-field approximation is know to be exact in the case of
``fully connected'' interactions. In terms of
equation~\ref{eq:partitionf}, this corresponds to a matrix $\mat{G}$
with the same constant in all its offdiagonal entries. The matrix
$\mat{G}$ is constant when $\mat{A}$ has constant offdiagonal
entries. We will consider therefore the case of $\mat{A}$'s diagonal
entries being equal to $-1$ and its offdiagonal entries to a constant
$E_1$. Using equation~\ref{eq:generatormat}, the $i$th component of
the $k$th generator is then
\begin{equation}
  g^k_i = - \frac{E_1}{1 + (S-1)E_1^2}
\end{equation}
for $i \neq k$, and
\begin{equation}
  g^k_k = \frac{1}{1 + (S-1)E_1^2} \ .
\end{equation}
Using equation~\ref{eq:Gmat}, we therefore obtain that the diagonal
entries of $\mat{G}$ are equal to $1$, while the offdiagonal ones are
constant and equal to
\begin{equation}
  G_{ij} = \frac{-2 E_1 + (S-2) E_1^2}{1 + (S-1)E_1^2} \ .
\end{equation}
We define the constant $\beta$ as
\begin{equation}
  \beta = S \frac{-2 E_1 + (S-2) E_1^2}{1 + (S-1)E_1^2} \ ,
  \label{eq:betadef}
\end{equation}
and therefore we have $G_{ii}=1$ and $G_{ij}=\beta/S$ for $i\neq
j$. The determinant of $\mat{G}$ in this case turns out to be
\begin{equation}
 \det(\mat{G}) = \left(1+ \frac{S-1}{S}\beta\right) \left(1 - \frac{\beta}{S} \right)^{S-1} \approx (1 + \beta) e^{-\beta} \ ,
  \label{eq:detmeanfield}
\end{equation}
where the last form holds for large $S$. In this case of constant
interactions, we obtain, from equation~\ref{eq:sstab_final_conto4},
\begin{equation}
  \begin{split}
    \Xi = & \sqrt{\det(\mat{G})} \left(\frac{2 }{\sqrt{\pi}}\right)^S
    \int_{\mathbb{R}^S} \ud^S \vect{z} \ \prod_{i=1}^S \left(
    \Theta( z_i ) \ e^{-z_i^2 } \exp\left( - z_{i} \frac{\beta}{S}
    \sum_{j\neq i} z_{j} \right) \right) = \\
    & = \sqrt{\det(\mat{G})} \left(\frac{2 }{\sqrt{\pi}}\right)^S
    \int_{\mathbb{R}^S} \ud^S \vect{z} \  \left(
    \prod_{i=1}^S \Theta( z_i ) \right) \
    \exp\left( - \sum_i z_i^2  - \frac{\beta}{S}
    (\sum_{i} z_{i} )^2  \right)  
     \ ,
    \label{eq:sstab_final_conto5}
  \end{split}
\end{equation}
up to subleading terms in $S$.

Equation~\ref{eq:sstab_final_conto5} can be written as
\begin{equation}
  \begin{split}
    \Xi = & \sqrt{\det(\mat{G})} \left(\frac{2 }{\sqrt{\pi}}\right)^S Z_h
    \Bigl\langle \exp\left(  - \frac{\beta}{S}
    (\sum_{i} z_{i} )^2 + h \sum_i z_i \right)
    \Bigr\rangle_h
     \ ,
  \end{split}
\end{equation}
where
\begin{equation}
  \begin{split}
Z_h & :=
 \int_{\mathbb{R}^S} \ud^S \vect{z} \  \left(
    \prod_{i=1}^S \Theta( z_i ) \right) \
    \exp\left( - \sum_i z_i^2  - h \sum_i z_i  \right) = \\
&   = \left( \int_0^\infty \ud z \ e^{-z^2 - h z} 
    \right)^S =  \left( \frac{\sqrt{\pi}}{2} e^{h^2/4} \erfc(h/2)
  \right)^S
     \ ,
    \label{eq:sstab_final_Zh}
  \end{split}
\end{equation}
where $\erfc(\cdot)$ is the complementary error function, defined as
\begin{equation}
  \erfc(x) = \frac{2}{\sqrt{\pi}} \int_x^\infty \ud t \ e^{-t^2} 
  \ .
  \label{eq:erfcdef}
\end{equation}
The average $\langle\cdot\rangle_h$ is defined as
\begin{equation}
  \begin{split}
    \Bigl\langle f(\vect{z})
    \Bigr\rangle_h :=
    \frac{ 1 }{ 
 Z_h} \int_{\mathbb{R}^S} \ud^S \vect{z} \  \left(
    \prod_{i=1}^S \Theta( z_i ) \right) \
    \exp\left( - \sum_i z_i^2  - h \sum_i z_i  \right) f(\vect{z}) 
     \ .
    \label{eq:sstab_final_conto7}
  \end{split}
\end{equation}
Using Jensen's inequality in equation~\ref{eq:sstab_final_conto7} we have that
\begin{equation}
  \begin{split}
    \Xi = & \sqrt{\det(\mat{G})} \left(\frac{2 }{\sqrt{\pi}}\right)^S Z_h
    \Bigl\langle \exp\left(  - \frac{\beta}{S}
    (\sum_{i} z_{i} )^2 + h \sum_i z_i \right)
    \Bigr\rangle_h \geq \\
    & \geq \sqrt{\det(\mat{G})} \left(\frac{2 }{\sqrt{\pi}}\right)^S Z_h
     \exp\left( \Bigl\langle - \frac{\beta}{S}
    (\sum_{i} z_{i} )^2 + h \sum_i z_i \Bigr\rangle_h \right) 
     \ .
    \label{eq:sstab_final_conto8}
  \end{split}
\end{equation}
In the following we will approximate the first expression with the second one.
It is possible to prove that, in the large $S$ limit, the second expression converges to the
first one.

Applying the mean-field approximation we neglect fluctuations of the variables,
i.e. we have
\begin{equation}
\Bigl\langle - \frac{\beta}{S}
    (\sum_{i} z_{i} )^2 + h \sum_i z_i \Bigr\rangle_h
= -\frac{\beta}{S} \Bigl\langle ( \sum_i  z_i )^2 \Bigr\rangle_h + h \sum_i \langle z_i \rangle_h
\approx
S \left(- \beta m^2 +  h m \right)\ ,
  \label{eq:sstab_final_conto9}
\end{equation}
where
\begin{equation}
m := \langle z_i \rangle_h = - \frac{1}{S} \frac{\partial }{\partial h} \log(Z_h) \ .
  \label{eq:meqdef}
\end{equation}
By introducing equation~\ref{eq:sstab_final_conto9} in equation~\ref{eq:sstab_final_conto8}
we have
\begin{equation}
  \Xi \approx \sqrt{\det(\mat{G})}  Z_h \left( \frac{2 }{\sqrt{\pi}}
\exp\left( - \beta m^2 + h m  \right)\right)^S = \Xi_{MF}
 \ .
  \label{eq:sstab_final_conto11}
\end{equation}
This equation is a function of $h$, which is a free parameter. Since it is a
lower bound for the actual value of $\Xi$, the best approximation would correspond
to the value of $h$ which maximizes the approximation.
We have therefore that $h$ is a solution of the following equation
\begin{equation}
 0 = \frac{\partial}{\partial h} \log( \Xi_{MF} ) =  \frac{\partial }{\partial h} \log(Z_h) +
S \frac{\partial }{\partial h} ( -  \beta m^2 + h m ) = 
S ( h - 2 \beta m ) \frac{\partial m }{\partial h} 
 \ ,
  \label{eq:sstab_final_selfconst}
\end{equation}
where $m$ is given by equation~\ref{eq:meqdef}. We obtain therefore $m = h/(2\beta)$ and then, by
neglecting sub-leading terms in $S$ and introducing $m = h/(2\beta)$ in equation~\ref{eq:sstab_final_conto11}
\begin{equation}
\frac{1}{S}\log\Xi_{MF} \approx \log \left( \erfc(h/2)
\exp\left( \frac{h^2}{4} \frac{1+\beta}{\beta} \right)\right)  
 \ .
  \label{eq:sstab_final_mfpap}
\end{equation}
By maximizing this equation respect to
$h$ we obtain
\begin{equation}
 0 = \frac{\partial}{\partial h} \log( \Xi_{MF} ) =  \frac{h}{2}\left(\frac{1}{\beta}+1\right) + \frac{\partial }{\partial h} \log\left( \erfc(h/2) \right)
=  \frac{h}{2}\left(\frac{1}{\beta}+1\right) - \frac{e^{-h^2/4}}{\sqrt{\pi}\erfc(h/2)}
 \ .
  \label{eq:sstab_selfconst2}
\end{equation}

Equation~\ref{eq:sstab_selfconst2} cannot be solved exactly.
By expanding around $h=0$ we obtain
\begin{equation}
 0 = 
\frac{h}{2}\left(\frac{1}{\beta}+1\right) - \frac{1}{\sqrt{\pi}} - \frac{h}{\pi}
 \ ,
  \label{eq:sstab_selfconst2exp}
\end{equation}
which is solved by
\begin{equation}
h =  \frac{2 \beta \sqrt{\pi}}{\pi + \beta(\pi-2)}
 \ .
  \label{eq:sstab_solution_appj}
\end{equation}
One can observe that the solution $h=0$ corresponds to $\beta=0$, i.e. to a
non-interacting ecosystem. Expanding around $h=0$ is therefore meaningful when
the interactions are not too strong. It is possible to verify that the approximate
solution~\ref{eq:sstab_solution_appj} is very close to the actual solution
obtained by solving numerically equation~\ref{eq:sstab_selfconst2} also for 
not too small values of $\beta$

\begin{figure}
  \begin{center}
    \includegraphics[width =
    0.45\linewidth]{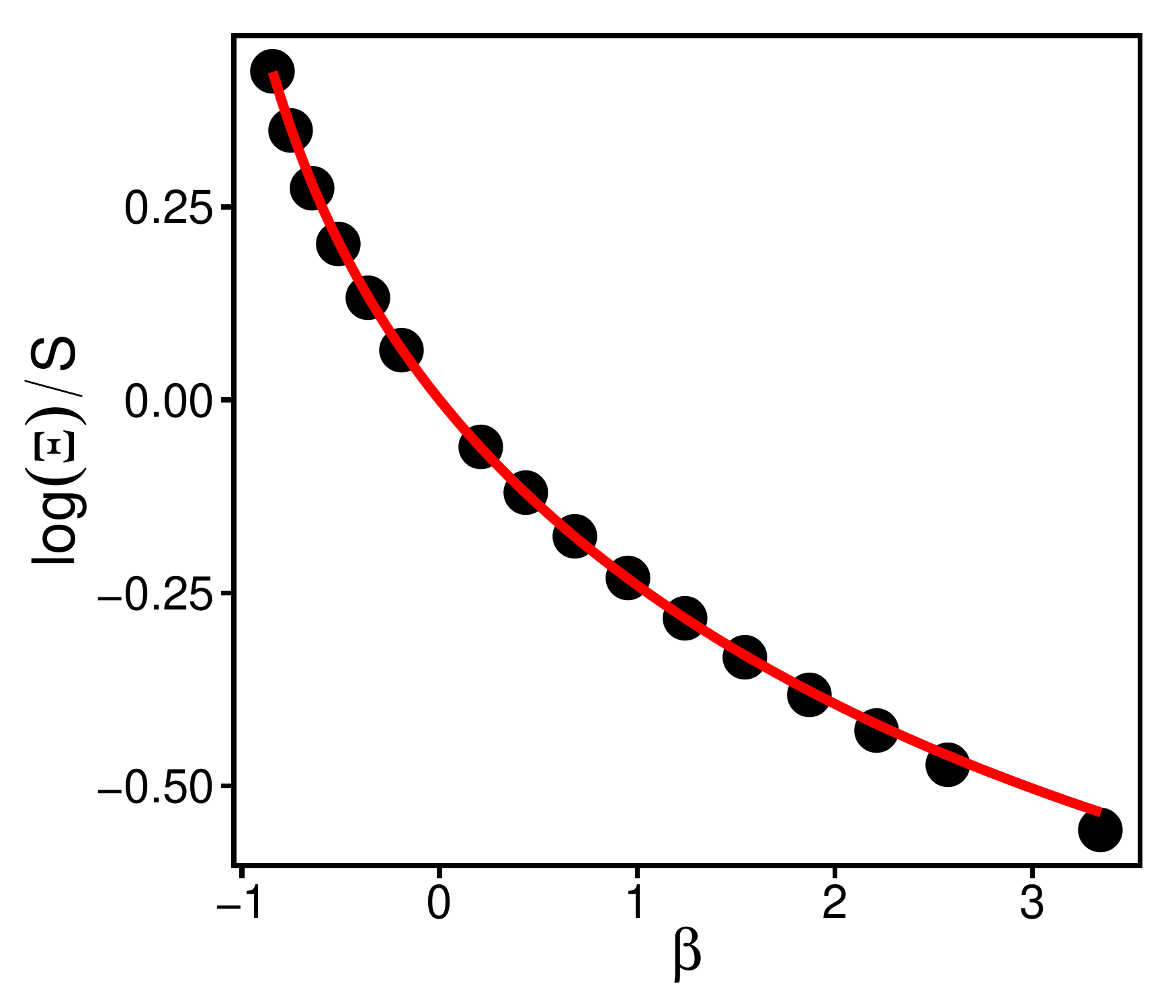}
  \end{center}
  \caption{Approximation of $\Xi$ using mean field theory. The black
    dots are numerical simulations obtained by integrating $\Xi$
    numerically (see section~\ref{sec:numint}) for a constant
    interaction matrix. The red curve is the analytical approximation
    obtained using the mean-field approximation (see
    equation~\ref{eq:sstab_final_conto9_2}). $\beta$ is a function of
    $E_1$ and $S$, and is defined in equation~\ref{eq:betadef}. The
    range of $\beta$ considered here is the same of the one appearing
    in figure 1 of the main text.}
  \label{fig:meanfieldapprox}
\end{figure}

Using equation~\ref{eq:sstab_solution_appj} into equation~\ref{eq:sstab_final_mfpap} we obtain
\begin{equation}
\frac{1}{S}\log\Xi_{MF} \approx  \frac{\beta(1+\beta)\pi}{\left(\pi+\beta(\pi-2)\right)^2}
+ \log \erfc\left(
 \frac{ \sqrt{\pi} \beta  }{ \pi + \beta(\pi-2) }
\right)
 \ ,
  \label{eq:sstab_final_mfpap2}
\end{equation}
which is our final result. In figure~\ref{fig:meanfieldapprox} we compare
this equation with the volume computed numerically in the case of constant interactions, finding
a very good match.

%By considering the leading term in $S$ we have that
%\begin{equation}
%  \frac{1}{S} \log(\Xi_{MF}) \sim 
%\log\left( \erfc(h/2)\left) \ ,
%  \label{eq:sstab_final_conto9}
%\end{equation}
%where only the leading term in $S$ was
%kept. 

In the most general case of an interaction matrix with nonconstant
offdiagonal entries, we can consider
equation~\ref{eq:sstab_final_conto9} as an approximation valid in the
case of $E_2 \to 0$. As $\beta$ was defined in terms of the
generators, we can extend the approximation to the case $E_2 > 0$ by
considering $\beta$ as the expected value of $\mat{G}$'s entries,
which corresponds to the average overlap of two rows of the
interaction matrix $\langle\cos(\eta)\rangle$, defined in
equation~\ref{eq:sidematmean}.  In this more general case the
mean-field value of $\Xi$ is expected to be a good approximation when
$\mathrm{var}(\cos(\eta))$ is small enough. By substituting
$\beta=\langle\cos(\eta)\rangle$, using equation~\ref{eq:sidematmeanfin},
into equation~\ref{eq:sstab_final_conto9} we obtain
\begin{equation}
\begin{split}
  \frac{1}{S} \log(\Xi) & \approx 
\frac{\pi  E_1 (2 d-E_1 S) \left(2 d E_1+d-S \left(2 E_1^2+E_2^2\right)\right)}{\left(d (2 (\pi -2) E_1+\pi )-S \left(2 (\pi -1) E_1^2+\pi  E_2^2\right)\right)^2} \\
&\log \left(\erfc\left(\frac{\sqrt{\pi } E_1 (E_1 S-2 d)}{S \left(2 (\pi -1) E_1^2+\pi  E_2^2\right)-d (2 (\pi -2) E_1+\pi )}\right)\right)
  \ .
\end{split}
  \label{eq:sstab_final_conto9_2}
\end{equation}
When $\mathrm{var}(\cos(\eta))$ is not small, we observed that the
empirical formula
\begin{equation}
\begin{split}
  \frac{1}{S} \log(\Xi) & \approx 
\frac{\pi  E_1 (2 d-E_1 S) \left(2 d E_1+d-S \left(2 E_1^2+E_2^2\right)\right)}{\left(d (2 (\pi -2) E_1+\pi )-S \left(2 (\pi -1) E_1^2+\pi  E_2^2\right)\right)^2} \\
&\log \left(\erfc\left(\frac{\sqrt{\pi } E_1 (E_1 S-2 d)}{S \left(2 (\pi -1) E_1^2+\pi  E_2^2\right)-d (2 (\pi -2) E_1+\pi )}\right)\right) + \\
& +   \log\left( 1 + \frac{ 3 S E_2^2 (1+E_c) }{2 \pi} \right)
  \ .
  \label{eq:empirical_formula}
\end{split}
\end{equation}
explains well the values obtained in simulations. This is the formula
we used to make figure 2 in the main text.

In order to simplify the expression and make it more readable, we can expand
equation~\ref{eq:sstab_final_mfpap2} around $\beta=0$, i.e., when the interactions
between species are small. By expanding $(\Xi_{MF})^{1/S}$ around $\beta=0$
and taking the logarithm of the expression, we obtain 
\begin{equation}
\frac{1}{S}\log\Xi_{MF} \approx   \log \left(
1 - \frac{\beta}{\pi}
\right)
 \ .
\end{equation}
Equation 2 of the main text was obtained by substituting $\beta=\langle\cos(\eta)\rangle$, using equation~\ref{eq:sidematmeanfin}, in the case of $E_2 = 0$.

\section{Feasibility of consumer-resource communities}

This section considers explicitly a community with two trophic levels and consumer-resource interactions. While empirical communities have a more complicated interaction structure, this example is particularly relevant to better understand how $\Xi$ should be interpreted.

We consider a system with $S_R$ resource and $S_C$ consumer ($S_R + S_C = S$) populations,
whose dynamics is described by equation~\ref{eq:dynmodel_LV} with the interaction matrix
\begin{equation}
\label{eq:consres_intmat}
\mat{A} = \left(
    \begin{array}{cc}
     -\mat{C} & -\mat{B} \\ 
    \mat{Z} \mat{B}^T \mat{W} & \mat{0} \\
  \end{array}
  \right) \ ,
\end{equation}
where $\mat{C}$ is an $S_R \times S_R$ nonnegative matrix, $\mat{B}$ is an $S_R \times S_C$ nonnegative matrix, while $\mat{Z}$ and $\mat{W}$ are two positive diagonal matrices of dimension $S_R \times S_R$ and $S_C \times S_C$, respectively.

If $\mat{C}$ is a positive diagonal matrix, any feasible fixed point is globally asymptotically stable~\cite{Case1979}. When $\mat{C}$ is not diagonal, one can prove that any feasible fixed point
is globally asymptotically stable if $\mat{C} \mat{W}^{-1}$ is positive definite (i.e., $-\mat{C} \mat{W}^{-1}$ is
negative definite). Assuming that this condition holds, stability of feasible fixed points is ensured and we can study feasibility alone.

Using equation~\ref{eq:fixed_point_eq}, we obtain the equations
\begin{equation}
r^R_i = \sum_{j=1}^{S_R} C_{ij} n^{R*}_j + \sum_{j=1}^{S_C} B_{ij} n^{C*}_j  \ ,
\end{equation}
\begin{equation}
- r^C_i = \sum_{j=1}^{S_R} Z_i B_{ji} W_j n^{R*}_j \ ,  
\end{equation}
where $\vect{r}^R$ and $\vect{r}^C$ are the intrinsic growth rates of resources and consumers, while $\vect{n}^{R*}$ and $\vect{n}^{C*}$ are their equilibrium abundances. Since all the matrices that appear in this equation are nonnegative, an intrinsic growth rate vector is contained in the feasibility domain only if $r^R_i > 0$ for all $i = 1,\dots,S_R$ and $r^C_i < 0$ for all $i = 1,\dots,S_C$. An intrinsic growth rate vector that does not respect these conditions is not in the feasibility domain. The feasibility domain is therefore fully contained in one orthant, implying that the maximum value of its size is $\Xi=1$.

The $S$-dimensional volume of the feasibility domain is nonzero
only if it is defined by $S$ linearly independent generators.
The generators of the feasibility domain are proportional to the columns of the interaction matrix.
If the interaction matrix has the form of equation~\ref{eq:consres_intmat}, $S_R$ generators will have the form
$\vect{g} = ( \vect{v}, \vect{0} )$, where $\vect{v}$ has $S_C$ components. These generators can be linearly independent only
if $S_R \geq S_C$, and therefore $\Xi > 0$ only if $S_R \geq S_C$.
More generally, if $\det(A)=0$, then $\Xi = 0$~\cite{Haerter2016}.

Assuming that the determinant of $A$ is different from zero, we can use equation~\ref{eq:sstab_change} obtaining
\begin{equation}
    \Xi =  \sqrt{\det(\mat{A})} \left(\frac{2 }{\sqrt{\pi}}\right)^S
    \int_{\mathbb{R}^S} \ud^S \vect{z} \ \left( \prod_{i=1}^S 
    \Theta( z_i ) \right)  \  \exp\left(  \sum_{ij} z_{i}
    A_{ij} z_{j} \right) \ . 
\end{equation}
Given the structure of the matrix $A$, it is convenient to write $\vect{z} = ( \vect{v}, \vect{u} )$,
where $\vect{v}$ and $\vect{u}$ are two vectors with $S_R$ and $S_C$ components respectively.
The argument of the exponential can be rewritten as
\begin{equation}
\sum_{ij} z_{i} A_{ij} z_{j} = - \sum_{i=1}^{S_R} \sum_{j=1}^{S_R} v_i C_{ij} v_j - \sum_{i=1}^{S_R} \sum_{j=1}^{S_C} v_i B_{ij} (1-Z_i W_j) u_j \ . 
\end{equation}
By integrating over the variables $\vect{u}$, we finally obtain
\begin{equation}
    \Xi =  \sqrt{\det(\mat{A})} \left(\frac{2 }{\sqrt{\pi}}\right)^S
    \int_{\mathbb{R}^S} \ud^{S_R} \vect{v} \ \left( \prod_{i=1}^{S_R} 
    \Theta( v_i ) \right)  \  \exp\left(  \sum_{ij} v_{i}
    C_{ij} v_{j} \right) \frac{1}{\prod_{j=1}^{S_C} \sum_{i=1}^{S_R} v_i B_{ij} (1-Z_i W_j)  }  \ . 
\end{equation}

%This expression can be simplified by neglecting inter-specific competition among resources (i.e. $C_{ij} = \delta_{ij}$).
%Assuming $Z_i W_j = \eta$ and $B_{ij} = \gamma L_{ij}$, the determinant of $A$ simply becomes
%\begin{equation}
%\det(A) =  \det( \gamma^2 \eta L L^T ) = \left( \gamma^2 \eta \right)^{S_C} \det( L L^T ) \ , 
%\end{equation}
%and therefore we obtain (CHECK IF CORRECT )
%\begin{equation}
%    \Xi = \gamma^{S_C} \sqrt{\eta^{S_C}} \sqrt{\det(\mat{L}^T \mat{L})} \left(\frac{2 }{\sqrt{\pi}}\right)^S
%    \int_{\mathbb{R}^S} \ud^{S_R} \vect{v} \ \left( \prod_{i=1}^S 
%    \Theta( v_i ) \right)  \  \exp\left( - \sum_{i} v_{i}^2  \right) \frac{1}{\gamma^{S_C} (1-\eta)^{S_C}\prod_{j=1}^{S_C} \sum_{i=1}^{S_R} v_i L_{ij} }  \ .   
%\end{equation}
%Note that the determinant of $\mat{L}^T \mat{L}$ (CHECK THIS) is different from zero only if $S_R \geq S_C$
%and therefore we obtain
%\begin{equation}
%    \Xi =  \left(\frac{\sqrt{\eta}}{(1-\eta)} \right)^{S_C} \sqrt{\det(\mat{L}^T \mat{L})} \left(\frac{2 }{\sqrt{\pi}}\right)^S
%    \int_{\mathbb{R}^S} \ud^{S_R} \vect{v} \ \left( \prod_{i=1}^{S_R} 
%    \Theta( v_i ) \right)  \  \exp\left( - \sum_{i} v_{i}^2  \right) \frac{1}{ \prod_{j=1}^{S_C} \sum_{i=1}^{S_R} v_i L_{ij} }  \ .   
%\end{equation}
%The size of the feasibility domain $\Xi$ is therefore independent of $\gamma$. 

\begin{figure}
  \begin{center}
    \includegraphics[width = 0.9\linewidth]{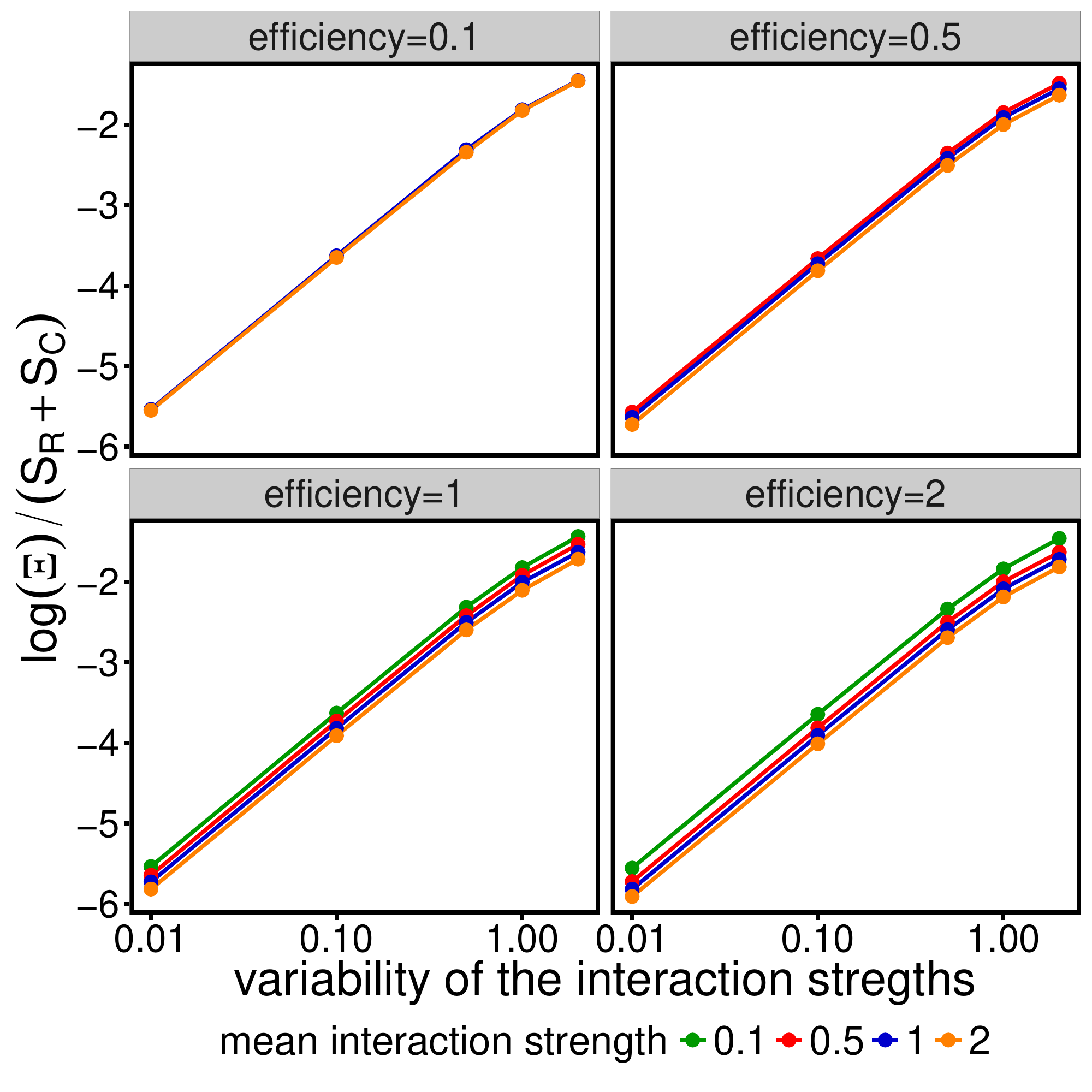}
  \end{center}
  \caption{Feasibility domain of consumer-resource community. We considered an interaction matrix of the form of equation~\ref{eq:consres_intmat},
with $S_R = 40$ and $S_C=30$, $C_{ij} = 1$ if $i=j$ and zero otherwise, $Z_i W_j = \eta > 0$ for any $i$ and $j$, and $\mat{B}$ with entries independently drawn from a
Lognormal distribution with mean $\mu$ and variance $c_v^2 \mu^2$. For each parameterization we computed the feasibility domain $\Xi$ using the method explained in
section~\ref{sec:numint}. The value of $\Xi$ is mostly determined by the coefficient of variation of the interaction, and it depends only weakly on
the mean interaction strength $\mu$ and the efficiency $\eta$.
}
  \label{fig:bipconsres}
\end{figure}

Figure~\ref{fig:bipconsres} shows the size of the feasibility domain of a consumer-resource community, computed using Monte Carlo integration as explained in section~\ref{sec:numint}.
We consider an interaction matrix with the structure of equation~\ref{eq:consres_intmat}, with a diagonal $\mat{C}$
(i.e., $C_{ij}=1$ if $i=j$ and zero otherwise) and scalar matrices $\mat{Z}$ and $\mat{W}$ (i.e., $Z_{ii}=W_{ii}=\eta$).
The elements of the rectangular matrix $\mat{B}$ were independently drawn from a lognormal distribution with mean
$\mu$ and variance $c_v^2 \mu^2$, where $c_v$ is the coefficient of variation. Since $\mat{C}$ is equal to the identity matrix,
then the interaction matrix is diagonally stable and therefore any feasible point is globally stable~\cite{Case1979}.
Figure~\ref{fig:bipconsres}
shows the effect of $\eta$, $\mu$ and $c_v$ on the size $\Xi$ of the feasibility domain. Interestingly, $\eta$ and $\mu$ have a
small effect on $\Xi$, while the coefficient of variation has a strong influence on it.
It is important to notice that, as explained above, as the interspecific interaction goes to zero (and therefore both $c_v$ and $\mu$ tend to zero), $\Xi \to 0$ as well.
Note that in this case not all the species are self-regulated ($A_{ii}=0$ for the predators), and therefore in absence of interspecific interactions, $\Xi \neq 1$.

\section{Empirical networks and randomizations}
\label{sec:empnet}

We considered 89 mutualistic networks and 15 food webs.  Empirical
networks are encoded in terms of adjacency matrices $\mat{L}$, with
$L_{ij} = 1$ if species $j$ affects species $i$ and zero otherwise.

\subsection{Mutualistic networks}

The 89 mutualistic networks (59 pollination networks and 30
seed-dispersal networks) were obtained from the Web of Life dataset
(\texttt{www.web-of-life.es}), where references to the original works
can be found.  When the original network was not fully connected, we
considered the largest connected component.

In the case of mutualistic networks, the adjacency matrix $\mat{L}$ is
bipartite, i.e., it has the structure
\begin{equation}\label{eq:adjblock}
  \mat{L} = \left(
    \begin{array}{cc}
      0 & \mat{L}_b \\
      \mat{L}_b^T & 0
    \end{array}
  \right) \ ,
\end{equation}
where $\mat{L}_b$ is a $S_A\times S_P$ matrix ($S_A$ and $S_P$ being
the number of animals and plants respectively).  The adjacency matrix
contains information only about the interactions between animals and
plants, but not about competition within plants or animals.

We parameterized the interaction matrix in the following way:
\begin{equation}\label{eq:interblock}
  \mat{A} = \left(
    \begin{array}{cc}
      \mat{W}^A & \mat{L}_b \circ \mat{W}^{AP} \\
      \mat{L}_b^T \circ \mat{W}^{PA} & \mat{W}^P
    \end{array}
  \right) \ ,
\end{equation}
where the symbol $\circ$ indicates the Hadamard or entrywise product
(i.e., $(\mat{A}\circ \mat{B})_{ij} = A_{ij} B_{ij}$), while
$\mat{W}^A$, $\mat{W}^{AP}$, $\mat{W}^{PA}$, and $\mat{W}^{P}$ are all
random matrices.  $\mat{W}^A$ and $\mat{W}^P$ are both square matrices
(of dimension $S_A\times S_A$ and $S_P\times S_P$), while
$\mat{W}^{AP}$ and $\mat{W}^{PA}$ are rectangular matrices of size
$S_A\times S_P$ and $S_P\times S_A$ respectively.  The diagonal
elements $W^A_{ii}$ and $W^P_{ii}$ were set to $-1$, while the pairs
$(W^{A}_{ij},W^{A}_{ji})$ and $(W^{P}_{ij},W^{P}_{ji})$ were drawn
from a bivariate normal distribution with mean $\mu_{-}$, variance
$\sigma^2_+ = c \mu_{-}^2$, and correlation $\rho \sigma_{+}^2$. Since
these two matrices represent competitive interactions, $\mu_{-}<0$.
The the pairs $(W^{AP}_{ij},W^{PA}_{ji})$ were extracted from a
bivariate normal distribution with mean $\mu_{+}$, variance
$\sigma_{-}^2 = c \mu_{+}^2$, and correlation $\rho \sigma_{-}^2$,
where $\mu_{+}>0$.

We analyze more than 600 parameterizations, obtained by considering
different values of $\mu_{-}$, $\mu_{+}$, $c$, and $\rho$.
For each network and parametrization we computed the size of feasibility domain
$\Xi$.
 The bottom
panel of Figure 2 in the main text was obtained by comparing $\Xi$ obtained in this
way with the analytical prediction obtained in equation~\ref{eq:sstab_final_conto9_2}.

\subsection{Food webs}

A summary of the properties and reference of the food webs can be
found in table~\ref{tab:fw}.  In the case of food webs the adjacency
matrix $L$ is not symmetric, and an entry $L_{ij}=1$ indicates that
species $j$ consumes species $i$. We removed all cannibalistic loops.
Since both $L_{ij}$ and $L_{ji}$ are never simultaneously equal to one
(there are no loops of length two), we parameterized the offdiagonal
entries of $\mat{A}$ as
\begin{equation}\label{eq:interfw}
  A_{ij} = W^{+}_{ij}  L_{ij} + W^{-}_{ji} L_{ji} \ ,
\end{equation}
while the diagonal was fixed at $-1$. Both $ \mat{W}^{+}$ and
$\mat{W}^{-}$ are random matrices, where the pairs
$(W^{+}_{ij},W^{-}_{ij})$ are drawn from a bivariate normal
distribution with marginal means $(\mu_{+},\mu_{-})$ and correlation
matrix
\begin{equation}\label{eq:corrmatfw}
  \left(
    \begin{array}{cc}
      c \mu_{+}^2 & \rho c \mu_{+}^2 \\
      \rho c \mu_{-}^2 & c \mu_{-}^2
    \end{array}
  \right)
\end{equation}

We analyzed more than 200 parameterizations, obtained by considering
different values of $\mu_{-}$, $\mu_{+}$, $c$, and $\rho$.
For each network and parametrization we computed the size of feasibility domain
$\Xi$.
 The bottom
panel of Figure 2 in the main text was obtained by comparing $\Xi$ obtained in this
way with the analytical prediction obtained in equation~\ref{eq:sstab_final_conto9_2}.
In this case the analytical prediction overestimate the actual value of $\Xi$, indicating that
there is a role of structure in determining structural stability.

\begin{table}
  \caption{References and properties of the 15 food webs analyzed in the work}\label{tab:fw}
  \begin{center}
    \begin{tabular}{ | l | l | l | l | }
      \hline
      Name & S & Number of links & Connectance \\ \hline
      Ythan Estuary~\cite{DataYthan} & 92 & 414 & 0.1 \\ \hline  
      St. Marks~\cite{DataStMarks} & 143 & 1763 & 0.17 \\ \hline  
      Grande Cari\c{c}aie~\cite{DataCaricaie} & 163 & 2048 & 0.16 \\ \hline  
      Serengeti~\cite{DataSerengeti} & 170 & 585 & 0.04 \\ \hline  
      Flensburg Fjord~\cite{DataFlens} & 180 & 1567 & 0.1 \\ \hline  
      Otago Harbour~\cite{DataOtago} & 180 & 1856 & 0.12 \\ \hline  
      Little Rock Lake~\cite{DataLittleRock} & 181 & 2316 & 0.14 \\ \hline  
      Sylt tidal basin~\cite{DataSylt} & 230 & 3298 & 0.12 \\ \hline  
      Caribbean Reef~\cite{DataReef} & 249 &  3293 & 0.11 \\ \hline  
      Kongs Fjorden~\cite{DataKongs} & 270 & 1632 & 0.04 \\ \hline  
      Carpinteria Salt Marsh~\cite{DataFWPar} & 273 & 3878 & 0.1 \\ \hline  
      San Quintin~\cite{DataFWPar} & 290 & 3934 & 0.09 \\ \hline  
      Lough Hyne~\cite{DataLough} & 349 & 5088 & 0.08 \\ \hline  
      Punta Banda~\cite{DataFWPar} & 356 & 5291 & 0.09 \\ \hline  
      Weddell Sea~\cite{DataWeddell} & 488 & 15435 & 0.13 \\ \hline  
    \end{tabular}
  \end{center}
\end{table}

\section{Randomization of empirical networks: assessing the role of structure}
\label{sec:rndmzation}

\subsection{Mutualistic networks}
\label{ssec:rndmutualism}

\begin{figure}
  \begin{center}
    \includegraphics[width = 0.9\linewidth]{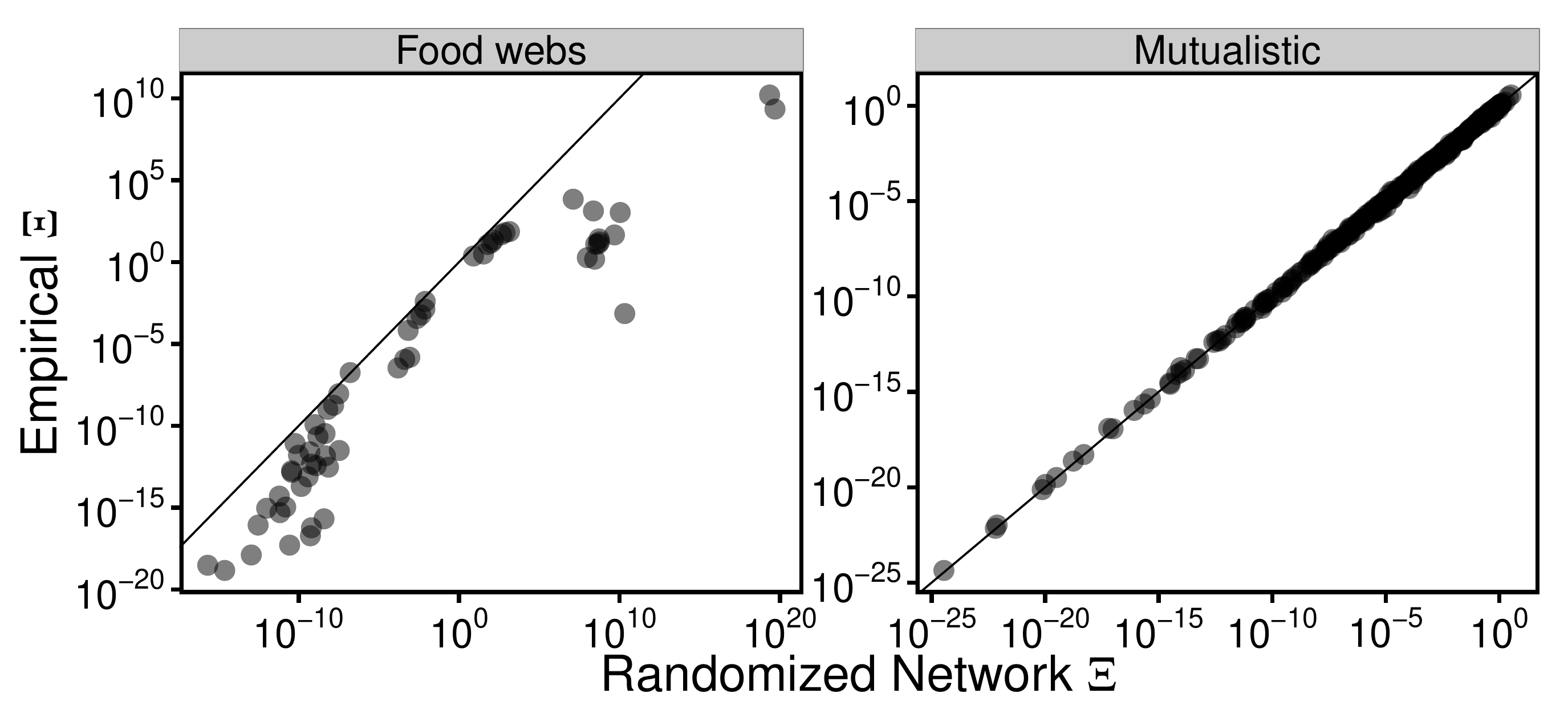}
  \end{center}
  \caption{Size of feasibility domain $\Xi$ in empirical networks and randomizations. Empirical networks and their randomizations
were parametrized as explained in section~\ref{sec:empnet}. Each empirical network was parametrized $100$ times and the average $\Xi$
was compared with the one obtained by averaging $100$ randomizations. Each point in this plot correspond therefore to a value of $\Xi$ of an empirical network
and its randomizations
averaged over the extraction of the interaction strenghts for a given combination of the parameters as explained in section~\ref{sec:empnet}. }
  \label{fig:rndz_mut_old}
\end{figure}

We compared
the size of the feasibility domain obtained for empirical networks with the corresponding
randomizations. For each network we randomized the block $\mat{L}_b$
100 times, by generating connected networks with same size and
number of links.  We parameterized each randomized network
independently as described in section~\ref{sec:empnet}, and we compared their properties
with those of the empirical network, parameterized independently
100 times.
Figure~\ref{fig:rndz_mut_old} shows the comparison between $\Xi$ of random and empirical networks.
As expected from the fact that the analytical prediction for random matrices works well,
the empirical values and the values obtained with randomizations are compatible.
Comparing this figure with figure 2 of the main text we observe that the empirical values and the ones obtained
with randomizations match also in the cases were the analytical approximation failed. This implies that the reason of
the mismatch is due to the difference between the analytical approximation and the randomizations, and it is not due
to the specific structure of the empirical interactions. There are two main sources of errors in this case. On one hand, ours analytical prediction
is expected to work is the number of species is large enough and if the variance of interactions is not to high (that is not always true for the parametrizations
used). On the other hand, our approximation was formulated for random matrices, while randomizations of mutualistic networks still conserve a
bipartite structure.

The randomization procedure explained above and figure~\ref{fig:rndz_mut_old} show that the size of the coexistence domain obtained
with empirical network structure is well predicted by the one obtained with random structure. This
result does not imply that structure has no effect on $\Xi$, but it shows that, if this effect exists,
it must be relatively small (compared for instance to the variation of $\Xi$ obtained by changing the interaction strengths), i.e.
the relative error made by approximating empirical networks with random structure must be small.

Since the effect of structure is small, it is also expected to be very sensible to the interaction strengths.
When we parametrized empirical networks and their randomizations to obtain figure~\ref{fig:rndz_mut_old}, we drawn
the interaction strengths several times from a given distributions. The realized coefficients were therefore different
across different networks, and the values of $\Xi$ shown in figure~\ref{fig:rndz_mut_old} were averaged over these independent
extractions. Since the difference between randomizations and empirical structure is small, it  might be impossible to detect any
difference with this procedure.

In order to explore and quantify the effect of the empirical structure on the size of feasibility domain, we adopted a different 
parametrization and randomization method.
Given an empirical network, we drawn the interaction strengths only once from a given distribution (as described in section~\ref{sec:empnet}).
Using this list of interaction
strenghts we parametrized $100$ times each empirical network. Different parametrization differ in the position
of the coefficients, but not in their values that are conserved across parametrizations. We then compared their size of feasibility domain
with the one obtained by parameterizing with the same list of coefficients $100$ randomized networks obtained as explained above. 

\begin{figure}
  \begin{center}
    \includegraphics[width =
    0.9\linewidth]{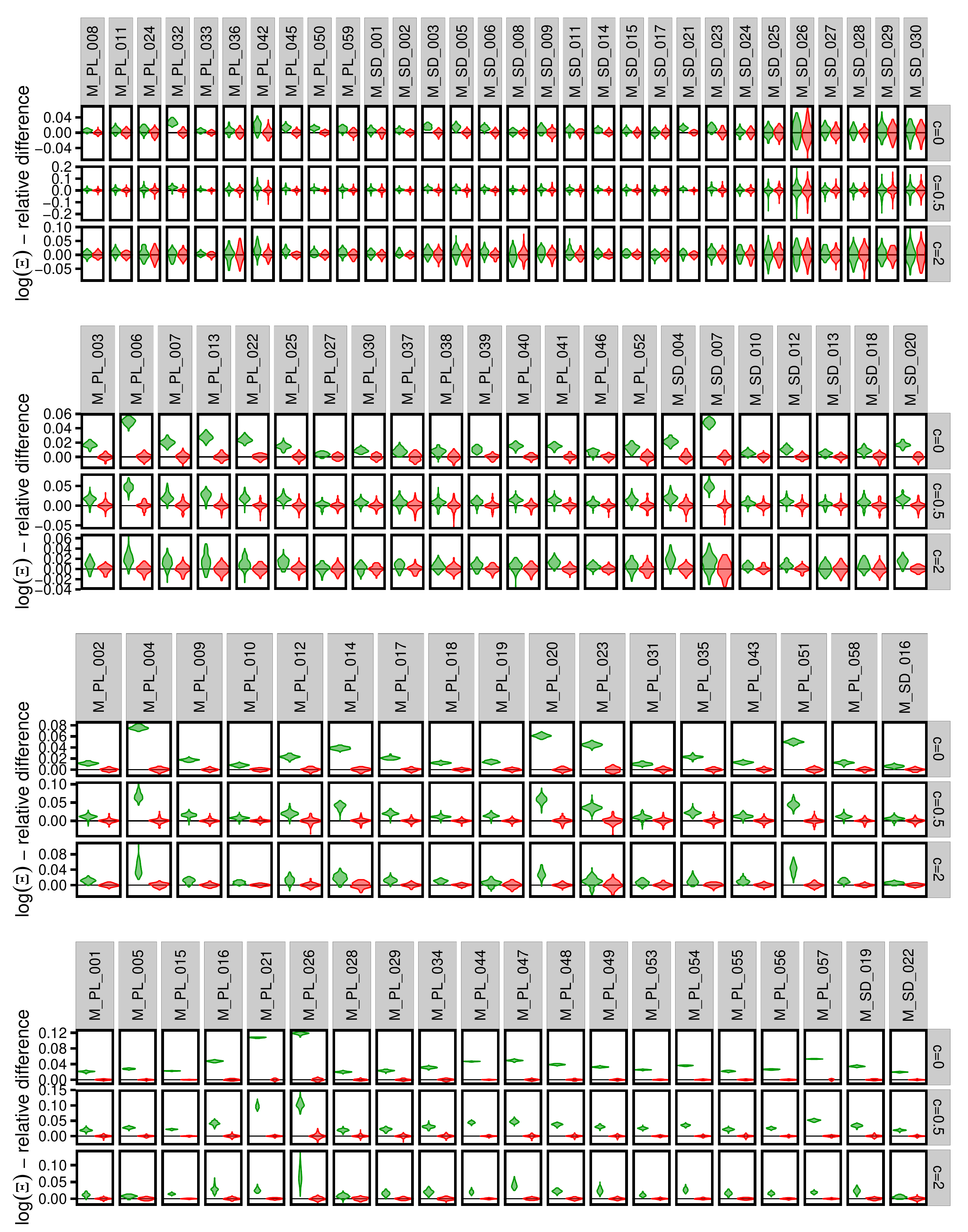}
  \end{center}
  \caption{We measure the effect of mutualistic network structure on the size of the feasibility domain
as described in section~\ref{ssec:rndmutualism}. Red violin plots are randomizations, green ones are empirical networks.
The empirical networks are grouped in four rows based on the number of species ($S<50$, $ 50 \leq S < 80$, $ 80 \leq S < 150$ and $S \geq 150$, respectively).
This figure was obtained with $\mu_{+} = 0.25 \mu_{max}$, $\mu_{-} = 0$ and for three different values of $c$.
 }
  \label{fig:rnd_0.25_0}
\end{figure}

\begin{figure}
  \begin{center}
    \includegraphics[width =
    0.9\linewidth]{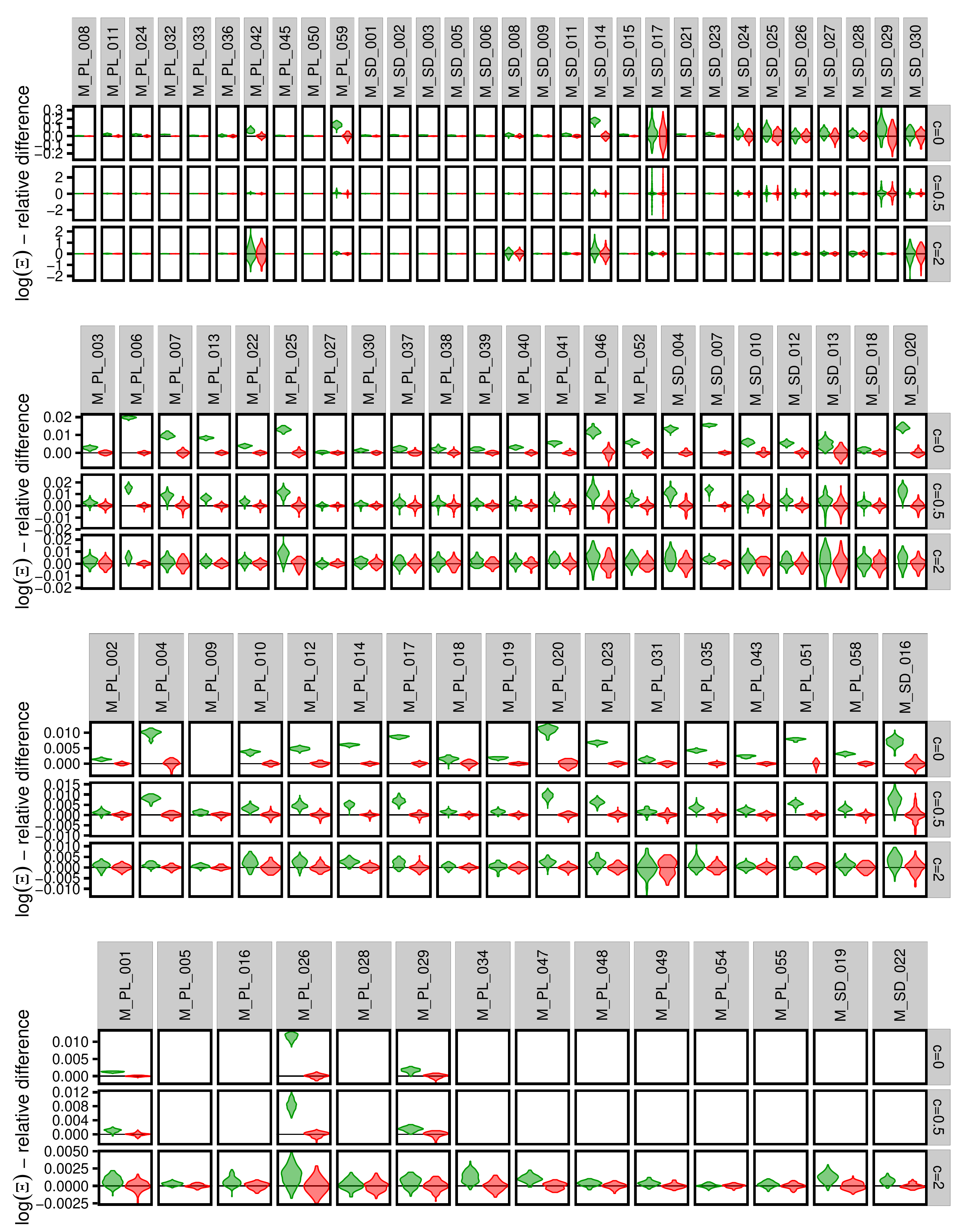}
  \end{center}
  \caption{Same as figure~\ref{fig:rnd_0.25_0} but with $\mu_{+} = 0.25 \mu_{max}$ and $\mu_{-} = 0.5 \mu_{+}$}
  \label{fig:rnd_0.25_0.5}
\end{figure}

\begin{figure}
  \begin{center}
    \includegraphics[width =
    0.9\linewidth]{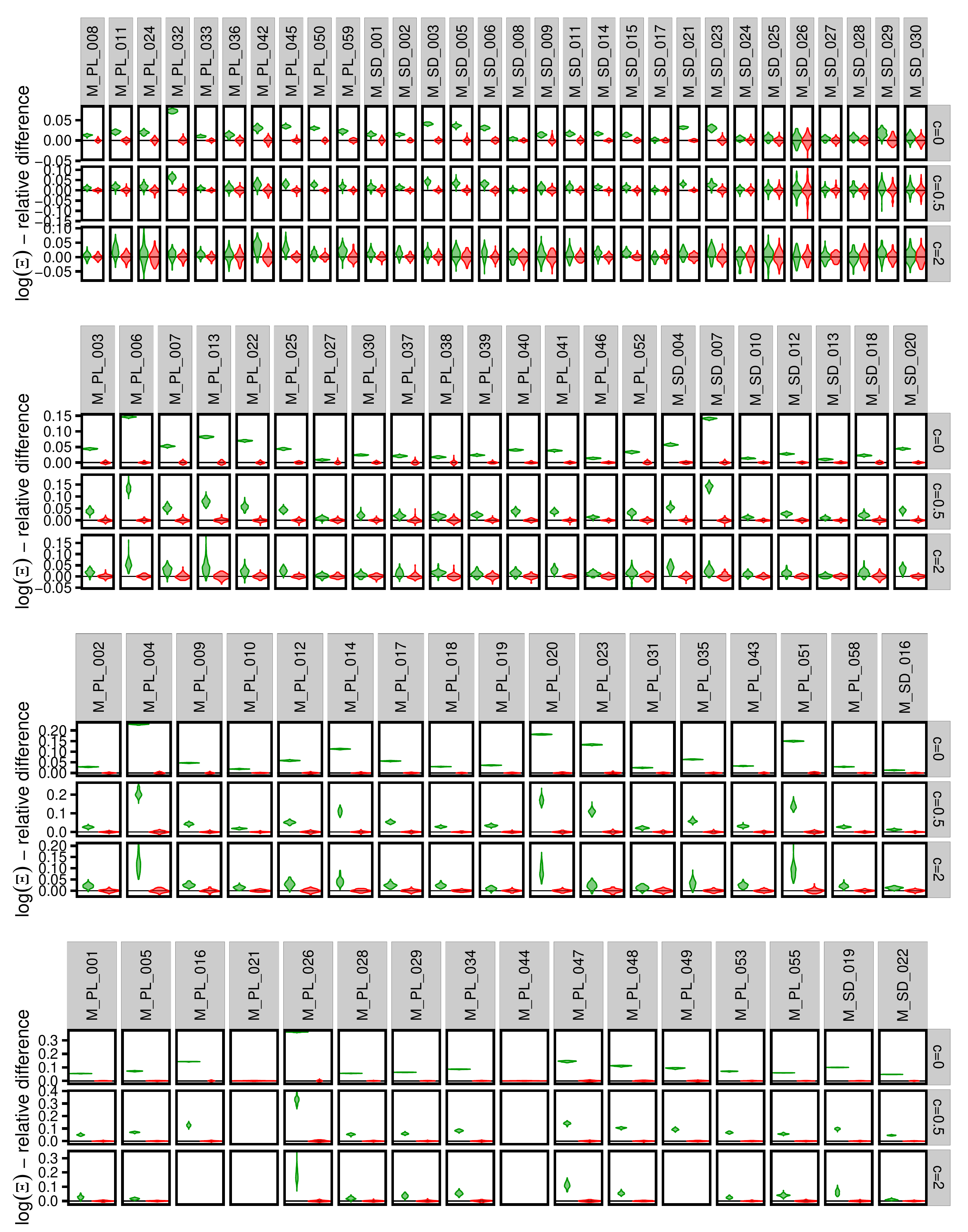}
  \end{center}
  \caption{Same as figure~\ref{fig:rnd_0.25_0} but with $\mu_{+} = 0.5 \mu_{max}$ and $\mu_{-} = 0$}
  \label{fig:rnd_0.5_0}
\end{figure}

\begin{figure}
  \begin{center}
    \includegraphics[width =
    0.9\linewidth]{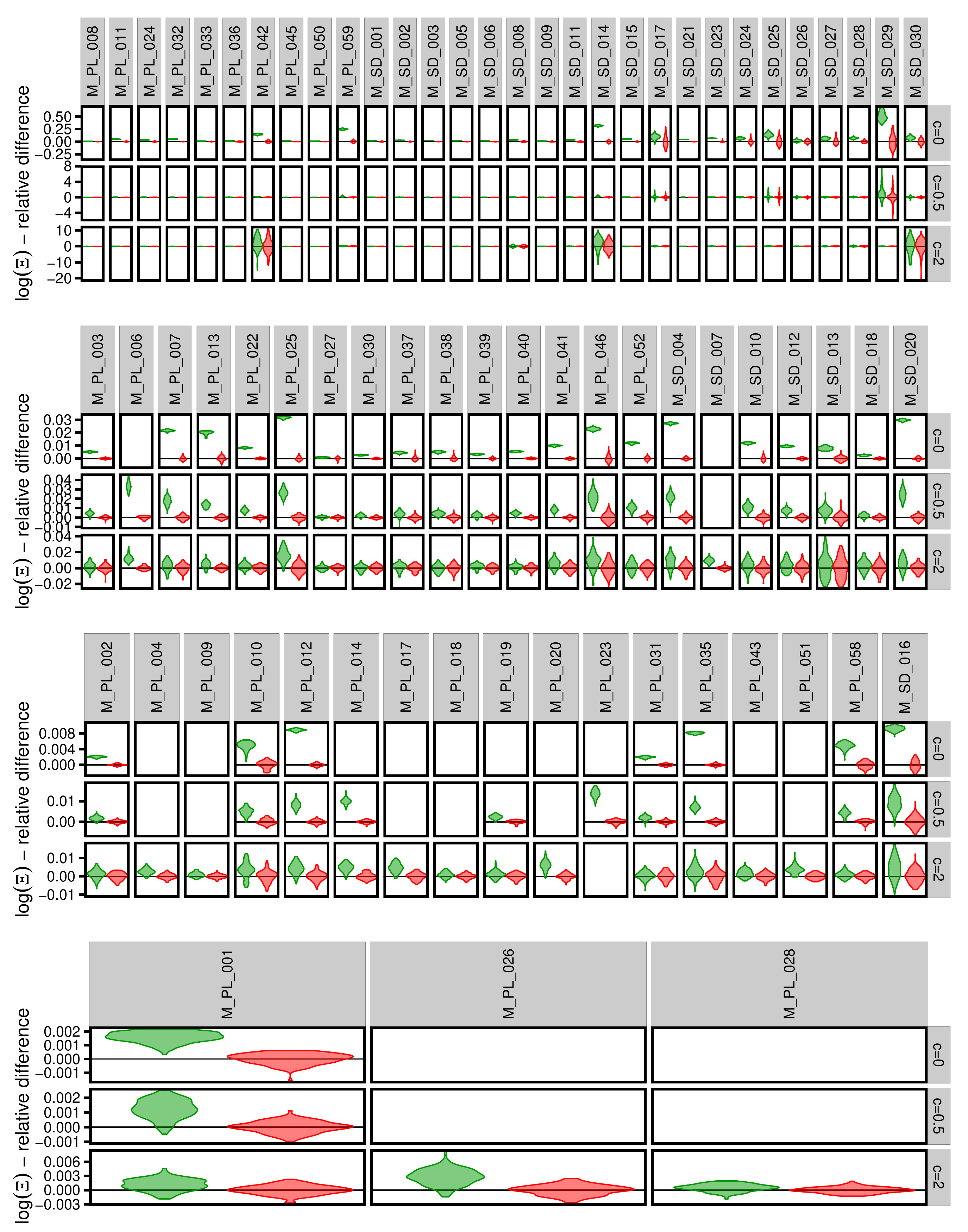}
  \end{center}
  \caption{Same as figure~\ref{fig:rnd_0.25_0} but with $\mu_{+} = 0.5 \mu_{max}$ = $\mu_{-} = 0.5 \mu_{+}$}
  \label{fig:rnd_0.5_0.5}
\end{figure}

Figures~\ref{fig:rnd_0.25_0},~\ref{fig:rnd_0.25_0.5},~\ref{fig:rnd_0.5_0} and~\ref{fig:rnd_0.5_0.5} show the results
obtained for different distributions of interaction strengths (parametrized as explained in section~\ref{sec:empnet}).
In absence of competition and in absence of variation in the interaction strengths, there is the maximum observable effect.
As the competition level is increased and once variation in the interaction strengths is introduced, the effect of the network topology
on the total size of feasibility domain becomes negligible.

\subsection{Food webs}
\label{ssec:rndfwebs}

We compared the 
size of feasibility domain of empirical networks with their
corresponding randomizations and a network generated accordingly
to the cascade model\cite{Cohenetal1990}.

For each network, we randomized the
adjacency matrix $\mat{L}$ 100 times, by generating connected
networks with the same size and number of links.

We also generated networks generated accordingly to the cascade model (using the same method explained in~\cite{Allesina2015a}).
In this case the adjacency matrix was obtained by generating connected networks with the same size and number of links,
 by assigning a link between species $i$ and $j$ only if $i > j$.

\begin{figure}
  \begin{center}
    \includegraphics[width =
    0.9\linewidth]{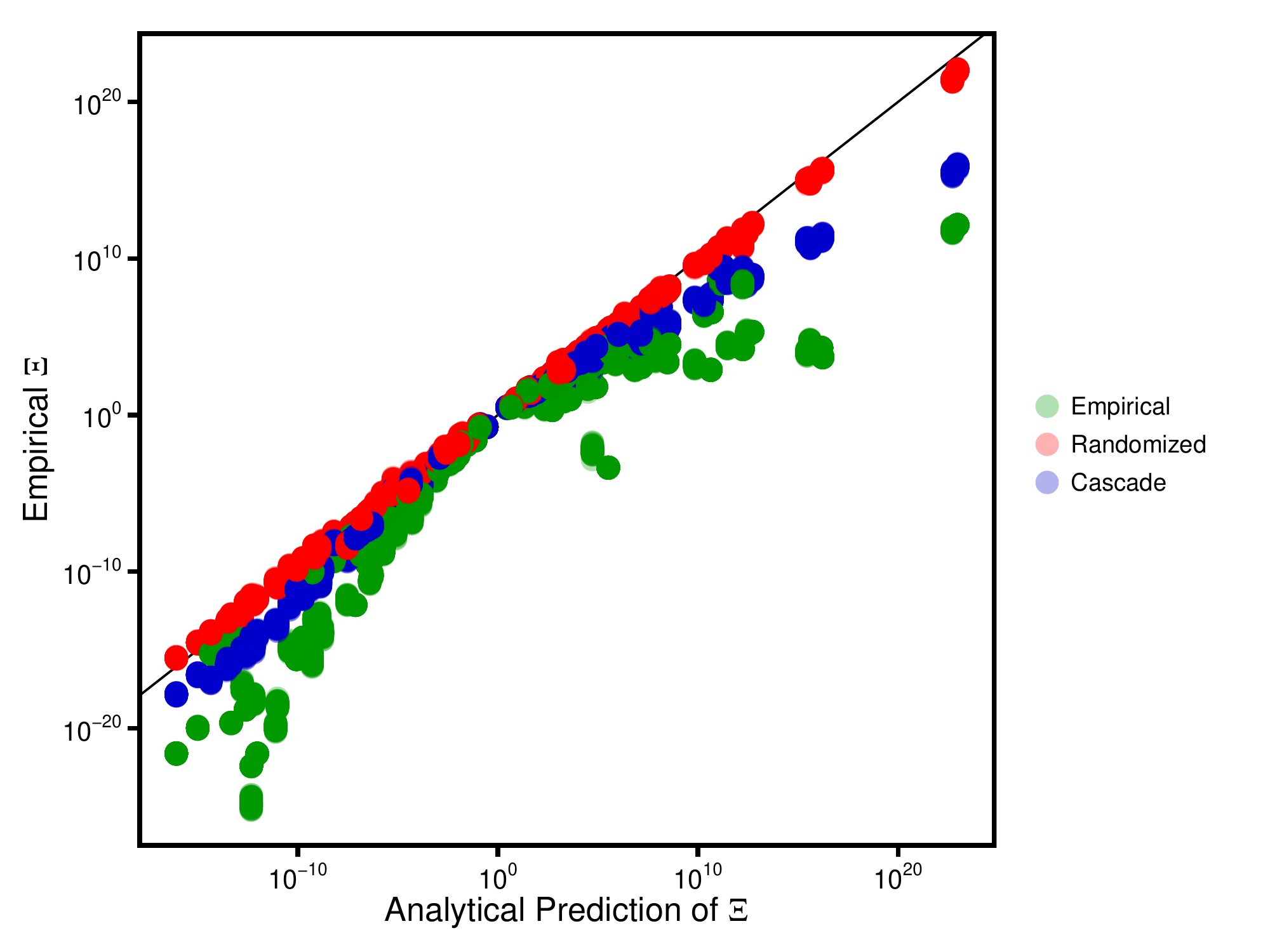}
  \end{center}
  \caption{In this figure we compared the analytical prediction of the feasibility domain obtained in section~\ref{sec:approxxi}
        with the numerical calculated values for random networks, empirical networks and networks generated via the cascade models.
        The feasibility domain of random networks is well predicted by our analytical approximation, which fails to predict
        the empirical one and the one obtained using the cascade model.    
}
  \label{fig:rndfw_allsim}
\end{figure}

\begin{figure}
  \begin{center}
    \includegraphics[width =
    0.9\linewidth]{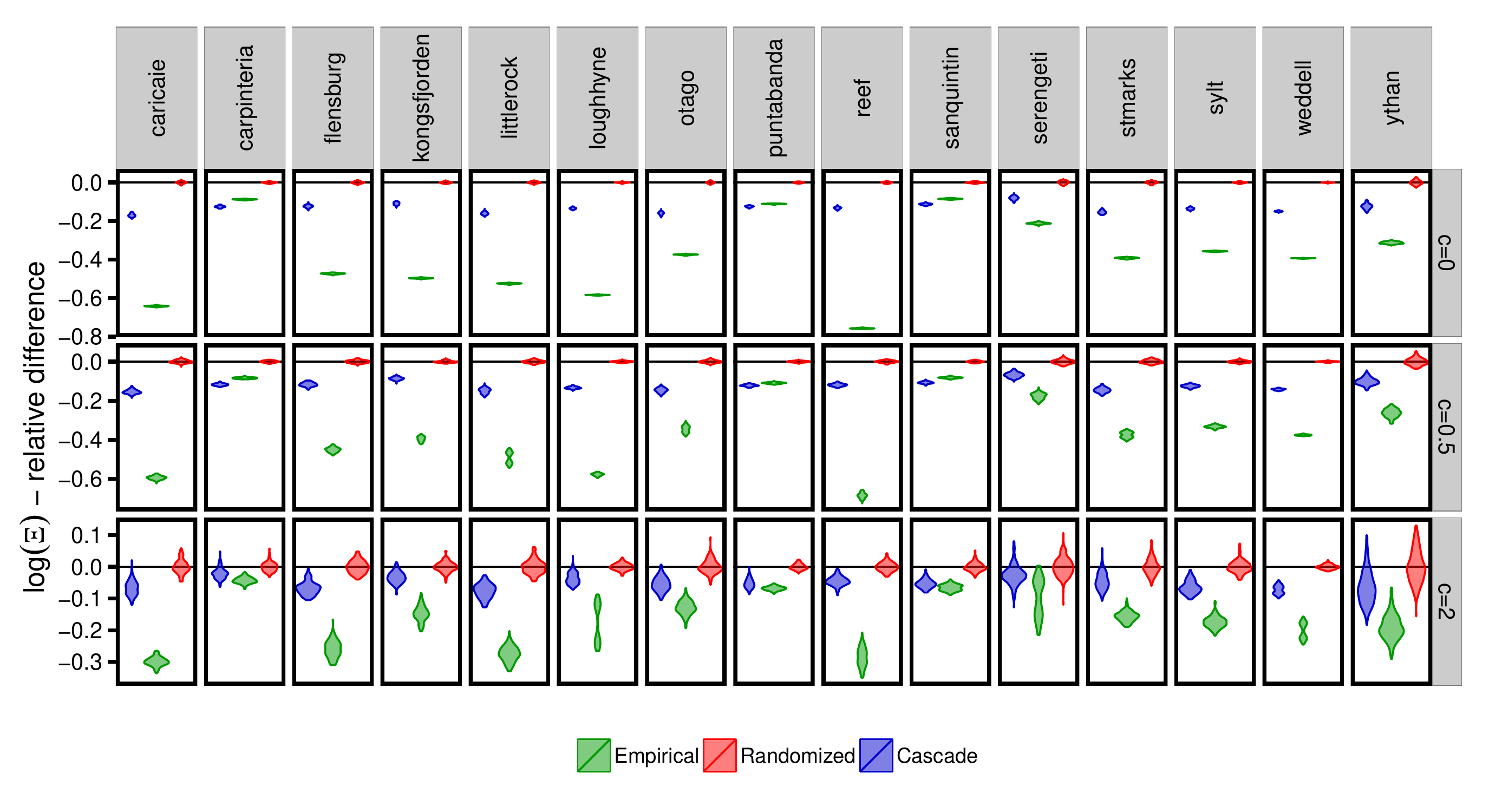}
  \end{center}
  \caption{We measure the effect of food web network structure on the size of the feasibility domain
        as described in section~\ref{ssec:rndfwebs}. Red violin plots are randomizations, green ones are empirical networks, while blue ones correspond to the cascade model.
        This figure was obtained as explained in section~\ref{ssec:rndfwebs} with $\mu_{-} = 0.25 \mu_{max}$, $\mu_{+} = 0.5 \mu_{-}$ and for three different values of $c$.
}
  \label{fig:rndfw_0.25_0.5}
\end{figure}

\begin{figure}
  \begin{center}
    \includegraphics[width =
    0.9\linewidth]{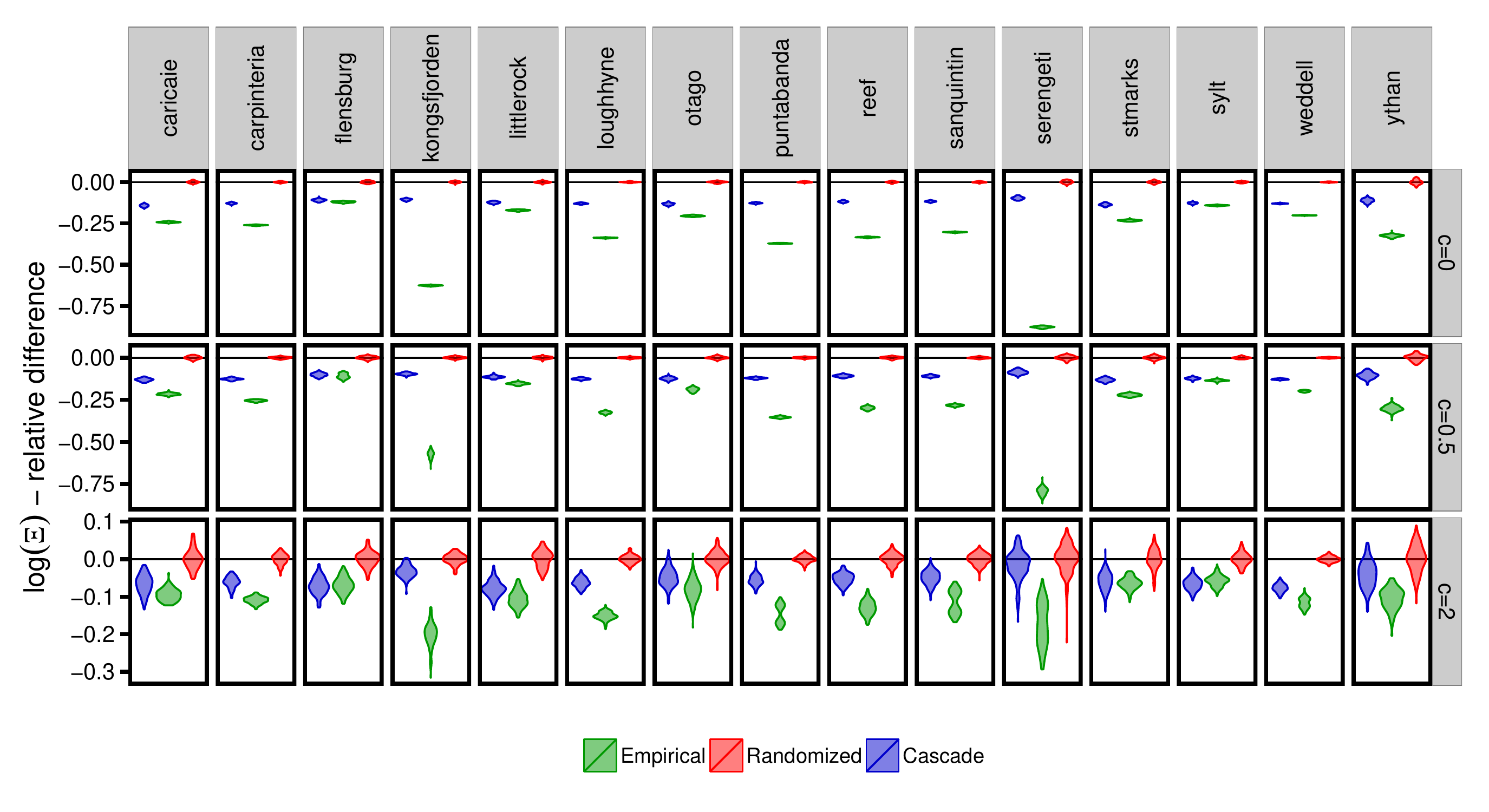}
  \end{center}
  \caption{Same as figure~\ref{fig:rndfw_0.25_2} but with $\mu_{-} = 0.25 \mu_{max}$ and $\mu_{+} = 2 \mu_{-}$}
  \label{fig:rndfw_0.25_2}
\end{figure}

Figure~\ref{fig:rndfw_allsim} is the same as figure 2 of the main text, with the addition of randomizations and networks generated
with the cascade model. As expected the analytical prediction works very well in describing random networks, while it fails significantly
to predict the size of the feasibility domain of cascade and empirical networks.
To better quantity the difference between those empirical structures and randomizations, we compared each network separately in
figure~\ref{fig:rndfw_0.25_0.5} and~\ref{fig:rndfw_0.25_2}. We observe that random networks have always larger feasibility domain than networks
generated by the cascade model and the empirical ones. Networks generated via the cascade model
almost always overestimate the empirical feasibility domains, showing that empirical network structure has a significant negative 
effect on the size of the feasibility domain.

\section{\ Possible biases in previous analysis of structural stability}
\label{sec:problemsinprevanal}

In section~\ref{sec:numint} we showed how to estimate the feasibility
domain numerically in a fast and reliable way. In previous
approaches~\cite{Rohr2014}, the feasibility domain (structural stability) was not directly
calculated, but approximately inferred using a regression method.
In this section we show that
the method used by Rohr et al.~\cite{Rohr2014} could be biased and is
not always applicable.

The Authors considered a bipartite mutualistic system described by the
dynamical model
\begin{equation}\label{eq:rohrdynmodel}
  \displaystyle
  \left\{
    \begin{array}{cc}
      \displaystyle \frac{d n^A_i}{dt} & = \ \displaystyle n^A_i \left( r_i^A - \sum_{j=1}^{S_A} \beta^A_{ij} n^A_j +
        \frac{ \sum_{j=1}^{S_P} \gamma^A_{ij} n^P_j }{ 1 + h_i^A \sum_{j=1}^{S_P} \gamma^A_{ij} n^{P}_j  }  \right) \\
      \displaystyle \frac{d n^P_i}{dt} & = \ \displaystyle n^P_i \left( r_i^P - \sum_{j=1}^{S_P} \beta^P_{ij} n^P_j +
        \frac{ \sum_{j=1}^{S_A} \gamma^P_{ij} n^A_j }{ 1 + h_i^P \sum_{j=1}^{S_A} \gamma^P_{ij} n^A_j  }  \right)
    \end{array}
  \right. \ ,
\end{equation}
where $S_A$ ($S_P$) is the number of animals (plants), and $n^A_i$
($n^P_i$) is the abundance of animal (plant) species $i$. For the
purposes of this section we consider the case of linear functional
responses $h_i^A = h_i^P = 0$, as all the methodology used in~\cite{Rohr2014}
was developed in this case. If the functional response is
linear, this equation reduces to equation~\ref{eq:dynmodel_LV}, where
the interaction matrix $\mat{A}$ is given by
\begin{equation}
  \mat{A} = \left(
    \begin{array}{cc}
      -\mat{\beta}^A & \mat{\gamma}^A \\
      \mat{\gamma}^P & -\mat{\beta}^P
    \end{array}
  \right) \ .
\label{eq:Arorh}
\end{equation}
Here $\mat{\beta}^A$ and $\mat{\beta}^P$ are $S_A\times S_A$ and
$S_P\times S_P$ matrices, respectively, while $\mat{\gamma}^A$ and
$\mat{\gamma}^P$ are $S_A\times S_P$ and $S_P\times S_A$ matrices. The
Authors used a constant parameterization for the competition
parameters, setting $\beta^A_{ii}=\beta^P_{ii}=1$ and
$\beta^A_{ij}=\beta^P_{ij}=\rho$ if $j\neq i$.  The mutualistic
benefits were parameterized as
\begin{equation}
  \displaystyle
  \begin{array}{cc}
    \gamma^A_{ij} & = \ \displaystyle \gamma_0 \frac{L_{ij}}{(k_i^A)^\delta} \\
    \gamma^P_{ij} & = \ \displaystyle \gamma_0 \frac{L_{ji}}{(k_i^P)^\delta} 
  \end{array}
  \ ,
\end{equation}
where $L_{ij}$ is the nonzero block of the adjacency matrix of the
interaction network, i.e., $L_{ij} = 1$ if there is an interaction
between animal $i$ and plant $j$, and zero otherwise.  The numbers
$k_i^A = \sum_{j=1}^{S_P} L_{ij}$ and $k_i^P = \sum_{j=1}^{S_A} L_{ji}$ are
the degree of animal/plant $i$. The two remaining parameters,
$\gamma_0$ and $\delta$, quantify the levels of mutualistic strength
and the mutualistic tradeoff~\cite{Saavedra2013}.

The method proposed by Rohr et al.~\cite{Rohr2014} was based on what
the Authors called the ``structural vector''.  It was defined as the
center of feasibility domain and was calculated by transforming the
mutualistic dynamics into an effective competitive one.  Using this
effective dynamics it was possible to calculate an effective
structural vector, which was then transformed back to the one of the
mutualistic system. Starting from the structural vector, the Authors
considered different perturbations of the growth rates by changing
their direction from that of the original structural vector by some
given angle. The dynamics was then integrated and the probability that
all species survived was calculated, given a particular
perturbation. Running this across several different perturbations and
parameterizations, it was possible to perform a regression between the
interaction parameters, the angle by which the growth rates were
perturbed, nestedness, and other parameters appearing in the
interaction matrix.  Using the coefficients obtained through the
regression, it was quantified the effect of nestedness and other properties
on the size of the feasibility domain.
Here we present some possible issues emerging from this approach.

\textbf{It is not always possible to find the structural vector}.  In
order to calculate the structural vector, one needs to transform the
mutualistic system into an effective competitive one. One can
define the matrix $\mat{T} = \mat{1} + \mat{\gamma} \mat{\beta}^{-1}$,
where $\mat{1}$ is the identity matrix and
\begin{equation}
  \mat{\beta} = \left(
    \begin{array}{cc}
      \mat{\beta}^A & 0 \\
      0 & \mat{\beta}^P
    \end{array}
  \right) \ ,
\end{equation}
and
\begin{equation}
  \mat{\gamma} = \left(
    \begin{array}{cc}
      0 & \mat{\gamma}^A \\
      \mat{\gamma}^P & 0
    \end{array}
  \right) \ .
\end{equation}
By multiplying both sides of
equation~\ref{eq:Arorh} by $\mat{T}$ one obtains the effective
interaction matrix
\begin{equation}
  \label{eq:Rohreffective}
  \mat{A}_{\text{eff}} = \left(
    \begin{array}{cc}
      -\mat{\beta}^A + \mat{\gamma}^P (\mat{\beta}^P)^{-1} \mat{\gamma}^A & 0 \\
      0 & -\mat{\beta}^P + \mat{\gamma}^A (\mat{\beta^A})^{-1} \mat{\gamma}^P 
    \end{array}   \right) =:
\left(
    \begin{array}{cc}
      \mat{B}_{\text{eff}}^{A} & 0 \\
      0 & \mat{B}_{\text{eff}}^{P} 
    \end{array}
  \right) \ ,
\end{equation}
In order to calculate the structural vectors, one has to assume that
the eigenvectors associated with the largest singular eigenvalues of
$(\mat{B}_{\text{eff}}^{A})^T \mat{B}_{\text{eff}}^{A}$ and $
\mat{B}_{\text{eff}}^{A} (\mat{B}_{\text{eff}}^{A})^T $ have only
positive components (and an equivalent condition on $
\mat{B}_{\text{eff}}^{P}$ ).  This is not generally true, as also
stated by the Authors~\cite{Rohr2014}. They therefore imposed the
extra assumption that $(\mat{B}_{\text{eff}}^{A})^T
\mat{B}_{\text{eff}}^{A}$ and $ \mat{B}_{\text{eff}}^{A}
(\mat{B}_{\text{eff}}^{A})^T $ indeed have only positive entries (and
the equivalent conditions on $\mat{B}_{\text{eff}}^{P}$).  In this
case, the Perron--Frobenius theorem allows all entries of the leading
eigenvector to be chosen positive; i.e., it necessarily points in some
feasible direction. The Authors then identified the structural vectors
with these eigenvectors.

However, the extra requirement that $(\mat{B}_{\text{eff}}^{A})^T
\mat{B}_{\text{eff}}^{A}$ and $ \mat{B}_{\text{eff}}^{A}
(\mat{B}_{\text{eff}}^{A})^T $ be strictly positive imposes
constraints on the interaction matrix that reduces the number of
parameterizations that can be analyzed with this method. Since this
assumption does not hold in general, there are cases in which the
structural vector does not exist. Using our approach, this vector is
not needed (see sections~\ref{sec:numint} and~\ref{sec:sidelength}).

\textbf{When the structural vector exists, it is not unique}. Under
what conditions would the matrices $(\mat{A}^{\text{eff}})^T
\mat{A}^{\text{eff}}$ and $\mat{A}^{\text{eff}}
(\mat{A}^{\text{eff}})^T$ satisfy the conditions of the
Perron--Frobenius theorem?  It is easy to show that this can never be
the case. From equation~\ref{eq:Rohreffective} we see that
$\mat{A}^{\text{eff}}$ is block-diagonal, therefore
$(\mat{A}^{\text{eff}})^T \mat{A}^{\text{eff}}$ and
$\mat{A}^{\text{eff}} (\mat{A}^{\text{eff}})^T$ are block-diagonal as
well.  This means that the Perron--Frobenius theorem does not hold
(the matrix is reducible); instead, the two diagonal blocks each have
an all-positive leading eigenvector (assuming that all the
coefficients are positive in the two blocks).  Any linear combination
of the two will have positive components. There is no reason to prefer
one linear combination over another, and while it is true that some
linear combinations may point closer to the center of the feasibility
domain, there is no way to determine using the Authors' methods which
combination does, if any.

\textbf{The structural vector is not the center of the feasibility
  domain}.  Let us assume now that the
structural vector exists and it points toward the center of the feasibility
domain of the effective competitive system.
To obtain the structural vector, one has to
transform it back to a vector of the original, mutualistic system. The
transformation from the effective to the original system is done by
multiplying with the matrix $\mat{T}^{-1}$. This matrix is not a
rotation, and therefore it does not preserve the angles between
vectors. Even if a vector is the center of the feasibility domain in
the effective system, it will not in general be the center of the
original domain. In particular, its distance to the actual center of
the original domain will be dependent on parameterization and network
structure, as the transformation matrix depends on these.

In contrast, the center of the feasibility domain can be easily
expressed with our approach in terms of the matrix $\mat{A}$ and its associated generators
(section~\ref{sec:geomprop}, equation~\ref{eq:barycenter3}). It is also
easy to check that the barycenter is different from the one obtained using
the method of Rohr et al.~\cite{Rohr2014}.

\textbf{The regression procedure can in principle produce biases}.
The relationship between network structure and the size of the
feasibility domain was obtained by calculating the probability of
coexistence $p(\theta_A, \theta_P)$, where $\theta_{A/P}$ is the angle
by which the direction of the growth rate vector of animals/plants was
changed with respect to the structural vector. The Authors then
performed a linear regression
\begin{equation}
\begin{split}
  \text{logit}(p(\theta_A, \theta_P)) & \sim \beta_1 \log \theta_A + \beta_2 \log \theta_B + \beta_3 \gamma_0 C +
  \beta_4 \gamma_0^2 C^2 \\ & + \beta_5 \gamma_0 C N + \beta_6 \gamma_0 C N^2 + \beta_7 \gamma_0 C \delta + \beta_8
  \gamma C \delta^2 \ ,
\end{split}
\end{equation}
where $C$ is the connectance of the mutualistic adjacency matrix and
$N$ is its nestedness (note that $\bar{\gamma}$, used by Rohr et
al.~\cite{Rohr2014}, is equal to $C \gamma_0$).  The fitted parameters
where then used to determine the effect of nestedness and other
quantities on the feasibility domain. The functional dependence
assumed above cannot be justified \emph{a priori}, and an incorrect
functional dependence can in principle lead to erroneous fitting
results. For instance, the effect of those properties could be
different depending not just on the raw angle of perturbation, but
also which direction that angle is taken in. We can imagine two
feasibility regions with the exact same size but different shapes: one
of the two is equally wide in all directions, while the other
stretches very wide in some directions but is extremely narrow in
others (see section~\ref{sec:sidelength}). For sufficiently small
values of $\theta_{A/P}$, one will never leave the feasible domain in
the first of these examples, but may do so in the second if the
perturbation is performed in one of the ``narrow'' directions. The
first of these cases will therefore appear more feasible than the
second, even though the total size of the two feasibility regions is
in fact the same.  On the other hand, if the values of $\theta_{A/P}$
are large enough, than the perturbed vector in the first case will
never be feasible, while it will be feasible in the second case
because of the ``wide'' directions.  Moreover, this method does not
allow one to calculate the feasibility domain for a given network and
parameterization, as one can calculate only the probability of
coexistence given an angle of perturbation.

\section{\ Distribution of side lengths}
\label{sec:sidelength}

In section~\ref{sec:geomprop} we showed that the feasibility domain is
a convex polyhedral cone in the space of intrinsic growth rates
$\vect{r}$. Since the stationary solution of
equation~\ref{eq:dynmodel_LV} is linear in $\vect{r}$, we can
study the feasibility domain considering only vectors on the unit
sphere's surface. In section~\ref{sec:numint} we defined $\Xi$, which
quantifies the volume of the feasibility domain.

The size of the feasibility domain, i.e., how many combinations of the
intrinsic growth rates correspond to a feasible fixed point, is not
the only interesting property. Two systems having the same number of
feasible combinations of growth rates (i.e., the same value of $\Xi$),
can respond very differently to perturbations of the growth rates. We
imagine here that a perturbation (e.g., a change of the abiotic
conditions) correspond to a change in the growth rate vector. Since we
can consider normalized growth rate vectors (because of the linearity
of the equations), the effect of a perturbation on feasibility depends
only on the angular change of the growth rate vector and not on its
length.

The volume $\Xi$ quantifies how many growth rate vectors are
compatible with coexistence. Let us consider a feasible growth rate
vector, and perturb it in a random direction. What is the probability
that the new vector is still feasible? This is not just a function of
the size $\Xi$ of the feasibility domain. Indeed, one can imagine that
the feasibility domain is about equally spread in every direction---or
that, for the exact same value of $\Xi$, the feasibility domain is
streched in some directions but is very narrow in some other ones. A
perturbation in one of the ``narrow'' directions is much more likely
to lead out of the feasibility domain in the latter case than in the
former.

To quantify this property, one strategy could be to measure the
different responses on the perturbation (i.e., the probability of
being feasible) depending on the direction of the perturbation (in
which direction we change the growth rate vector). This choice has the
big disadvantage of depending not only on the properties of
interactions (the interaction matrix $\mat{A}$), but also on the
strength of the perturbation (the angular displacement between the
initial and the final growth rate vector) and the growth rate vector
before the perturbation (e.g., if the initial vector is close or far
from the edge of the feasibility domain). We propose instead a purely
geometrical method to quantify the response to different perturbations
(see figure 1 of the main text).

The feasibility domain, when restricted to the surface of a
hypersphere, can be imagined as the generalization of a triangle on a
sphere (see section~\ref{sec:feas3D}). The natural, geometric
quantities bounding the maximal perturbation that will leave the
system feasible, are the lengths of the triangle's sides. When $S$
species are considered, there are $S(S-1)/2$ sides. Their lengths
measure the maximum permissible perturbation of the growth rates in
the corresponding direction if one is to retain feasibility. This
property has the advantage of being purely geometrical, depending only
on the interactions (via the interaction matrix) and not, for
instance, on any choice of the initial conditions.

We can measure the distribution of the side lengths. Imagine we have
two interaction matrices with the same $\Xi$, but with very different
distributions of side lengths. One of them has all sides of equal
length, while the other one has a more heterogeneous distribution. In
the first case any direction of the perturbation is expected to have a
similar effect, and there are no particularly dangerous directions.
In the second case there are some directions of the perturbation that
are much more dangerous than others, and even a small change of
conditions along one of those dangerous direction can lead to the
extinction of one or more species.

\label{ssec:sidef}

We know that the feasibility domain is a convex polyhedral cone (see
section~\ref{sec:geomprop}).  Its ``corners'' are identified by its
generators and its sides are determined by all pairs of generators
(see section~\ref{sec:feas3D} for the $S=3$ case).

Since we are considering growth rates on the unit (hyper)sphere, and
the generators are normalized to one, any pair of generators will lie
on the sphere's surface. The scalar product of two generators is the
cosine of the angle between the two. Since the two generators are on
the unit ball's surface, the arc between the two (which is the side
length) is equal to the angle. We have therefore that the length of
the side of the feasibility domain corresponding to a pair of
generators $\vect{g}^i$ and $\vect{g}^j$ is
\begin{equation}
  \eta_{ij} = \arccos \left( \vect{g}^i \cdot \vect{g}^j \right) \ .
\end{equation}
Using equation~\ref{eq:generatormat}, we can express the $S(S-1)/2$
side lengths of the convex polytope explicitly in terms of the
interaction matrix:
\begin{equation}
  \displaystyle
  \eta_{ij} = \arccos \left( \frac{\sum_k A_{ki}A_{kj} }{ \sqrt{\sum_{k} A_{ki}A_{ki}\sum_{l} A_{lj}A_{lj}} }  \right)  \ .
  \label{eq:sidemat}
\end{equation}
We are interested in the distribution of the side lengths, and in
particular in its heterogeneity.  In the following section we will
calculate these quantities for random matrices.

\subsection{\ The distribution of side lengths in random matrices}

In this section we obtain the distribution of sides length for large
random matrices, whose entries are distributed accordingly to an
arbitrarily bivariate distribution.

We assume that the diagonal elements of $\mat{A}$ are all equal to
$-d$ (this hypothesis can be easily generalized), while the
offdiagonal pairs $(A_{ij},A_{ji})$ are random variables with
distribution $q(x,y)$. Our goal is to find the distribution of the
side lengths $\eta$ in the large $S$ limit, defined as
\begin{equation}
  \begin{split}
    P(\eta) =& \lim_{S \to \infty} \frac{1}{S(S-1)} \sum_{i \neq j}
    \int \prod_{m>n} \Bigl( \ud A_{mn} \ud A_{nm} q(A_{mn},A_{nm})
    \Bigr) \\ \times & \ \delta\left( \eta - \arccos \left(
        \frac{\sum_k A_{ki}A_{kj} }{ \sqrt{\sum_{k} A_{ki}A_{ki}\sum_{l} A_{lj}A_{lj} }  } \right) \right) ,
  \end{split}
\end{equation}
%\rem{Explain this definition a bit?}
Since we are summing over all $i$
and $j$, and all the rows are identically distributed, we can remove
the sum and consider just two rows:
\begin{equation}
  \begin{split}
    P(\eta) =& \lim_{S \to \infty} \int \prod_{m>n} \Bigl( \ud A_{mn}
    \ud A_{nm} q(A_{mn},A_{nm}) \Bigr) \\ \times & \ \delta\left( \eta
      - \arccos \left( \frac{\sum_k A_{k1}A_{k2} }{ \sqrt{\sum_{k}
            A_{k1}A_{k1}\sum_{l} A_{l2}A_{l2}} } \right) \right) ,
  \end{split}
\end{equation}
Since we are interested in the large $S$ limit, we have that
\begin{equation}
  \begin{split}
    \sum_{k} A_{k1}A_{k1} =& A_{11} + \sum_{k> 1} (A_{k1})^2 \approx
    -d + (S-1) \int \ud x \ud y \ q(x,y) \ x^2 \\ =& -d + (S-1) (
    E_1^2 + E_2^2 ) ,
    \label{eq:sideden}
  \end{split}
\end{equation}
where $E_1$ and $E_2$ are the first and second marginal moments of $q$
(equations~\ref{eq:E1def} and~\ref{eq:E2def}). Let us call this
quantity $Z$. In this limit we therefore obtain
\begin{equation}
  \begin{split}
    P(\eta) =& \lim_{S \to \infty} \int \prod_{m>n} \Bigl( \ud A_{mn}
    \ud A_{nm} q(A_{mn},A_{nm}) \Bigr) \
    \delta\left( \eta - \arccos \left( \frac{\sum_k A_{k1}A_{k2} }{Z  }  \right)  \right) \\
    =& \lim_{S \to \infty} \int \prod_{m>n} \Bigl( \ud A_{mn} \ud
    A_{nm} q(A_{mn},A_{nm}) \Bigr) \
    Z \ |\sin(\eta)| \ \delta\left( Z \cos(\eta) - \sum_k A_{k1}A_{k2} \right) \\
    =& \ Z |\sin(\eta)| \lim_{S \to \infty} \int \prod_{m>n} \Bigl(
    \ud A_{mn} \ud A_{nm} q(A_{mn},A_{nm}) \Bigr) \
    \delta\left( Z \cos(\eta) - \sum_k A_{k1}A_{k2} \right) \\
    =& \ Z |\sin(\eta)| \lim_{S \to \infty} \int \prod_{m>n} \Bigl(
    \ud A_{mn} \ud A_{nm} q(A_{mn},A_{nm}) \Bigr) \\ & \times
    \delta\left(Z \cos(\eta) - A_{11} A_{21} - A_{22} A_{12} - \sum_{k>2} A_{k1} A_{k2} \right) \\
    =& \ Z |\sin(\eta)| \lim_{S \to \infty} \int \prod_{m>n} \Bigl(
    \ud A_{mn} \ud A_{nm} q(A_{mn},A_{nm}) \Bigr) \\ & \times
    \delta\left( Z \cos(\eta) +d ( A_{12} + A_{21} ) - \sum_{k>2} A_{k1} A_{k2} \right) \\
    =& \ Z |\sin(\eta)| \int \ud t \int \ud s \int \ud A_{12} \ud
    A_{21} q(A_{12},A_{21}) \delta( t - A_{12} - A_{21} ) \\ & \times
    \int \prod_{k > 2} \ud A_{k1} \ud A_{k2} q(A_{k1}) q(A_{k2})
    \delta\left( s - \sum_{k>2} A_{k1} A_{k2} \right) \\ & \times
    \delta\left( Z \cos(\eta) +d t - \sum_{k>2} A_{k1} A_{k2} \right) \\
    =& \ Z |\sin(\eta)| \int \ud t \int \ud s \int \ud x \ud y \
    q(x,y) \delta( t - (x + y) ) \\ & \times \int \left(\prod_{k =
        1}^{S-2} d z_k d w_k q(z_k) q(w_k) \right) \delta\left( s -
      \sum_{k = 1}^{S-2} z_k w_k \right) \delta( Z \cos(\eta) +d t - s
    ) \ ,
  \end{split}
\end{equation}
where $q(z)$ is the marginal distribution of $q(x,y)$:
\begin{equation}
  q(z) = \int \ud x \ q(x,z) = \int \ud x \ q(z,x) .
  \label{eq:marginal}
\end{equation}
We can introduce the distribution of the sum:
\begin{equation}
  q_{s}(t) = \int \ud x \ud \ y q(x,y) \delta( t - (x + y)  ) .
\label{eq:distrsum}
\end{equation}
The term
\begin{equation}
  \int \left(\prod_{k = 1}^{S-2} \ud z_k \ud w_k \ q(z_k) q(w_k) \right) \ \delta\left( s - \sum_{k = 1}^{S-2} z_k w_k \right)
\end{equation}
is the distribution of a sum of $S-2$ uncorrelated random
variables. These random variables are the product $zw$ of two random
variables whose distribution is $q$. Since the second moment of $q(x)$
is finite, the central limit theorem holds and this distribution
converges, in the large $S$ limit, to a Gaussian distribution with
mean
\begin{equation}
  S\int \ud x \ud y \ q(y) q(x) \ xy = S E_1^2
\end{equation}
and variance
\begin{equation}
  S \left( \int \ud x \ud y \ q(y) q(x) \ (xy)^2 - E_1^2 \right) = S E_2^4 \ .
\end{equation}
We have therefore
\begin{equation}
  \begin{split}
    P(\eta) =& \ Z |\sin(\eta)| \int \ud t \ud s \ q_s(t)
    \frac{\exp\left( \frac{-(s-SE_1^2)^2}{2 S E_2^4}
      \right)}{\sqrt{2S\pi}E_2^2} \delta( Z \cos(\eta) +d t - s ) = (
    S ( E_1^2 + E_2^2 ) - d ) \\ & \times \frac{ |\sin(\eta)| }{
      \sqrt{2S\pi}E_2^2 } \int \ud t \ q_s(t) \exp \left(
      -\frac{\big(S ( E_1^2 + E_2^2 ) \cos(\eta) - d \cos(\eta)-SE_1^2
        + d t \big)^2}{2 S E_2^4} \right) \ .
    \label{eq:sidefinal}
  \end{split}
\end{equation}
The distribution of $\eta$ is not universal as it depends on $q_s(t)$,
which depends on the distribution of the coefficients. On the other
hand, the dependence is explicit, and it is possible to calculate
$P(\eta)$ for any distribution $q(x,y)$.

We show explicitly the case of $q(x,y)$ being a bivariate normal
distribution, i.e.,
\begin{equation}
  q(x,y) = \frac{1}{2 \pi E_2^2 \sqrt{1-E_c^2}} \exp \left(  -
    \frac{(x-E_1)^2 + (y-E_1)^2- 2E_c(x-E_1)(y-E_1)}{2 E_2^2}
  \right) \ .
  \label{eq:normalbiv}
\end{equation}
In this case $q_s(t)$ is a normal distribution, and can be obtained
from eq~\ref{eq:distrsum}
\begin{equation}
  \begin{split}
    q_s(t) & = \frac{1}{2 \pi E_2^2 \sqrt{1-E_c^2}} \int \ud y \exp
    \left( - \frac{(t-y-E_1)^2 + (y-E_1)^2- 2E_c(t-y-E_1)(y-E_1)}{2
        E_2^2}
    \right) \\
    & = \exp\left(-\frac{(1-E_c) (t-2 E_1)^2}{4 E_2^2} \right)
    \frac{1}{2 \sqrt{\pi } E_2 (1+E_c) \sqrt{1-E_c } } \ .
    \label{eq:normalbivsum}
  \end{split}
\end{equation}
Substituting into equation~\ref{eq:sidefinal}, we see that $P(\eta)$
has the form of a convolution of two Gaussians, and turns out to be
equal to
\begin{equation}
  \begin{split}
    P(\eta) = \frac{ |\sin(\eta)| }{ \sqrt{2\pi} \, \mathrm{var}(\cos(\eta))
    } \exp \left( - \frac{\big(\cos(\eta) - \langle \cos(\eta) \rangle
        \big)^2}{2 \, \mathrm{var}(\cos(\eta))} \right) \ .
    \label{eq:sidefinal_normal}
  \end{split}
\end{equation}
The mean $\langle \cos(\eta) \rangle$ and variance
$\mathrm{var}(\cos(\eta))$ will be computed in the next section in the
most general case of an arbitrary interaction distribution.

\subsection{\ Moments for random matrices}

As explained in the previous section, the distribution of the side
lengths is not a universal quantity, as it depends on the distribution
of interaction strengths. In this section we compute the mean and the
variance in the general case, showing that they depends only on $E_1$,
$E_2$ and $E_c$.

Here and in the main text we do not report the moments of the side
length $\eta$, but the moments of its cosine.  The cosine of the side
length measures the overlap between two rows of the interaction matrix
(or the scalar product of two generators of the convex polytope). As
its value gets close to one, the side length approaches zero.

Starting from equation~\ref{eq:sidemat}, we have that
\begin{equation}
  \big<\cos(\eta)\big> = \frac{1}{S(S-1)} \sum_{i \neq j} \cos(\eta_{ij}) =
  \frac{1}{S(S-1)} \sum_{i \neq j} \left( \frac{\sum_k A_{ik}A_{jk} }{ \sqrt{\sum_{k} A_{ik}A_{ik}\sum_{l} A_{jl}A_{jl}} }  \right)  ,
  \label{eq:sidematmean}
\end{equation}
Since we are interested in the large $S$ limit, we can write the
denominator as in equation~\ref{eq:sideden} and obtain
\begin{equation}
  \big<\cos(\eta)\big> = 
  \frac{1}{S(S-1)} \sum_{i \neq j} \left( \frac{\sum_k A_{ik}A_{jk} }{ -d + S(E_1^2 + E_2^2) }  \right) \ ,
  \label{eq:sidematnorm}
\end{equation}
and then
\begin{equation}
  \big<\cos(\eta)\big> = 
  \frac{1}{S(S-1)} \sum_{i \neq j} \left( \frac{ A_{ii}A_{ji} + A_{ij}A_{jj} +  \sum_{k\neq i, j} A_{ik}A_{jk} }{ -d + S(E_1^2 + E_2^2) }  \right) \ .
  \label{eq:sidematnorm2}
\end{equation}
In the large $S$ limit, this becomes
\begin{equation}
  \big<\cos(\eta)\big> = 
  \frac{ -2dE_1 + S E_1^2 }{ -d + (S-2)(E_1^2 + E_2^2) }
\label{eq:sidematmeanfin}
\end{equation}
to leading order in $S$.

In a similar way, we can write the second moment as
\begin{equation}
  \big<\cos(\eta)^2\big> = \frac{1}{S(S-1)} \sum_{i \neq j} \cos(\eta_{ij})^2 =
  \frac{1}{S(S-1)} \sum_{i \neq j} \left( \frac{\sum_k A_{ik}A_{jk} }{ \sqrt{\sum_{k} A_{ik}A_{ik}\sum_{l} A_{jl}A_{jl}} }  \right)^2  .
  \label{eq:sidevar}
\end{equation}
In the large $S$ limit we obtain
\begin{equation}
  \begin{split}
    \big< & \cos( \eta)^2\big> = \frac{1}{S(S-1)} \sum_{i \neq j}
    \frac{ \Bigl( \sum_k A_{ik}A_{jk} \Bigr)^2 }{ \Bigl( -d + S(E_1^2
      + E_2^2) \Bigr)^2 } =
    \frac{1}{S(S-1)} \sum_{i \neq j}  \frac{ \sum_k \sum_l A_{ik}A_{jk}A_{il}A_{jl}  }{ \Bigl( -d + S(E_1^2 + E_2^2)  \Bigr)^2 } \\
    & = \frac{1}{S(S-1)} \sum_{i \neq j}  \frac{ \big( A_{ii}A_{ji} + A_{ij}A_{jj} +  \sum_{k\neq i,j} A_{ik}A_{jk} \big) \big( A_{ii}A_{ji} + A_{ij}A_{jj} +  \sum_{l\neq i, j} A_{il}A_{jl} \big) }{ \Bigl( -d + S(E_1^2 + E_2^2)  \Bigr)^2 } \\
    & = \frac{1}{S(S-1)} \sum_{i \neq j} \frac{ d^2 (A_{ij}+A_{ji})^2
      - 2 d (A_{ij}+A_{ji}) \sum_{k\neq i,j} A_{ik}A_{jk} + (
      \sum_{k\neq i, j} A_{ik}A_{jk} )^2 }{ \Bigl( -d + S(E_1^2 +
      E_2^2) \Bigr)^2 } \ .
    \label{eq:sidevar2}
  \end{split}
\end{equation}
We can compute the averages of the different terms, obtaining
\begin{equation}
  \begin{split}
    \frac{1}{S(S-1)} \sum_{i \neq j} (A_{ij}+A_{ji})^2 = & \
    \frac{1}{S(S-1)} \sum_{i \neq j} (A_{ij}^2+A_{ji}^2 + 2
    A_{ji}A_{ji} ) \\ = & \ 2(E_1^2+E_2^2) + 2 ( E_c E_2^2 + E_1^2 ) =
    4 E_1^2 + 2 (1+E_c)E_2^2 \ ,
  \end{split}
\end{equation}
\begin{equation}
  \begin{split}
    \frac{1}{S(S-1)} \sum_{i \neq j} (A_{ij}+A_{ji}) \sum_{k\neq i
      \neq j} A_{ik}A_{jk} & = \frac{1}{S(S-1)} \sum_{i \neq j}
    (A_{ij}+A_{ji}) (S-2) E_1^2 \\ & = 2 (S-2) E_1^3 \ ,
  \end{split}
\end{equation}
and
\begin{equation}
  \begin{split}
    \frac{1}{S(S-1)} \sum_{i \neq j} \left( \sum_{k\neq i \neq j}
      A_{ik}A_{jk} \right)^2 & =
    \frac{1}{S(S-1)} \sum_{i \neq j} \sum_{k\neq i ,j}\sum_{l\neq i,j} A_{ik}A_{il} A_{jk} A_{jl} \\
    & = \frac{1}{S(S-1)} \sum_{i \neq j} \sum_{k\neq i ,j} \left(
      \sum_{l\neq i,j,k} ( A_{ik}A_{il} A_{jk} A_{jl} ) + A_{ik}^2
      A_{jk}^2
    \right) \\
    & = (S-2)(S-3)E_1^4 + (S-2)(E_1^2+E_2^2)^2 \ .
    \label{eq:sidevar4}
  \end{split}
\end{equation}
We finally get that, in the large $S$ limit,
\begin{equation}
  \mathrm{var}(\cos(\eta)) = \big<\cos(\eta)^2\big> - \big<\cos(\eta)\big>^2 = \frac{ 2 d^2 (1+E_c) E_2^2 + S(E_2^2+E_1^2)^2 - S E_1^4 }{ \left( -d + S(E_1^2 + E_2^2)  \right)^2 }  \ .
  \label{eq:sidevarfinal}
\end{equation}

\section{\ Side heterogeneity for different structures and empirical networks}

\begin{figure}
  \begin{center}
    \includegraphics[width =
    0.7\linewidth]{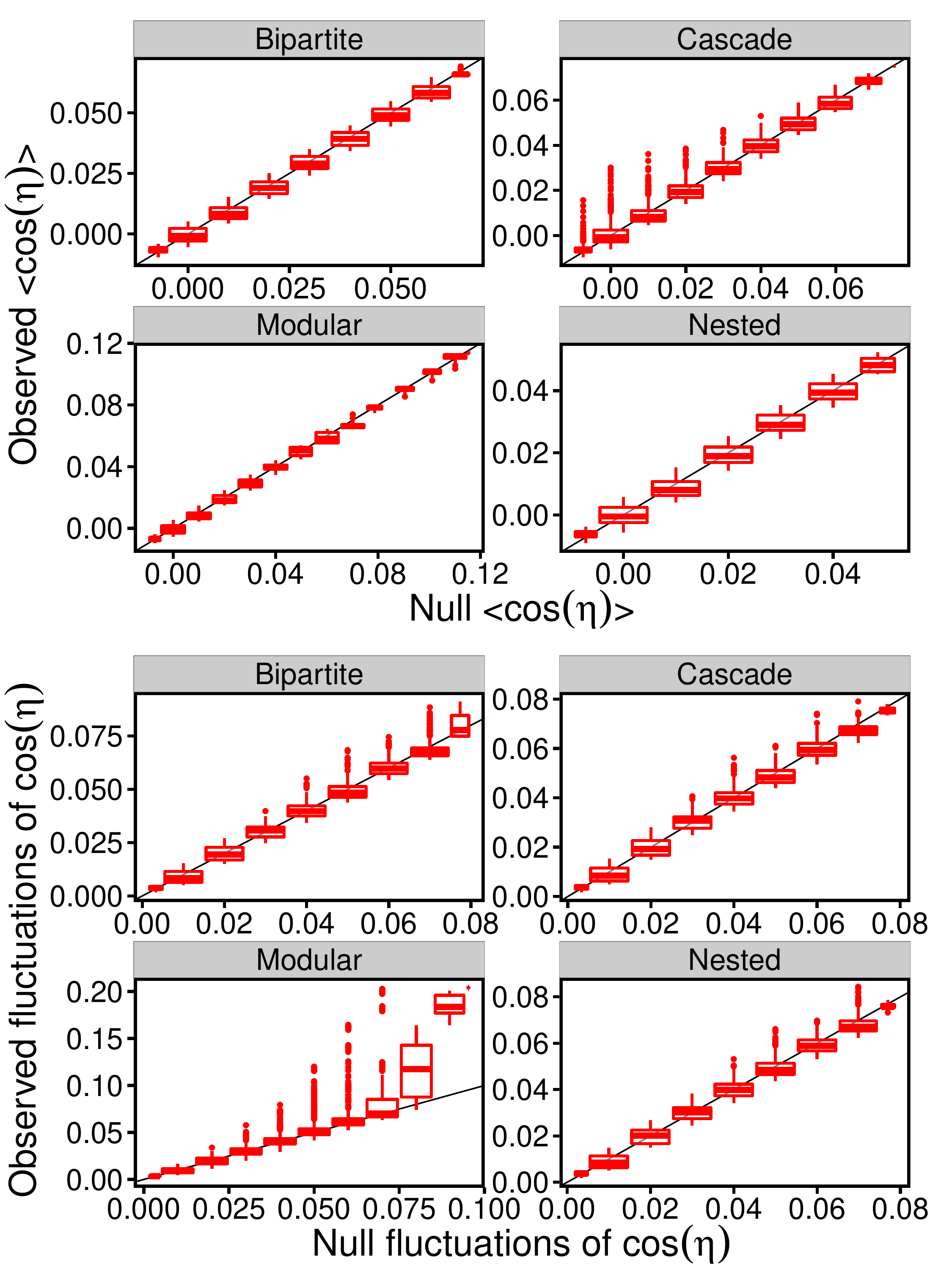}
  \end{center}
  \caption{We measure the effect of non random structures of mean and variance of side lengths. With the exception of the cascade model,
    all the structures considered do not have an important effect on $\langle \cos(\eta) \rangle$. On the other side, all the non-random structures
    considered have a positive effect on the fluctuations of $\cos(\eta)$.
    All the networks considered had a connectance $C=0.2$.
}
  \label{fig:sidestruct}
\end{figure}

In figure~\ref{fig:sidestruct}  we considered the effect of four nonrandom structures
on the mean and variance of the side lengths. The
interaction strengths were drawn from a normal distribution with given
mean, variance, and correlation.  For some structures we considered
multiple interaction types and therefore multiple means (one positive
and one negative), in which case the coefficient of variation of the
interactions and the correlation was constant and independent of the
mean. Networks were parametrized as explained in section~\ref{sec:empnet}.

\begin{itemize}
\item \textbf{Modular}. In this case we considered interaction
  matrices with a perfect block structure (to generate figure 3 we
  considered four blocks of equal size).
\item \textbf{Bipartite}. In this case we considered an interaction
  matrix with two bipartite blocks of equal size. The mean interaction
  of the offdiagonal blocks was set to be negative, while the one of
  the in-diagonal blocks was positive.
\item \textbf{Nested}. The interaction matrix had a bipartite
  structure. The diagonal blocks had a random structure with negative
  mean interaction strength. In the offdiagonal blocks, we consider a
  connectance equal to one half and we built a perfectly nested
  matrix. The mean interaction strength was positive in the
  offdiagonal blocks.
\item \textbf{Cascade}. We build a matrix using the cascade model, and
  parameterize it with a positive and a negative mean depending on the
  role of the species in the interaction.
\end{itemize}

\begin{figure}
  \begin{center}
    \includegraphics[width =
    0.9\linewidth]{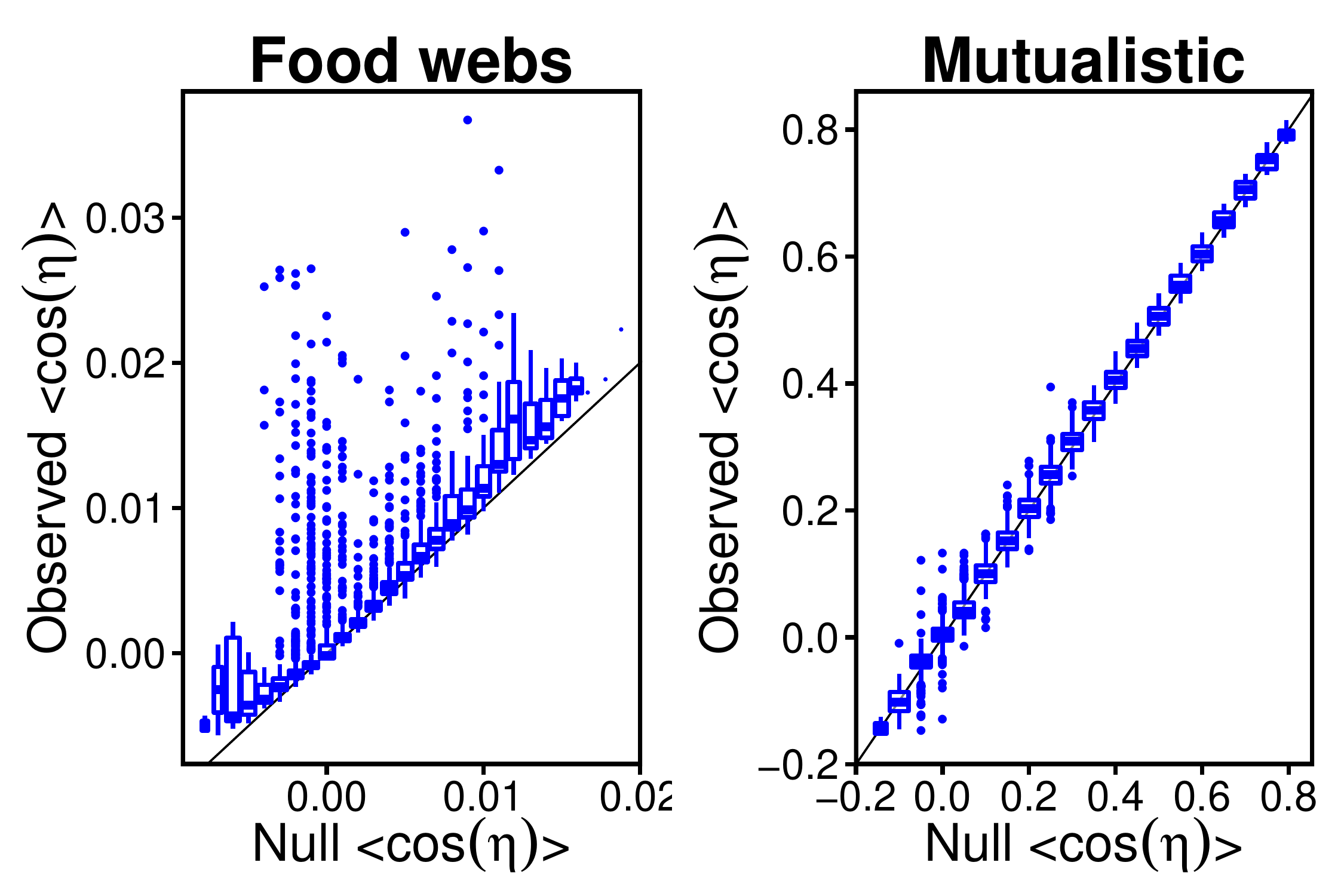}
  \end{center}
  \caption{Comparision between $\big<\cos(\eta)\big>$ obtained for empirical networks and its null expectation for empirical food webs and
    mutualistic networks. This figure was realized with the same parametrization of figure 3 of the main text and as described in section~\ref{sec:empnet}.
}
  \label{fig:sidemeanemp}
\end{figure}

In the case of empirical structures,
figure 3 of the main text, was obtained considering the same networks and the same parameterizations
considered in section~\ref{sec:empnet}.  We compared $\mathrm{var}(\cos(\eta))$ with the values
expected in the random case. Figure~\ref{fig:sidemeanemp} shows the comparison between $\big<\cos(\eta)\big>$ obtained for empirical networks
with the null prediction. Its value is well predicted by the null expectation for mutualistic networks, while the null expectations
underestimates this value for food webs. This is consistent with the fact that the size of feasibility domain of random networks is larger that
the one of empirical networks.

\section{\ Feasibility domain for $S=3$}
\label{sec:feas3D}

\begin{figure}
  \begin{center}
    \includegraphics[width =
    0.9\linewidth]{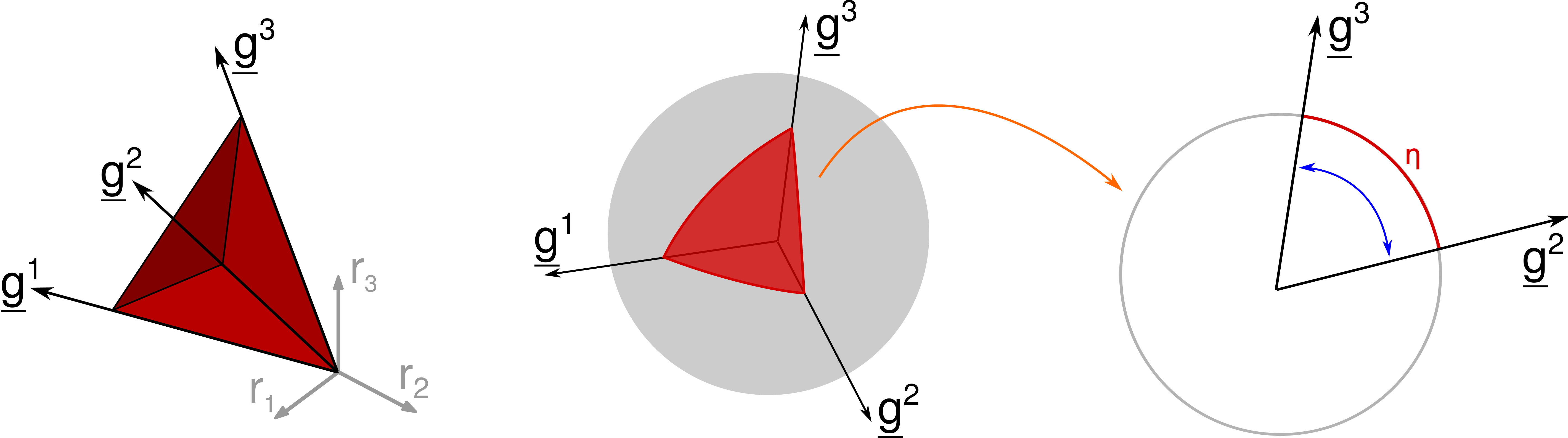}
  \end{center}
  \caption{Convex polyhedral cone and its section on a sphere. Left:
    the feasibility domain is a convex polyhedral cone, which is
    completely determined by its $S$ generators (when $S=3$ we have
    $3$ generators $\vect{g}^1$, $\vect{g}^2$, and
    $\vect{g}^3$). Center: since we consider a linear equation we
    can focus the analysis only on the intersection between the convex
    polyhedral cone and the unit sphere's surface, which in three
    dimensions results in a spherical triangle. Right: each side of
    the convex polyhedral cone can be determined from a pair of
    generators as an arc $\eta$ of the sphere's surface. Since we are
    considering the unit sphere, the arc length $\eta$ is equal to the
    angle between the two generators.}
  \label{fig:fig3d}
\end{figure}

When $S=3$, it is possible to visualize in three dimensions a convex
polyhedral cone and the feasibility domain~\cite{Svirezhev1978}. In figure~\ref{fig:fig3d}
we show a convex polyhedral cone in three dimensions and its
generators.

An important feature of convex polyhedral cones is that if
$\vect{r}$ belongs to the cone, then so does $c\vect{r}$ for
any positive constant $c$. As explained in section~\ref{sec:geomprop},
this is a consequence of the linearity of
equation~\ref{eq:dynmodel_LV}. It is relevant therefore to limit our
analysis to the growth rate vectors on the unit sphere, i.e., to
vectors $\vect{r}$ such that
\begin{equation}
  \| \vect{r} \| = \sqrt{ r_1^2 + r_2^2 + r_3^2 } = 1 \ .
  \label{eq:3dnorm}
\end{equation}
When we consider the vector in the feasibility domain on the surface
of a unit sphere we obtain the areas of figure 1 in the main text. In
this case, the quantity $\Xi$ is the area of the triangle, while the
side lengths are the three sides of the triangle. Note that the
polygon is not a triangle (as it lies on a sphere), but rather a
spherical triangle. Its sides are arcs of a circumference, while its
corners are identified by the three generators of the convex
polyhedral cone.

In the $S=3$ case it is possible to obtain a closed expression for the area $\Xi$ \cite{Gourion2010}:
\begin{equation}
  \Xi = \frac{8}{\pi} \arctan \Bigl( \frac{ |\det(\mat{G})| }{1 + \vect{g}^1\cdot \vect{g}^2 + \vect{g}^2\cdot \vect{g}^3 + \vect{g}^1\cdot \vect{g}^3} \Bigr) 
  + \Theta \Bigl(
  - 1 - \vect{g}^1\cdot \vect{g}^2 - \vect{g}^2\cdot \vect{g}^3 - \vect{g}^1\cdot \vect{g}^3
  \Bigr) \ ,
  \label{eq:3dvolume}
\end{equation}
where the second term adds one to the first term when the argument of
the arctangent is negative, while the matrix $\mat{G}$ is defined as
\begin{equation}
  G_{ij} = \vect{g}^i \cdot \vect{g}^j \ .
\end{equation}
Equation~\ref{eq:3dvolume} can be expressed directly in terms of the
matrix $\mat{A}$ using equation~\ref{eq:generatormat}.

\section{\ Nonlinear per capita growth rates}
\label{sec:nonlin}

In general, the effect of a species on the per capita growth rate of
other species is not linear. Equation~\ref{eq:dynmodel_LV} assumes
this to be linear and the results presented in this paper were
obtained under this assumption.  Nonlinearity of the per capita growth
rates can be thought of as a dependence of the interaction matrix
$\mat{A}$ on $\vect{n}$:
\begin{equation}\label{eq:dynmodel_LVnl}
  \displaystyle
  \frac{\ud n_i}{\ud t} = n_i \left( r_i + \sum_{j=1}^S A_{ij}(\vect{n}) n_j \right) 
  \ .
\end{equation}
For instance, in the case of predator-prey interactions with a Holling
type II functional response, it would have the form
\begin{equation}\label{eq:mat_LVnl}
  \displaystyle
  A_{ij}(\vect{n}) = \frac{A^0_{ij}}{1+\sum_j h_{ij} A^0_{ij} n_j }
  \ ,
\end{equation}
where the $h_{ij}$ are the handling times.

The presence of nonlinearity has strong consequences for both
feasibility and stability. It is no longer possible to disentangle
feasibility and stability with a simple condition on $A^0_{ij}$. This
means that feasibility will depend not only on the direction of
$\vect{r}$, but also on its length.

The results presented here are a necessary stepping stone for
assessing the feasibility of nonlinear systems. When the degree of
nonlinearity is small (e.g., $h_{ij}\approx 0$), one can use our
results, valid for the case $h_{ij} = 0$, to find the center of the
feasibility domain and the generators. One can then treat the
departure from $h_{ij} = 0$ as a small perturbation, and therefore,
instead of having to explore the full vast parameter space, use the
solution of the linear case as a starting point for numerical
calculations to converge on the actual, nonlinear feasibility
domain. On the other hand, in the limit of very large $h_{ij}$ values,
It is possible to show that the nonlinear form in
equation~\ref{eq:mat_LVnl} is approximately linear, and so again it is
possible to use our method. The effect of intermediate values of $h_{ij}$ on the feasibility domain is, however, still an open question.

\section*{References}
\bibliographystyle{unsrt}
\bibliography{StructStabSI}

\end{document}